\documentstyle[prb,aps,epsfig,eqsecnum]{revtex}

\draft

\begin{document}

\def\begeq{\begin{equation}}
\def\endeq{\end{equation}}
\def\begeqar{\begin{eqnarray}}
\def\endeqar{\end{eqnarray}}

\title{Quantum Brownian Motion on a Triangular Lattice and c=2 Boundary
Conformal Field Theory}
\author{Ian Affleck$^a$, Masaki Oshikawa$^b$, Hubert Saleur$^{c,}$\cite{Saleuraddress}}
\address{$^a$ Institute for Theoretical Physics, University of California, 
Santa Barbara, CA93106-4030, U.S.A. and \\ Canadian Institute for Advanced
Research  and Department of Physics and Astronomy, University of
British
Columbia,
Vancouver, B.C., Canada, V6T 1Z1 \\
$^b$Department of Physics, Tokyo Institute of Technology, \\
Oh-okayama,
Meguro-ku,
Tokyo 152-8551, Japan\\
$^c$ S.Ph.T., C.E.N. Saclay, 91191 Gif-sur-Yvette, Cedex, France}

\date{September 6, 2000}

\maketitle

\begin{abstract}
We study a single particle diffusing on a triangular lattice and interacting
with a heat bath, using boundary conformal field theory (CFT) and exact
integrability techniques.  We
derive a correspondence between the phase diagram of this problem and that
recently obtained for the 2 dimensional 3-state Potts model with a boundary.
Exact results are obtained on phases with intermediate mobilities.
These correspond to non-trivial
boundary states in a conformal field theory with 2 free bosons which we
explicitly construct for the first time.  These conformally invariant
boundary conditions are {\it not} simply  products of Dirichlet and Neumann
ones and
unlike those trivial boundary conditions, are not invariant under a
Heisenberg algebra.
\end{abstract}


\section{Introduction}
Conformal field theory (CFT) with boundaries
finds applications both to open string theory and to various quantum
impurity problems in condensed matter physics.  These generally describe
gapless bulk excitations  interacting with some localized degrees of
freedom.  In these problems the gapless bulk excitations can be
represented by a conformal field theory in (1+1) space-time dimensions,
often simply free bosons.  It is generally found that the boundary
dynamics renormalize, at low energies, to a conformally invariant
boundary condition (b.c.).

Quantum Brownian motion (QBM)
provides an intriguing example of such a problem. Here one considers a
heavy particle moving in a d-dimensional periodic potential and interacting
with a heat bath.   A  simplified
model for the heat bath\cite{Schmid,Guinea,Fisher}
 is an infinite set of harmonic oscillators coupled
linearly to the particle co-ordinate.  (This is the obvious generalization of
the Caldeira-Leggett model\cite{Caldeira} for a particle in a double well potential.)

When the oscillator spectral
weight vanishes {\it linearly} at low frequencies, corresponding to
ohmic dissipation, the set of oscillators may be represented by
a (1+1) dimensional quantum conformal field theory of free massless bosons living
on a fictitious half-line with the heavy particle at the origin.  The
number of bosonic fields required is given by the dimensionality of
the space in which the heavy particle is diffusing (normally three).
  The boson fields at the origin, $\vec \phi (0)$,
 correspond to the
momentum of the particle and the dual fields, $\vec{\tilde \phi}(0)$ to
the particle's position.  The boson CFT may be
regarded as compactified on the lattice on which the heavy particle
is diffusing.  The dimensionless compactification radius (scaled by
the coupling to the bath) is a crucial parameter for QBM.
The fictitious extra dimension is analogous to
the coordinate along the
string, $\sigma$, in open  string theory, while
the field $\vec{\phi}$
plays the role of the string coordinates in $D$
dimensional space time,
$X^\mu$. Conformal invariant boundary conditions
obtained in the dissipative
quantum mechanics framework have therefore
immediate applications to 
open string theory, and potential interpretations
in terms of D-branes.\cite{Callan}

Until recently it was generally believed that, depending on the strength
of the dissipation relative to the period of the potential, the particle
could only be in either  localized or freely diffusing phases corresponding
to Dirichlet (D):$ \vec {\tilde \phi} (0)=$ constant, or Neumann (N):
$\vec{ \phi }(0)=$ constant,
 boundary conditions in the CFT, respectively.  However, it
was recently shown\cite{YiKane} that, for certain lattices, other phases are possible
with intermediate mobility, in which the particle is neither perfectly
localized nor freely diffusing.  These correspond to non-trivial
conformally invariant boundary conditions in the free boson CFT
which are neither D nor N.  In fact, the existence of such phases
was shown earlier in another quantum impurity problem: tunneling though
a single impurity in a  quantum wire.\cite{Furusaki,Kane-Fisher}

While these interesting phases seem to cry out for a general boundary CFT
solution this has, so far, eluded us.  The problem of classifying
all (or all physically relevant) conformally invariant boundary conditions
in $c=2$ conformal field theory (2 free bosons) remains very much open.
In the much simpler case of 1 free boson, $c=1$, it is widely believed
that only D and N phases generally occur.  On the other hand, for
conformal field theories with a finite number of conformal towers, an
elegant way of generating non-trivial conformal boundary conditions is
provided by fusion.  This method can be used to generate an apparently
complete set of conformally invariant b.c.'s in, for example, the Ising
model\cite{Cardy} or 3-state Potts model.\cite{AOS}
In the case of Wess-Zumino-Witten (WZW)
models, one can obtain by fusion the boundary conditions that describe
the low temperature phases of various generalized
Kondo problems.\cite{Kondo-review}
However, the infinite number of conformal towers
present in $c=2$ CFT makes the generalization of the fusion technique
rather subtle.  Indeed, it is not even clear in general how to
understand the simple D and N b.c.'s this way.  Whether or not
some generalized notion of fusion will ultimately provide a complete
solution to this problem for $c=2$ is unclear at present.  Needless
to say, the generalization to larger values of c, in particular
c=3 corresponding to 3-dimensional QBM, is also an open problem.

The original argument\cite{Furusaki,Kane-Fisher} for
the existence of non-trivial phases came
from a study of the phase diagram in the quantum wire
problem.  This was studied\cite{Kane-Fisher}
 by  a type of ``$\epsilon$-expansion''.
By fine-tuning the compactification radius the
non-trivial fixed points can be moved arbitrarily close to either
N or D fixed points, allowing a perturbative expansion.  The other
solved case for c=2 was obtained in the context of QBM on a
triangular lattice.  For a special value of
dissipation strength (compactification radius) a
 beautiful exact mapping of the problem onto the critical point
 of the 3-channel Kondo problem\cite{Kondo-review}
   was obtained by Yi and Kane,\cite{YiKane}
  allowing use
 of results from that problem\cite{Kondo-review} based on
 boundary CFT in the $SU(2)_3$ WZW model.  Although
 it can be seen that non-trivial phases exist for a range of
 dissipation strength on the triangular lattice, no general
 solution has been found.

 Several years after the Caldeira-Leggett type model was proposed for
QBM\cite{Schmid} it was argued\cite{Itai,Zimanyi}
 that this model {\it does not} provide a valid
description of a heavy particle interacting with phonons or electron-hole
pairs.  In particular, the localized phase does not occur in physical models
of quantum Brownian motion.\cite{Castro}
Thus physical applications of the model
may only be to the quantum wire and related problems.  Nevertheless,
the Caldeira-Leggett type QBM model is perfectly consistent as an ultraviolet
(UV) regularization of a boundary conformal field theory.

 In this paper we re-examine the soluble triangular lattice
 QBM Yi-Kane model.  Our purpose is both to understand
 this fascinating system better and also to shed light
 on the general boundary CFT problem.  Rather than relating
 this problem to the $SU(2)_3$ WZW model we instead relate
 it to the 3-state Potts model.  This is done via a
 conformal embedding whereby the bulk degrees of freedom
 are represented by a direct sum of the 3-state Potts
 model, an Ising model and a tri-critical Ising model,
 satisfying:
 \begin{equation}
 c=c_{\rm \scriptsize Potts}+c_{\rm \scriptsize Ising}
 + c_{\rm \scriptsize tricritical}=4/5+1/2+7/10=2.\label{cadd}
 \end{equation}
 The $Z_3$ symmetry of the Potts model corresponds to the
 point group symmetry of the appropriate triangular lattice
 model.  This seems like a more natural formulation of
 the QBM problem since it does not contain an $SU(2)$ symmetry.
 The 4 conformally invariant boundary conditions
 in the Potts model: free, fixed, mixed and ``new'' are
 shown to correspond to the 4 phases that can occur in
 the QBM problem.  In this way it is possible to obtain
 the various fixed points by fusion in the Potts sector
 of the theory.  These four fixed points correspond to
 localized and freely diffusing fixed points, the non-trivial
 fixed point of Yi and Kane and one new fixed point not
 previously known.  The mobilities of these phases and
 also the groundstate entropies can be calculated from
 the fusion approach.  We also show that it is possible
 to use an alternative conformal embedding to Eq. (\ref{cadd})
 where the Ising and tricritical Ising sectors are replaced
 by a $Z_3^{(5)}$ conformal field theory, which is a sort
 of multicritical Potts model.\cite{FZ2}
 We demonstrate that some of
 the crossovers between these fixed points induced by
 relevant boundary operators can be studied exactly
 using integrability techniques.  This would allow, for
 example, exact calculations of universal quantities
 at all temperatures for a system which is crossing
 over between freely diffusing behavior at higher temperatures
 and localized or ``non-trivial'' behavior at $T=0$.  The
 integrability also helps to understand the phase diagram
 of the model by determining all RG flows from a given
 fixed point.

 In the next section we review the fusion approach  to
 boundary CFT and the standard D and N b.c.'s for free
 bosons.  We also show there that it is possible to obtain
 the D b.c. from the N b.c. by fusion with a twist field
 for the c=1 case of a single periodic boson, for certain
 rational radii. In Sec. III  we introduce the QBM problem and its
 connection with boundary CFT, with careful attention to
 the boundary conditions and the groundstate degeneracy.
 We show that this degeneracy scales as the size of the
 space on which the particle moves. We  derive, apparently
 for the first time, the connection between the particle
 co-ordinate and the value of the dual field at the
 origin. In Section IV we introduce a model of
 QBM on a triangular lattice which generalizes the model
 of Yi and Kane,  and conjecture its phase diagram,
 suggesting the analogy with the Potts model.  In Section
 V we discuss our conformal embedding and show how fusion
 in the Potts sector gives the various fixed points conjectured
 in Sec. IV.  We
 also calculate the  groundstate entropies and mobilities
 of these fixed points  and discuss the corresponding
 boundary states.  We also discuss some additional
 fixed points, that occur for more general Hamiltonians,
 and are obtained by fusion in the Ising, tricritical Ising
 or $Z_3^{(5)}$ sectors. In Section VI we discuss the integrable
 flows between the various fixed points. In Section VII we
 discuss generalizations of this model to $(n-1)$ bosons,
 corresponding to QBM in $(n-1)$ dimensions and also discuss
 the relationship with the n-channel Kondo problem.

\section{Boundary Conformal Field Theory}
As will be shown in the next section, QBM can be formulated in terms of
(1+1) dimensional massless bosons on the half-line, $x>0$,
with interactions only at the boundary,
$x=0$.  As such it is related to a number of other problems in particle
physics and condensed matter physics that have been successfully
studied using boundary conformal field theory (BCFT).  The basic
assumption in this approach is that the boundary dynamics renormalize,
at low energies, to an effective conformally invariant b.c.  A constructive
approach to classifying possible fixed points is then to attempt to
enumerate all possible conformally invariant b.c.'s corresponding to
a given critical bulk theory.  This has met with great success in
situations where the bulk theory can be formulated in terms of a finite
number of conformal towers, with the set of conformally invariant b.c.'s
being generally in one to one correspondence with the bulk conformal
towers.  In this case, the ``fusion'' technique is very useful in
constructing the conformally invariant b.c.'s.  Unfortunately, many
interesting quantum impurity problems apparently cannot be formulated
in terms of a finite number of bulk conformal towers, generally corresponding
to integer values of $c$.  Consequently the problem of studying the
fixed points in these cases remains very much open. As we will show in Sec. V,
for a special choice of parameters,  QBM on a triangular lattice can
be reduced, using a conformal embedding, to a finite number of conformal
towers.

In this section we review two quite different approaches to
boundary conformal field theory.  The fusion approach of Cardy
which has been very successful for CFT's with a finite number of
conformal towers and the straightforward D and N b.c.'s for
free bosons.  It is not known, in general, how to apply the
fusion approach for free bosons but we demonstrate that this is
possible for a single free boson at a special (discrete, infinite) set of
compactification radii.

\subsection{Fusion Approach to Boundary CFT}

We review briefly some important results due to Cardy on the case of
a finite number of conformal towers.

Precisely what is meant by a conformally invariant boundary
condition?  Without boundaries, conformal transformations are
analytic mappings of the complex plane: \begin{equation} z\equiv
\tau+ix,\end{equation} into itself:  \begin{equation} z\to
w(z).\end{equation}
 We may Taylor
expand an arbitrary conformal transformation around the origin:
\begin{equation} w(z) = \sum_n
a_n z^n,\label{CTexp}\end{equation} where the $a_n$'s are arbitrary
complex coefficients. 

They label the various generators of the
conformal group.  It is the fact that there is an infinite number
of generators (i.e. coefficients) which makes conformal invariance
so powerful in (1+1) dimensions.
 Now suppose that we have a boundary at $x=0$, the real axis.  At
best, we might hope to have invariance under all transformations
which leave the boundary fixed.  This implies the condition:
\begin{equation} w(\tau )^* = w(\tau )\end{equation}
where $\tau$ denotes imaginary time.  

We see that
there is still an infinite number of generators, corresponding to
the $a_n$'s of Eq. (\ref{CTexp}) except that now we must impose the
conditions: \begin{equation} a_n^* = a_n.\end{equation}  We have
reduced the (still $\infty$) number of generators by a factor of
1/2.  The fact that there is still an $\infty$ number of
generators, even in the presence of a boundary, means that this
boundary conformal symmetry remains extremely powerful.

Boundary conformal invariance implies  that the momentum density
operator, $T-\bar T$ vanishes at the boundary.  This amounts to a
type of unitarity condition.
Since $T(t,x)=T(t+x)$ and $\bar T(t,x)=\bar T(t-x)$, it follows that
\begin{equation} \bar T(t,x)=T(t,-x).\label{TTbar}\end{equation}
i.e. we may
regard $\bar T$ as the analytic continuation of $T$ to the negative
axis.  Thus instead of working with
left and right movers on the half-line we may work with left-movers
only on the entire line.

When the bulk theory contains a conserved current, $J$,
it is possible to impose additional symmetry requirements on
the boundary conditions of the form:
\begin{equation}
J(0)=\pm \bar J(0).\label{JJbar}\end{equation}
In the case of free bosons the generic symmetry consists of
a produce of U(1) current algebras (one for each boson) known as
a Heisenberg algebra.  It is then possible, more generally, to
impose an invariance of the form:
\begin{equation}
\vec J(0)={\cal R}\bar{\vec J}(0),\label{JJbarR}\end{equation}
where the vector $\vec J$ contains the U(1) currents and
${\cal R}$ is a rotation matrix. Since the chiral energy density 
operators, $T$ and $\bar T$ can be written in Sugawara form:
\begin{equation}
T={1\over 4\pi} \vec J^2, \ \ \bar T={1\over 4\pi} \bar{\vec J}^2,
\end{equation}
it follows that Eq. (\ref{TTbar}) is still obeyed. 
 One of the lessons of this
paper is that, while it is fairly straightforward to construct 
boundary conditions in free boson theories that are invariant
with respect to the full Heisenberg algebra, it is much more
difficult, apparently, to find the other boundary conditions
which  obey only the Virasoro condition of Eq. ({\ref{TTbar})
and not the additional constraints of Eq. (\ref{JJbarR}).

To exploit this symmetry, following Cardy, it is very convenient to
consider a conformally invariant system defined on a cylinder of
circumference $\beta$ in the $\tau$-direction (imaginary time)
and length $l$ in the
$x$ direction, with conformally invariant boundary conditions $A$
and $B$ at the two ends.  [See Figure (\ref{fig:cyl}).] From the
quantum mechanical point of view, this corresponds to a finite
temperature, $T=1/\beta$.  The partition function for this system
is: \begin{equation} Z_{AB} = \hbox{tr}e^{-\beta
H^l_{AB}},\label{ZAB1}\end{equation} where we are careful to label
the Hamiltonian by the boundary conditions as well as the length of
the spatial interval, both of which help to determine the
spectrum.  (We sometimes 
refer to this as the open string channel).  Alternatively, we may make 
a modular transformation,
$\tau \leftrightarrow x$.  Now the spatial interval, of length,
$\beta$, is periodic (the closed string channel). 
 We write the corresponding Hamiltonian as
$H^\beta_P$.  The system propagates for a time interval $l$ between
initial and final states $A$ and $B$.  Thus we may equally well
write: \begin{equation} Z_{AB} =
\langle A|e^{-lH^\beta_P}|B\rangle .\label{ZAB2}\end{equation}
Equating these two
expressions, Eq. (\ref{ZAB1}) and (\ref{ZAB2}) gives powerful
constraints which allow us to determine the conformally invariant
boundary conditions.

\begin{figure}
\epsfxsize=10 cm
\centerline{\epsffile{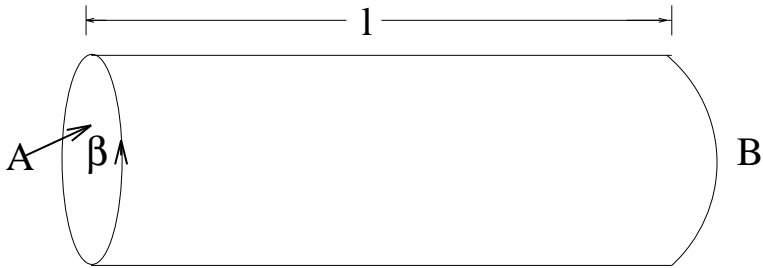}}
\caption{Cylinder of length $l$, circumference $\beta$ with
boundary conditions $A$ and $B$ at the two ends.}
\label{fig:cyl}
\end{figure}

Due to the condition of Eq. (\ref{TTbar}), in the purely
left-moving formulation, the energy momentum density,
$T$ is basically unaware of the boundary condition. Hence, in calculating
the spectrum of the system with boundary conditions $A$ and $B$
introduced above, we may regard the system as being defined
periodically on a torus of length $2l$ with left-movers only.  The
conformal towers of $T$ are unaffected by the boundary conditions,
$A$, $B$.  However, which conformal towers occur {\it does} depend
on these boundary conditions.  We introduce the characters of the
Virasoro algebra, for the various conformal towers:
\begin{equation} \chi_a (e^{-\pi \beta /l})\equiv \sum_ie^{-\beta
E^a_i(2l)},\end{equation} where $E^a_i(2l)$ are the energies in the
$a^{\hbox{th}}$ conformal tower for length $2l$.  i.e.:
\begin{equation} E^a_i(2l) = {\pi \over l}x_i^a-{\pi c\over 24
l},\end{equation} where the $x_i^a$'s correspond to the (left)
scaling dimensions of the operators in the theory and $c$ is the
conformal anomaly.

  The spectrum of $H^l_{AB}$ can only consist of
some combination of these conformal towers.  i.e.: \begin{equation}
Z_{AB}=\sum_an^a_{AB}\chi_a(e^{-\pi \beta
/l}),\label{naAB}\end{equation} where the $n^a_{AB}$ are some
non-negative integers giving the multiplicity with which the
various conformal towers occur.
 For minimal  conformal field theories, 
 $a$ runs over the finite set of  irreducible representations of the 
Virasoro algebra. For more complicated theories like the ones of
 interest in this paper, the corresponding set is infinite, 
and it is not always clear what kind or representations to expect on
 the 
right hand side of this equation. We stress that we are  interested in 
 boundary conditions restricted by conformal invariance only,
 so characters of generalized
 symmetry algebras are not relevant here.

  Importantly, only these
multiplicities depend on the boundary conditions, not the
characters, which are a property of the bulk left-moving system.
Thus, a specification of all possible multiplicities, $n^a_{AB}$
amounts to a specification of all possible boundary conditions
$A$.  The problem of specifying conformally invariant boundary
conditions has been reduced to determining sets of integers,
$n^a_{AB}$. For minimal conformal field theories, where the number
of conformal towers is finite, only a finite number of integers
needs to be specified.

Now let us focus on the boundary states, $|A\rangle $.  These must obey
the operator condition:
\begin{equation} [T(x)-\bar T(x)]|A\rangle =0 \ \
(\forall x).\end{equation}
(Note that, after the modular transformation, $x$ denotes the 
periodic co-ordinate.)
Imposing periodic boundary conditions along the boundary and 
Fourier transforming with respect to
$x$, this becomes:
\begin{equation} [L_n-\bar
L_{-n}]|A\rangle =0.\label{bstatecondition}\end{equation}

This implies that all boundary states,
$|A\rangle $ must be linear combinations of the ``Ishibashi
states'':\cite{Ishibashi}
\begin{equation} |a\rangle  \equiv
\sum_m|a;m\rangle \otimes \overline{|a;-m\rangle }.
\label{Ishibashi}\end{equation}
Here $m$ labels all states in the $a^{th}$ conformal tower.  The
first and second factors in Eq. (\ref{Ishibashi}) refer to the left
and right-moving sectors of the Hilbert Space. 
Here, $a$ belongs to the set of representations of the Virasoro
  algebra which appear simultaneously in the right and left sector of
  the theory with periodic boundary conditions. It is in general a
  smaller set than the one appearing in Eq. (\ref{ZAB2}).
   Thus we may write:
\begin{equation} |A\rangle  = \sum_a|a\rangle \langle a0|A\rangle .\end{equation} Here,
\begin{equation} |a0\rangle \equiv |a;0\rangle \otimes
\overline{|a;0\rangle }.\end{equation} (Note that while the states,
$|a;m\rangle \otimes \overline{|b;n\rangle }$ form a complete orthonormal set,
the Ishibashi states, $|a\rangle $ do not have finite norm.)  Thus,
specification of boundary states is reduced to determining the
matrix elements, $\langle a0|A\rangle $.  For minimal conformal field theories,
there is a finite number of such matrix elements.  Thus the
partition function becomes:
\begin{equation} Z_{AB} =
\sum_a\langle A|a0\rangle \langle a0|B\rangle \langle a|e^{-lH^\beta_P}|a\rangle .
\end{equation}
From the
definition of the Ishibashi state, $|a\rangle $ we see that:
\begin{equation} 
\langle a|e^{-lH^\beta_P}|a\rangle =\sum_me^{-2lE_m^a(\beta )},\end{equation}
the factor of 2 in the exponent arising from the
equal contribution to the energy from $T$ and $\bar T$.  This can be
written in terms of the characters:
\begin{equation}
\langle a|e^{-lH^\beta_P}|a\rangle =\chi_a (e^{-4\pi l/\beta}).
\end{equation}

 We are now in a position to equate these two expressions for
$Z_{AB}$: \begin{equation} Z_{AB} =
\sum_a\langle A|a0\rangle \langle a0|B\rangle \chi_a(e^{-4\pi
l/\beta})=\sum_an^a_{AB}\chi_a(e^{-\pi \beta
/l}).\label{ZAB12}\end{equation} This equation must be true for all
values of $l/\beta$.  It is very convenient to use the modular
transformation of the characters:\cite{Kac,Cardy2} \begin{equation}
\chi_a(e^{-\pi \beta /l})=\sum_bS^b_a\chi_b(e^{-4\pi
l/\beta}).\end{equation}  Here we refer to $S^a_b$ is as the matrix 
of modular transformations.   We thus obtain a set of
equations relating the multiplicities, $n^a_{AB}$ which determine
the spectrum for a pair of boundary conditions and the matrix
elements $\langle a0|A\rangle $ determining the boundary states: \begin{equation}
\sum_bS^a_bn^b_{AB} = \langle A|a0\rangle \langle a0|B\rangle .\label{Cardy}\end{equation} We
refer to these as Cardy's equations;
  they basically allow a
determination of the boundary states and spectrum.
 The problem is to find a
  set of boundary states $A$ (defined by the coefficients 
$\langle A|a0\rangle$) satisfying Cardy's equations, 
such that, for any pair $A,B$ in this set, the
$n_{AB}^b$ are non-negative integer coefficients, the identity
representation appearing at most once, $n_{AB}^0\leq 1$.  An
important point in deriving (\ref{Cardy}) is the linear independence 
of the characters. This property may actually not hold for
``non-diagonal'' theories. In such cases, there will
generally be some other quantum number that allows one to distinguish 
the corresponding operators, and still prove Eq. (\ref{Cardy}).
The study of these equations  has given rise to considerable
theoretical activity recently.\cite{PSS,BPPZ,FSI} In the case of
rational
conformal field theories, and with the additional key hypothesis of
{\sl completeness}, it has been shown that  finding sets of
 conformally invariant boundary conditions
is equivalent to finding integer valued representations of the fusion
algebra. The latter problem is under reasonable control, and as a
result, conformal boundary conditions for minimal models have for
instance 
been classified in Ref. (\onlinecite{BPPZ}).

In the present paper, we are interested in a more difficult problem:
in fact, since we look for boundary conditions respecting
only conformal 
invariance, the two boson problem has the complexity of an irrational
theory,
even for rational values of the radius. Very little is known about
this case, and we rely heavily on a not entirely systematic
but highly efficient method called fusion, which is inspired 
by Cardy's seminal paper,\cite{Cardy} and was used with success in 
the 
Kondo  problem.\cite{Kondo-review}

 Generally, boundary states
corresponding to trivial boundary conditions can be found by
inspection.  i.e., given $n^b_{AA}$ we can find $\langle a|A\rangle $.  We can
then generate new (sometimes non-trivial) boundary states by {\it
fusion}.  i.e. given any conformal tower, $c$, we can obtain a new
boundary state $|B\rangle $ and new spectrum $n^a_{AB}$ from the ``fusion
rule coefficients'', $N^c_{ab}$.  These non-negative integers are
defined by the operator product expansion (OPE) for (chiral)
primary operators, $\phi_a$.  In general the (OPE) of $\phi_a$ with
$\phi_b$ contains the operator $\phi_c$ $N^c_{ab}$ times.  We denote
by $B$ the new
boundary condition and boundary state, generated by fusion from 
the boundary condition $A$ by fusion with the operator $c$.
The partition function, $Z_{BD}$, with an arbitrary boundary condition
$D$ at the other end of the system, is given by the multiplicities:
given by a fusion with a primary field $c$ 
has multiplicities  given by the fusion rule
 coefficients $N^a_{bc}$ as
\begin{equation}
n^a_{DB} =  \sum_b N^a_{bc} n^b_{AD},
\label{nfusion}
\end{equation}
hence the name ``fusion construction''.
Because $N^a_{bc}$  are non-negative integers,
the new multiplicities $n^a_{DB}$
are also non-negative integers, and thus physical. Alternatively, in
the closed string channel, the new boundary state $|B>$ is defined through
\begin{equation}
\langle a0|B\rangle  = \langle a0|A\rangle {S^a_c\over S^a_0},
\label{bsfusion}
\end{equation}
where $0$ labels
the conformal tower of the identity operator.

Let us now  show
 that the boundary state~(\ref{bsfusion})  gives
the multiplicity(\ref{nfusion}), 
i.e. they satisfy Cardy's equation(\ref{Cardy}).
The right-hand side of Eq.  (\ref{Cardy}) becomes:
\begin{equation}
\langle D|a0\rangle \langle a0|B\rangle  = 
\langle D|a0\rangle \langle a0|A\rangle {S^a_c\over S^a_0}.
\end{equation}
The left-hand side becomes:
\begin{equation}
\sum_bS^a_bn^b_{DB} = \sum_{b,d}S^a_bN^b_{dc}n^d_{DA}.
\end{equation}
We now use a remarkable identity relating the matrix of modular
 transformations to
the fusion rule coefficients, known as the Verlinde
formula:\cite{Verlinde} \begin{equation}
\sum_bS^a_bN^b_{dc}={S^a_dS^a_c\over S^a_0}.\end{equation}  This
gives:
 \begin{equation} \sum_bS^a_bn^b_{DB} = {S^a_c\over
S^a_0}\sum_dS^a_dn^d_{DA}={S^a_c\over
S^a_0}\langle D|a0\rangle \langle a0|A\rangle =
\langle D|a0\rangle \langle a0|B\rangle ,\end{equation} proving that fusion
does indeed give a new solution of Cardy's equations. The
multiplicities, $n^a_{BB}$ are given by double fusion:
\begin{equation} n^a_{BB} =
\sum_{b,d}N^a_{bc}N^b_{dc}n^d_{AA}.\end{equation}   [Recall that
$|B\rangle $ is obtained from $|A\rangle $ by fusion with the primary operator
$c$.]  It can be checked that the Cardy equation with $A=B$ is then
obeyed.

As mentioned before, the key problem in boundary CFT is the construction of a complete
set of boundary states (and b.c.'s), that is the largest
possible set of
boundary states, $|A_i\rangle $ such that $Z_{A_iA_j}$ is a physical
partition function.
By physical we mean that the partition function  with any two
boundary conditions is always given by  non-negative integer multiplicities
$n^a_{A_i A_j}$.
Noting that any linear
combination of physical boundary states with non-negative integer coefficients,
\begeq \sum_i n_i|A_i\rangle ,\endeq
also gives a physical partition function in this sense, we see that
we must impose an additional condition to eliminate such states.
The lowest energy state with the same b.c. at both ends
of the system is independent of the b.c. corresponding to the
absolute finite size groundstate, with $x_i=0$.  (This follows from
making a conformal mapping to the half-plane and using the fact
that the identity boundary operator exists with any b.c.)
We may choose to impose
an additional condition that $Z_{A_i A_i}$ contains the zero energy
state exactly once, $n^0_{A_i A_i}=1$.
This eliminates linear combination
states.  Of course, once we have found a complete set, we may
always form linear combinations with non-negative integer coefficients.
Actually, the fusion construction sometimes gives such linear
combination states.
These can also arise from integrable flows\cite{Lesage,Chim,Recknagel}
and have
physical applications\cite{tricritical} in some cases.  (The existence
of extra dimension 0 boundary operators is often associated with
first order phase transitions.)

 In the case of minimal diagonal theories, conformal boundary
conditions
are in one to one correspondence with the set of Virasoro
representations.\cite{Cardy,BPPZ} The corresponding boundary states
can be generated by fusion
starting from the the boundary state $|\tilde 0\rangle $ defined by
its diagonal partition function, $n^a_{\tilde 0 \tilde 0}=\delta^{a0}$.

In more complicated cases, fusion has proven an invaluable tool.
 One of the key advantages of the
method is that one can use fusion with fields that are not in the
spectrum of the bulk theory. Indeed, as already explained above,  
while the
 the boundary state (in the closed
string channel) must be compatible with the bulk spectrum,  
the conformal towers appearing in the open string channel {\sl do not}
 have to be in
the bulk spectrum. In many cases, it is then possible 
to use not the fusion algebra of the original bulk theory but another
one,
still compatible with the modular transformation matrix, to generate
boundary states. For instance, in the case of the three state $(A_4,D_4)$ Potts
model, the characters appearing in the bulk partition function are a
subset of all the possible characters at $c={4\over 5}$. Using the
full matrix of modular transformations
 and fusion algebra for the associated diagonal
tetracritical Ising model $(A_4,A_5)$ led to the discovery of 
the new boundary condition in Ref. (\onlinecite{AOS}).
 We will discuss in the next
 section
how the fusion method can successfully be applied to the case of a free boson.

Before leaving this section, 
we should
emphasize that complete sets of boundary states form compatible
sets of states which may be overlapping.  For example, in the Ising
model the three states corresponding to spin up, spin down and free b.c.'s
are a complete mutually compatible set.  However, one could also
consider imposing an up or down b.c. on the {\it dual} spins in the Ising
model.    Such states should also be compatible with the free b.c.
but not with the fixed b.c.'s on the original (non-dual) spins.  This
non-compatibility is presumably related to the non-local nature of
the dual spin variables when expressed in terms of the original spins.
This implies that one couldn't define a physical partition
function with the original spins pointing up at one boundary and
the dual spins pointing up (or down) at the other.  In a physical
application one typically begins with a particular b.c. and then
wants to find all other ones which are compatible with it.  For instance,
we may be interested in adding arbitrary local interactions near the
boundary and finding all b.c.'s to which the system can renormalize.

\subsection{Dirichlet Neumann, and Rotated Boundary Conditions For
Free Bosons}

We review the simplest conformally invariant boundary conditions,
namely Dirichlet and Neumann boundary conditions and
some of their generalizations,
for multicomponent compactified free bosonic field theory.\cite{Callan1}
The Lagrangian density of the free boson theory is given by
\begin{equation}
        {\cal L} = \frac{1}{2} (\partial_{\mu} \vec{\phi})^2,
\end{equation}
where $\vec{\phi}$ is a $c$-component boson field.
We identify the boson field as
\begin{equation}
        \vec \phi \sim \vec \phi + 2 \pi \vec u
\end{equation}
where $\vec u$ is any vector in the compactification lattice
$\Gamma$, which is a $c$-dimensional Bravais lattice.

Let us consider the finite size system of length $l$, with 
periodic boundary conditions.
The standard canonical quantization gives
the canonical equal-time commutation relations
\begin{equation}
[ \phi^j (t, x) , \Pi^k (t,x') ] =
        i \delta^{jk} \sum_m \delta(x - x' - ml),
\end{equation}
where $\vec{\Pi} = \dot{\vec{\phi}}$.
Together with the equation of motion
$\partial_{\mu} \partial^{\mu} \vec{\phi} = 0$,
we obtain the mode expansion of the field $\vec \phi$:
\begin{equation}
\vec \phi(t,x) = \vec{\phi}_0 + \frac{2 \pi x}{l} \vec{u}
        + \frac{ \vec{p} t}{l}
+ \sum_{n=1}^{\infty} \left( \frac{\vec{a}_{nL}}{\sqrt{4 \pi n}}
        e^{-i \frac{2\pi n}{l} (x+t)} +\mbox{h.c.} \right)
+ \sum_{n=1}^{\infty} \left( \frac{\vec{a}_{nR}}{\sqrt{4 \pi n}}
        e^{i \frac{2\pi n}{l} (x - t)} +\mbox{h.c.} \right),
\end{equation}
where $\vec{u}$ belongs to $\Gamma$ and
$\vec{p}$ is the conjugate momentum of $\vec{\phi}_0$. (h.c.
denotes Hermitean conjugate.)
Because $\vec{\phi}_0$ should be identified as
$\vec{\phi}_0 \sim \vec{\phi}_0 + 2 \pi \Gamma$, the eigenvalues of
$\vec{p}$, which we label $\vec v$,
 belong to the lattice $\Gamma^*$ which is the dual of
$\Gamma$.  This is defined by the condition:
\begeq \vec u\cdot \vec v \in {\cal Z},\endeq
for all vectors $\vec u \in \Gamma$ and $\vec v \in \Gamma^*$.
Each component of the vector $\vec{a}_{nL}$ is an annihilation operator;
the $a^j_{nL}$ satisfies the commutation relation
$[ a^j_{nL} , {a^k}^{\dagger}_{mL} ] = \delta^{jk} \delta_{nm} $
and all the other commutators involving $a^j_{nL}$ vanish.
A similar definition applies to $\vec{a}_{nR}$, which commutes with
$a^j_{mL}$ and ${a^j}^{\dagger}_{mL}$.

Now we can decompose the field $\vec{\phi}$ into chiral components,
namely left-mover $\vec{\phi}_L$ and right-mover $\vec{\phi}_R$:
\begin{equation}
\vec{\phi}_L(x^+) = \frac{\vec{\phi}_0}{2} + \frac{\tilde{\vec{\phi}}_0}{2} +
\frac{1}{2l}(2 \pi \vec{u} + \vec{p}) x^+ +
\sum_{n=1}^{\infty}
\left( \frac{\vec{a}_{nL}}{\sqrt{4 \pi n}} e^{-i \frac{2\pi}{l} n x^+}
        +\mbox{h.c.} \right),
\label{eq:phiLexpand}
\end{equation}
where $x^+ = t + x$.
The dual field $\tilde{\vec{\phi}} = \vec{\phi}_L - \vec{\phi}_R$
has the mode expansion:
\begeq \tilde{\vec \phi}(t,x)=\tilde{\vec \phi}_0+{2\pi t\over l}\vec u
+{x\over l}\vec v+
\sum_{n=1}^{\infty} \left( \frac{\vec{a}_{nL}}{\sqrt{4 \pi n}}
        e^{-i \frac{2\pi n}{l} (x+t)} +\mbox{h.c.} \right)
-\sum_{n=1}^{\infty} \left( \frac{\vec{a}_{nR}}{\sqrt{4 \pi n}}
        e^{i \frac{2\pi n}{l} (x - t)} +\mbox{h.c.} \right).
\endeq
The chiral component of the energy-momentum tensor is
given by $ T(x^+) = (\partial_+ \vec{\phi})^2$.
The mode expansion of the energy-momentum tensor gives
the generators $L_m$ of the Virasoro algebra:
\begin{equation}
T(x^+) = \frac{2\pi}{l^2} \sum_m L_m e^{-i m x^+ \frac{2 \pi}{l}} .
\end{equation}
Using the mode expansion~(\ref{eq:phiLexpand}), we obtain
\begin{equation}
L_m = \frac{1}{2} \sum_l : \vec{\alpha}_{m-l,L} \cdot \vec{\alpha}_{lL} :,
\label{eq:Lmalpha}
\end{equation}
where $::$ denotes the normal ordering and
\begin{equation}
\vec{\alpha}_{nL} \equiv
\left\{ 
\begin{array}{cc} - i \sqrt{n} \vec{a}_{nL} & (n>0) \\
        \frac{1}{\sqrt{4 \pi}}(2 \pi \vec{u} + \vec{p}) & (n=0) \\
        + i \sqrt{n} \vec{a}^{\dagger}_{-nL} & (n<0).
        \end{array}
\right.   \end{equation}
The generators $\bar{L}_m$ for the other chirality are
given by
\begin{equation}
\bar{L}_m = \frac{1}{2} \sum_l
: \vec{\alpha}_{m-l,R} \cdot \tilde{\vec{\alpha}}_{lR} :,
\end{equation}
where
\begin{equation}
\vec{\alpha}_{nR} \equiv
\left\{ \begin{array}{cc}
        - i \sqrt{n} \vec{a}_{nR} & (n>0) \\
        \frac{1}{\sqrt{4 \pi}}( - 2 \pi \vec{u} + \vec{p}) & (n=0) \\
        + i \sqrt{n} \vec{a}^{\dagger}_{-nR} & (n<0).
        \end{array}
\right. 
\end{equation}
An oscillator vacuum $| \mbox{vac} \rangle$
of this theory,
which satisfies $a^j_{nL/R} | \mbox{vac} \rangle =0$,
is thus characterized by two sets of zero-mode quantum numbers
$\vec{u} \in \Gamma$ and $\vec{v} \in \Gamma^*$,
where $\vec{v}$ is the eigenvalue of the operator $\vec{p}$.
We will denote the oscillator vacuum with these quantum
numbers as $|(\vec{u},\vec{v}) \rangle$.

Now let us consider the possible conformally invariant
b.c.'s.
As was discussed in Section IIA, the conformal invariance
of the b.c., Eq. (\ref{bstatecondition}), implies that these
are linear combinations of the Ishibashi states, defined in
Eq. (\ref{Ishibashi}).  Of course, the primary states must
be contained within the Hilbert space of the system (with periodic
b.c.'s.)  A physical boundary state must satisfy
Cardy's equation~(\ref{Cardy}),
which gives a strong constraint on the coefficients of the Ishibashi
states.

The (multicomponent) free boson field theory, which we discuss
in the present paper, has an infinite number of primary fields.
The classification of the boundary states for this case
is much more difficult than that for
CFTs with a finite number of conformal towers discussed
in Section~IIA.
In fact, at present we are far from the complete classification.
Even for the single-component $c=1$ theory, for which
the Dirichlet/Neumann b.c. and their generalizations are generally
believed to form the complete set,
we know no complete proof of this.
Here we just present some simple boundary states for
a multicomponent free boson theory.
Some examples of more non-trivial boundary states for $c=2$
will be constructed
using a conformal embedding in Section~V.

An oscillator vacuum $| (\vec{u},\vec{v})\rangle$
is naturally a primary state with respect to the Virasoro algebra.
However, there are an infinite number of Virasoro primaries
which are not oscillator vacua.
The most general boundary states would include
linear combinations
of Ishibashi states based on these complicated Virasoro primaries.
Nevertheless, as we will show in the following, the simplest
boundary states can be written in terms of oscillator
vacua and creation operators of oscillator bosons.

A sufficient (but not necessary)
condition for the free boson theory to satisfy the conformal
invariance~(\ref{bstatecondition}) is given by
\begin{equation}
( \vec{\alpha}_{nL} - \vec{\alpha}_{-nR} ) | B \rangle =0,
\label{eq:bcinv-d}
\end{equation}
for all integer $n$.  This corresponds to the imposition of
invariance under the Heisenberg algebra, as discussed in the
previous subsection, near Eq. (\ref{JJbar}).
For $n=0$, it reads $\vec{u} |B \rangle= 0$.
It means that, to satisfy eq.~(\ref{eq:bcinv-d}), we can
use only the states built on the oscillator vacua with $\vec{u}=0$.
It can be shown that the condition~(\ref{eq:bcinv-d}) is
satisfied by the state
\begin{equation}
| ( \vec{0}, \vec{v} ) \rangle \rangle
= \exp{( - \sum_n \vec{a}^{\dagger}_{nL} \cdot\vec{a}^{\dagger}_{nR} )}
        | ( \vec{0} , \vec{v}) \rangle .
\label{eq:d-base}
\end{equation}
This might be regarded as an Ishibashi state with respect
to the $U(c)$ current algebra.
It is a linear combination of (an infinite number of) Ishibashi
states with respect to the Virasoro algebra.

While each state~(\ref{eq:d-base})
satisfies~(\ref{eq:bcinv-d}) and hence the conformal invariance,
it does not satisfy Cardy's consistency condition.
Let us consider a linear combination of~(\ref{eq:d-base})
\begin{equation}
| N(\vec{\phi}_0) \rangle =
g_N \sum_{\vec{v}} \exp{(i \vec{v} \cdot \vec{\phi}_0 )}
        | ( \vec{0}, \vec{v} ) \rangle \rangle,
\label{eq:d-state}
\end{equation}
for a constant $g_N$ and constant vector $\vec{\phi}_0$.
The diagonal partition function on the strip is given by
\begin{equation}
Z_{NN}(\tilde{q}) =
{g_N}^2 \left( \frac{1}{\eta(\tilde{q})}\right)^c \sum_{\vec{v}}
        \tilde{q}^{\frac{\vec{v}^2}{8\pi}},
\end{equation}
where the summation is over the the entire dual lattice $\Gamma^*$.
Modular transforming to the open string channel, it reads
\begin{equation}
Z_{NN}(q) =
(4 \pi)^{c/2} V_0(\Gamma) g_N^2
\left( \frac{1}{\eta(q)}\right)^c \sum_{\vec{u}} q^{2 \pi \vec{u}^2},
\label{eq:ZDDopen}
\end{equation}
where $V_0(\Gamma)$ is the volume of the unit cell of the
compactification lattice $\Gamma$.  We note that the volume
of the unit cell of the dual lattice is given by:
\begeq V_0(\Gamma^*)=1/V_0(\Gamma ).\endeq
The factor $q^{\vec{u}^2}/(\eta(q))^2$ is the character of the $U(c)$
current algebra, and is a superposition of the Virasoro
characters with non-negative integer coefficients.
Choosing
\begeq
g_N = \frac{1}{ (4 \pi)^{c/4}  \sqrt{V_0(\Gamma)}},
\label{g_N}
\endeq
the state~(\ref{eq:d-state}) satisfies Cardy's condition
and thus is a physical boundary state.
The constant $g_D$ actually
represents the (generally fractional)
``ground-state degeneracy'' due to the boundary~\cite{g-theorem}.
From the diagonal partition function~(\ref{eq:ZDDopen})
in the open string channel, we can read off the scaling dimensions
of boundary operators occurring with  the boundary condition.
Namely, the scaling dimensions of the boundary operators are given by
\begin{equation}
\Delta_N = 2 \pi \vec{u}^2 + \mbox{(non-negative integer)},
\label{DeltaN}\end{equation}
where $\vec{u}$ is an element of the lattice $\Gamma$.

The physical meaning of the boundary state
$|N(\vec{\phi}_0)\rangle$ turns out to be
the Dirichlet boundary condition on $\vec \phi$ namely
$\vec{\phi} = \vec{\phi}_0 $ at the boundary.
This can be checked by the calculation of the partition
function in the open string channel imposing this
boundary condition.  This b.c. is
equivalent to the Neumann boundary condition on the dual
field, $d\tilde{\vec \phi} /dx|_0=0$.  We label the state
with respect to the dual field for convenience in the following
section.
We note that it is not clear whether the N boundary state
is the only solution of the Cardy's condition
even within the linear combinations of the bosonic Ishibashi
states~(\ref{eq:d-base}).

A similar construction of the boundary state is possible,
starting from the condition
\begin{equation}
( \vec{\alpha}_{nL} + \vec{\alpha}_{-nR} ) | B \rangle =0,
\label{eq:bcinv-n}
\end{equation}
instead of~(\ref{eq:bcinv-d}).  Again we are imposing invariance
under the Heisenberg algebra, Eq. (\ref{JJbar}).
The bosonic Ishibashi state is given by
\begin{equation}
| ( \vec{u}, \vec{0} ) \rangle \rangle
= \exp{( + \sum_n \vec{a}^{\dagger}_{nL} \cdot \vec{a}^{\dagger}_{nR} )}
        | ( \vec{u} , \vec{0}) \rangle .
\label{eq:n-base}
\end{equation}
A physical boundary state is given
\begin{equation}
| D(\tilde{\vec{\phi}}_0) \rangle =
g_D \sum_{\vec{u}} \exp{(i \vec{u} \cdot \tilde{\vec{\phi}}_0 )}
        | ( \vec{u}, \vec{0} ) \rangle \rangle,
\label{eq:n-state}
\end{equation}
where the summation is over the entire lattice $\Gamma$
and
\begeq g_D = \pi^{c/4} \sqrt{V_0(\Gamma)}.\label{g_D}\endeq
The diagonal partition function reads, in the closed string channel,
\begin{equation}
Z_{DD}(\tilde{q}) =
{g_D}^2 \left( \frac{1}{\eta(\tilde{q})}\right)^c \sum_{\vec{u}}
\tilde{q}^{\frac{\pi \vec{u}^2}{2}},
\end{equation}
and in the open string channel
\begin{equation}
Z_{DD}(q) =
\left( \frac{1}{\eta(q)}\right)^c \sum_{\vec{v}} q^{\vec{v}^2/(2 \pi)},
\label{eq:ZNNopen}
\end{equation}
The scaling dimensions of the boundary operators are
\begin{equation}
\Delta_D = \frac{\vec{v}^2}{2\pi} + \mbox{(non-negative integer)},
\label{DeltaD}\end{equation}
where $\vec{v}$ is an element of the dual lattice $\Gamma^*$.
The physical meaning of this boundary state is
the Neumann boundary condition for the field $\vec{\phi}$,
or equivalently the Dirichlet boundary condition
for the dual field $\tilde{\vec{\phi}}$,
i.e. $\tilde{\vec{\phi}} = \tilde{\vec{\phi}}_0$ at the boundary.

In the simplest case, $c=1$,
\begeq
\Gamma = \{nR|n\in {\cal Z}\},\ \ \Gamma^*=\{ m/R|m\in {\cal Z}\}
\label{gammac=1} \endeq
where $R$ is the compactification radius.  In this case the degeneracies
are given by:
\begeq g_N=1/[\pi ^{1/4}\sqrt{2R}],\ \  g_D=\pi^{1/4}\sqrt{R}.
\label{gNDc=1}\endeq

For the multicomponent ($c>1$) case, there is a simple generalization
of the above construction of N and D boundary states.
Since the generators of the Virasoro algebra~(\ref{eq:Lmalpha})
are bilinears in the creation/annihilation operator $\alpha_n$,
a sufficient condition for the conformal invariance~(\ref{bstatecondition})
is given by
\begin{equation}
( \vec{\alpha}_{nL} - {\cal R} \vec{\alpha}_{-nR} ) | B \rangle =0,
\label{eq:bcinv-r}
\end{equation}
where ${\cal R}$ is an orthogonal $c\times c$ matrix.  This corresponds
to another sort of covariance under the Heisenberg algebra, Eq. (\ref{JJbarR}).
This includes the N case~(\ref{eq:bcinv-d}) which
corresponds to ${\cal R}=1$ (identity matrix) and
the D case~(\ref{eq:bcinv-n}) which amounts to ${\cal R}=-1$.
Other choices of the orthogonal matrix gives
different boundary states.
For example, when ${\cal R}$ is a diagonal matrix with diagonal elements
$+1$ or $-1$, it represents a mixture of the D and N b.c.'s.
When ${\cal R}$ is a rotation matrix, it gives a b.c. corresponding
to the QBM under an external magnetic field.\cite{Callan2}
We postpone the discussion of such a problem to
a later publication.

The zero modes of the boson are restricted by the
condition~(\ref{eq:bcinv-r}) with $n=0$, namely
\begin{equation}
(2 \pi \vec{u} + \vec{v}) = {\cal R} ( - 2 \pi \vec{u} + \vec{v}) .
\end{equation}
Its solution is given as a linear combination of basic solutions
\begin{equation}
        (\vec{u},\vec{v}) =  \sum_n \lambda_n (\vec{u}_n,\vec{v}_n),
\label{eq:uvsol}
\end{equation}
where $\lambda_n$'s are integer coefficients.
For such $(\vec{u},\vec{v})$,
$2 \pi \vec{u} + \vec{v}$ form a new lattice $2 \pi \tilde{\Gamma}$.
Only the oscillator vacua with these zero modes
can contribute to the boundary state.

A possible boundary state is given as
\begin{equation}
| {\cal R} \rangle =
g_{\cal R} \sum_{(\vec{u},\vec{v})} | ( \vec{u}, \vec{v} ) \rangle \rangle,
\label{eq:r-state}
\end{equation}
where the sum is taken over~(\ref{eq:uvsol}) and
\begin{equation}
| ( \vec{u}, \vec{v} ) \rangle \rangle =
\exp{
(\sum_{n=1}^{\infty} - \vec{a}_{nL}^{\dagger} {\cal R}  \vec{a}_{nR}^{\dagger})}
| ( \vec{u}, \vec{v} ) \rangle .
\end{equation}
The diagonal partition function
for this state $Z_{{\cal R}{\cal R}}$ turns out to be identical to
that for $Z_{NN}$,
if the original lattice $\Gamma$ is replaced by $\tilde{\Gamma}$.
The scaling dimensions of the boundary operators and $g$-factors are
also easily given by this replacement.

For the multicomponent free boson,
we have thus constructed an infinite variety of b.c.'s
by the generalization of D/N as in eq.~(\ref{eq:bcinv-r}),
for the infinite possible different choices of ${\cal R}$.
The diagonal partition function for these b.c.'s has
a structure similar to that for D or N b.c. and can be
written by a non-negative linear combination of
bosonic (Heisenberg) characters $q^h/(\eta(q))^c$.
We emphasize that we have just given some simple examples of physical boundary
states for the free boson theory; we have not exhausted all the possible
conformally invariant b.c.'s.

In fact, more non-trivial ``non-bosonic'' boundary states
are possible in the multicomponent free boson, at least
in some cases.
In Section V we will construct such  highly
non-trivial boundary states and will demonstrate that
they are indeed not bosonic;
they cannot be understood by a generalization
of D and N  as described above.

\subsection{Dirichlet and Neumann Boundary Conditions From Fusion: c=1 Case}

The problem of finding out all possible conformal
invariant boundary conditions is still an open one for all
but minimal theories. Even the case of the free boson $c=1$
is not fully understood yet. In this section, we discuss this problem further,
and  demonstrate how the twist field (of dimension $1/16$)
allows one to  connect D and N boundary conditions in some simple rational
cases. This gives some justification for the fusion procedure to be implemented
in the following sections in the case $c=2$. Some of the following results
are related to the more abstract approach in Ref. (\onlinecite{FS}).

We thus consider a single compact free boson, corresponding to the one
dimensional limit of the model reviewed in the previous section with
the compactification lattice $\{nR|n\in {\cal Z}\}$.
We restrict to the case $R^2=p'/2\pi$, $p'$ an integer. ($p'=1$ is the
$SU(2)$ symmetric point and
$p'=2$ is the square of the Ising model.)
The torus partition function of the periodic boson
can be expressed in terms of the generalized characters  \cite{DVVV,DFMS}
\begin{equation}
K_\lambda^{(M)}=K_{-\lambda}^{(M)}=K_{M-\lambda}^{(M)}=
{1\over\eta}\sum_{n=-\infty}^\infty q^{(nM+\lambda)^2/2M},
\end{equation}
as
\begin{equation}
Z_{per}=\sum_{\lambda=0}^{M-1} \left|K_\lambda^{(M)}\right|^2,\label{perioz}
\end{equation}
with $M\equiv 2p'$. This decomposition reflects the symmetry of the periodic
boson at this particular radius under
a chiral algebra \cite{DVVV}(traditionally
denoted by ${\cal A}_{p'}$) which is larger than the Virasoro algebra, and
is generated by, besides the stress energy tensor $T$,
the current $j$ and a pair of vertex operators of integer spin $p'$.
The primary fields of the algebra ${\cal A}_{p'}$ are vertex operators
$V_\lambda$, $\lambda=0,\ldots,M-1$, with fusion coefficients
\begin{equation}
N_{\lambda\mu}^\nu=\delta_{\lambda+\mu,\nu}\hbox{  mod }N\label{fusionalg}
\end{equation}
while charge conjugation takes $V_\lambda\rightarrow V_{M-\lambda}$.

For the same periodic boson, we now consider partition functions in the open
string channel. Two kinds of conformal boundary conditions are known
beforehand: the Dirichlet boundary conditions, where the value of the field
on the boundary is fixed modulo the compactification lattice,
$\phi=\phi_0+2\pi nR$,
and the Neumann boundary conditions, where the value of the dual field on the
boundary
is fixed modulo the dual compactification lattice,
$\tilde{\phi}=\tilde{\phi}_0+ m/R$.  These boundary conditions
preserve the Virasoro symmetry, but do not, in general,
preserve the chiral symmetry ${\cal A}_{p'}$. In fact, only the set of
$M$ Dirichlet boundary conditions with  $\phi_0={\lambda\over 2R}$,
$\lambda=0,\ldots, M-1$ do so, with
 \begin{equation}
Z_{D( \lambda/2R)D( \mu/2R)}=K_{\lambda-\mu}
\end{equation}
These Dirichlet boundary conditions are presumably the only ${\cal A}_{p'}$
invariant boundary conditions. Introducing the Ishibashi states
$|\lambda \rangle \rangle$
associated with the module $\lambda$ of this algebra, the corresponding
boundary states read
\begin{equation}
|\tilde{\lambda} \rangle =
\sum_{\mu=0}^{M-1} {S_\lambda^\mu\over \sqrt{S_0^\mu}} |\mu \rangle \rangle
=\sum_{\mu=0}^{M-1} {e^{2i\pi\lambda\mu/M}\over M^{1/4}}|\mu \rangle \rangle
\end{equation}
where the matrix of modular transformations is given by
\begin{equation}
S_\lambda^\mu={1\over M^{1/2}} e^{2i\pi \lambda\mu/M}.
\end{equation}

The general DD partition functions $Z_{D(\lambda/2R)D(\mu/2R)}$ can
naturally be obtained by fusion from the  partition function $Z_{D(0)D(0)}$ using the  fusion algebra coefficients (\ref{fusionalg}).

It is possible to write some of  the
Neumann partition functions in terms of the $K_\lambda$'s. For instance
\begin{equation}
Z_{N(0)N(0)}=\sum_{\lambda=0,\lambda\hbox{ even}}^{M-1} K_\lambda^{(M)}
\end{equation}
Similarly,
\begin{equation}
Z_{N(0)N(\pi/2R)}=\sum_{\lambda=1,\lambda\ odd}^{M-1} K_\lambda^{(M)}
\end{equation}
All the Neumann boundary conditions {\sl do} however
break the symmetry ${\cal A}_{p'}$.
This is most clearly seen by writing the Dirichlet-Neumann partition function,
which  is independent of the radius, and well known to be
\begin{equation}
Z_{ND}={1\over 4\eta} \sum_{n} q^{(2n-1)^2/16}.
\end{equation}
This partition function  cannot be expressed in terms of generalized
characters. The corresponding closed string partition function  cannot either,
but of course, it {\sl is} expressible in terms of the bulk $c=1$ Virasoro characters, as it should be since both $D$ and $N$ are possible conformal boundary conditions for the free periodic boson:
\begin{equation}
Z_{ND}={1\over\sqrt{2}\eta (\tilde q)}\sum_n (-1)^n \tilde{q}^{\ n^2}
\end{equation}

To relate D and N boundary conditions by fusion, it is necessary to relax the
constraint on invariance under ${\cal A}_{p'}$. Of course, what one should
really do, since one is only interested in general in imposing conformal
invariance on the boundary conditions, not any higher symmetry, is to simply
use the fusion algebra of the $c=1$ theory for arbitrary radius; this however
poses technical  problems - in particular, continuous modular transformations
-  which are, for the moment, insurmountable. To proceed, the empirical
strategy we use is to consider ``variants'' of the theory, with different,
finite, extended algebras, for which fusion can be implemented and, hopefully,
meaningful  boundary conditions for our original free boson found. The natural
candidate that comes to mind is the $Z_2$ orbifold, with partition function
(recall $R^2=p'/2\pi;\ M=2p'$)

\begin{eqnarray}
Z_{orb}=\sum_{\lambda=1}^{p'-1} \left|K_\lambda^{(2p')}\right|^2
+2\left|{1\over 2} K_{p'}^{(2p')}\right|^2 +2\sum_{\lambda=1,3} \left|K_\lambda
^{(8)}\right|^2\nonumber\\
+\left|{1\over 2\eta}\sum_n q^{p'n^2}+(-1)^nq^{n^2}\right|^2
+\left|{1\over 2\eta}\sum_n q^{p'n^2}-(-1)^nq^{n^2}\right|^2
\end{eqnarray}
The operator content of this theory is quite different from that of the periodic free boson. The most obvious feature is that, due to the
 $\phi\rightarrow -\phi$ identification, the multiplicity two of vertex operators has now disappeared: we get instead operators with multiplicity one, the first $p'-1$ terms,
which  correspond to operators which we call $\Phi_\lambda$,
of dimension ${\lambda^2\over 4p'}$. The next term in the partition function
 stands for two degenerate
operators $\Phi_{p'}^{(i)}$ of dimension $p'/4$ (the two energy operators in
the Ising square case), the next for the twice degenerate
twist $\sigma^{(i)}$ and excited twist $\tau^{(i)}$ operators
of dimensions $1/16,9/16$ (the spins and the tensor product of spin and energy for the
Ising square case). Finally the last two terms are respectively
the characters of the identity and the current (which we denote $\theta$),
which is primary in the orbifold algebra. We will denote the corresponding 
characters (which appear as moduli squares in the foregoing formula) $K_I$
and $K_\theta$.

To proceed, observe that, presumably, the
boundary states of the periodic boson
and the orbifold boson are not mutually compatible in general: for instance, a
Dirichlet boundary state for the orbifold
$\left|D_o(\phi_0)\right \rangle $ should be a
linear combination of Dirichlet boundary states for the non orbifold boson at
values $\pm\phi_0$, and the normalization  will make
$\left|D_o(\phi_0)\right \rangle $
incompatible with $\left|D(\phi_0)\right \rangle$.
There are thus  various families of
boundary states we can envision constructing. To start - and get quickly to the
point -  we consider the particular cases of $\left|D(\phi_0=0)\right \rangle$,
$|N(\tilde{\phi}_0=0) \rangle $ for the periodic boson.
The corresponding partition functions, in the open string channel, 
can be fully
written in terms of the generalized characters of the  orbifold algebra:
\begin{eqnarray}
Z_{D(0)D(0)}=&K_I+K_\theta\nonumber\\
Z_{N(0)N(0)}=&K_I+K_\theta+2\left[K_2^{(2p')}+K_4^{(2p')}+\ldots+{1\over 2}
 K_{p'}^{(2p')}\right]\nonumber\\
Z_{DN}=&K_1^{(8)}+K_3^{(8)}\label{bdrz}
\end{eqnarray}
($p'$ even).
Formulas (\ref{bdrz})  suggest that it is now possible to connect
 D and N by fusion, which we now demonstrate.

To do so, we need the fusion algebra of the orbifold theory: it can be found for
instance in Ref.( \onlinecite{DFMS}) with the following results ($p'$ even):
\begin{eqnarray}
\sigma^{(1)}\times\ \theta=&\tau^{(1)}\nonumber\\
\sigma^{(1)}\times\sigma^{(1)}=&I+\Phi_{p'}^{(1)}+\sum_{\lambda \ even}\Phi_\lambda
\nonumber\\
\sigma^{(1)}\times\sigma^{(2)}=&\sum_{\lambda \ odd}\Phi_\lambda
\nonumber\\
\sigma^{(1)}\times \Phi_{\lambda\ even}=&\sigma^{(1)}\nonumber\\
\sigma^{(1)}\times \Phi_{\lambda\ odd}=&\sigma^{(2)}\nonumber\\
\sigma^{(1)}\times \Phi_{p'}^{(i)}=&\sigma^{(1)}\nonumber\\
\Phi_{p'}^{(i)}\times \Phi_{p'}^{(i)}=&I \nonumber\\
\Phi_{p'}^{(1)}\times \Phi_{p'}^{(2)}=&\theta\nonumber\\
\Phi_\lambda\times\Phi_\mu=&\Phi_{\lambda-\mu}+\Phi_{\lambda+\mu}\nonumber\\
\Phi_\lambda\times\Phi_\lambda=&1+\theta+\Phi_{2\lambda}\nonumber\\
\Phi_{2p'-\lambda}\times\Phi_\lambda=&\Phi_{2\lambda}+\Phi_{p'}^{(1)}+\Phi_{p'}^{(2)}\nonumber\\
\label{fusion}
\end{eqnarray}
where the fusion of $\Phi_\lambda$ operators holds only for $\mu\neq
\lambda,2p'-\lambda$, and $\lambda$ is defined modulo $p'$.

We now start with $Z_{DD}$ and  fuse with the $\sigma^{(1)}$ operator.  From
$\sigma^{(1)}\times I=\sigma^{(1)}$ and $\sigma^{(1)}\times\ \theta=\tau^{(1)}$
we get the partition function $Z_{ND}$, since the two terms in the last
equation  of (\ref{bdrz}) are the characters of the twist and excited twist
operators respectively.

For $p'$ even, we fuse again with $\sigma^{(1)}$. We need the basic fusion rule
$ \sigma^{(1)}\times\sigma^{(1)}=I+\phi_{p'}^{(1)}+\sum_{\lambda \
even}\phi_\lambda$, and also the rule deduced from associativity
$\sigma^{(1)}\times\tau^{(1)}=\theta+ \phi_{p'}^{(2)} +\sum_{\lambda \ even}
\phi_\lambda$; the fusion gives therefore $Z_{NN}$. To
summarize: by fusion with $\sigma^{(1)}$ we thus found $Z_{DD}\rightarrow
Z_{ND}\rightarrow Z_{NN}$.

Just to illustrate the difference with $p'$ odd,
we remark that in the latter case, while the first fusion is the same,  the next one has to be
with the {\sl other} twist operator, $\sigma^{(2)}$. Using the equivalent of
(\ref{fusion}) in this case
\begin{eqnarray}
\sigma^{(2)}\times\sigma^{(1)}=I+\sum_{\lambda\ even}\phi_\lambda\nonumber\\
\sigma^{(2)}\times\tau^{(1)}=\theta+\sum_{\lambda\ even}\phi_\lambda
\end{eqnarray}
we recover $Z_{NN}$ indeed. This time therefore we have
$Z_{DD}\rightarrow_{\sigma^{(1)}} Z_{ND}\rightarrow_{\sigma^{(2)}} Z_{NN}$.

To express the boundary states, it is better now to carry out a more  complete
analysis within the boundary conditions of the orbifold model.  We thus
consider  the proper orbifold Dirichlet boundary states, which are generally
defined by ($ \phi_0\neq 0,\pi R$)
\begin{equation}
\left|D_o(\phi_0)\right\rangle ={1\over\sqrt{2}}\left[\left|D(\phi_0)\right\rangle 
+\left|D(-\phi_0)\right\rangle \right]
\end{equation}
The corresponding Dirichlet partition functions are
\begin{eqnarray}
Z_{D_o(\lambda/2R)D_o(\mu/2R)}&=&
K_{\lambda-\mu}+K_{\lambda+\mu},\lambda\neq\mu,
\ \ \lambda\neq 2p'-\mu\nonumber\\
&=&K_I+K_\theta+K_{2\lambda},\ \ \lambda=\mu\nonumber\\
&=&K_{2\lambda}+2 {1\over 2}K_{p'}^{(2p')},\ \ \lambda=2p'-\mu
\end{eqnarray}
The only orbifold Neumann boundary states compatible with the orbifold algebra
are the ``end point'' ones: $\tilde{\phi}_0=0,{\pi\over 2R}$.
We will discuss them later.

We now  need the modular transformation matrix, which reads:
\begin{equation}
S={1\over \sqrt{8p'}}\left(\begin{array}{cccccc}
1&1&1&2&\sqrt{p'}&\sqrt{p'}\\
1&1&1&2&-\sqrt{p'}&-\sqrt{p'}\\
1&1&1&2(-1)^\mu &(-1)^{i-j'}\sqrt{p'}&-(-1)^{i-j''}\sqrt{p'}\\
2&2&2(-1)^\lambda&4\cos{\pi\lambda\mu\over 2p'}&0&0\\
\sqrt{p'}&-\sqrt{p'}&(-1)^{i'-j}\sqrt{p'}&0&\delta_{i'j'}\sqrt{2p'}&-\delta_{i'j''}\sqrt{2p'}\\
\sqrt{p'}&-\sqrt{p'}&(-1)^{i''-j}\sqrt{p'}&0&-\delta_{i''j'}\sqrt{2p'}&\delta_{i''j''}\sqrt{2p'}\end{array}\right)
\end{equation}
where the conventions are that the matrix maps $\left(I,\theta,\Phi_{p'}^{(i)},\Phi_\lambda,\sigma^{(i')},\tau^{(i'')}\right)$ onto the same quantities but with $i,j$ exchanged.

After modular transformations
\begin{equation}
Z_{D_o(\lambda/2R)D_o(\mu/2R)}={1\over \sqrt{2p'}}\left[2\tilde{\chi}_I
+2\tilde{\chi}_\theta+
4 {1\over 2}(-1)^{\lambda\pm\mu} \tilde{K}_{p'}^{(2p')}+4\sum_{\nu=1}^{p'-1}
\cos{\pi\lambda\nu\over 2p'}\cos{\pi\mu\nu\over 2p'}\tilde{K}_\nu^{(2p')}\right]
\end{equation}
Expressions of the various boundary states
in terms of Ishibashi states
of the orbifold algebra follows:
\begin{equation}
\left|D_o(\lambda/2R)\right\rangle ={1\over (2p')^{1/4}}\left\{
\sqrt{2}|I\rangle +\sqrt{2}|\theta\rangle +\sqrt{2}(-1)^\lambda\left(|\phi_{p'}^{(1)}\rangle +|\phi_{p'}^{(2)}\rangle \right)
+2\sum_{\nu=1}^{p'-1}\cos{\pi\lambda\nu\over 2p'}|\phi_\nu\rangle \right\}
\end{equation}
Setting $\lambda=0$ gives the periodic Dirichlet boundary state, up to a factor of $\sqrt{2}$:
\begin{equation}
|D(0)\rangle ={1\over (2p')^{1/4}}\left\{
|I\rangle +|\theta\rangle +{1\over\sqrt{2}}\left(|\phi_{p'}^{(1)}\rangle +|\phi_{p'}^{(2)}\rangle \right)
+\sqrt{2}\sum_{\lambda=1}^{p'-1}|\phi_\lambda\rangle \right\}
\end{equation}
Modular transformation gives the other partition functions in the closed string channel
\begin{eqnarray}
Z_{D(0)N(0)}={1\over \sqrt{2p'}}\left[\sqrt{p'}\tilde{K}_I-\sqrt{p'}\tilde{
K}_\theta\right]\nonumber\\
Z_{N(0)N(0)}={2\over \sqrt{2p'}}\left[\tilde{K}_I+\tilde{K}_\theta+
2 {1\over 2} \tilde{K}_{p'}^{(2p')}\right]
\end{eqnarray}
From this, we obtain  the Neumann boundary state
\begin{equation}
|N(0)\rangle ={1\over (2p')^{1/4}}\left\{
\sqrt{p'}|I\rangle -\sqrt{p'}|\theta\rangle +{\sqrt{p'}
\over\sqrt{2}}\left(|\phi_{p'}^{(1)}\rangle -|\phi_{p'}^{(2)}\rangle \right)
\right\}
\end{equation}
Observe that, because of the particular radius we have chosen, the Ishibashi
states which appear in $|N\rangle $
all appear in  $|D\rangle $ also: this is a necessary condition for N
 to be obtained
from D by fusion, as follows from Eq. (\ref{bsfusion})
 and of course would not hold in general.

So far, the boundary states we considered are all in the periodic sector - the
Hilbert space in the closed string channel is the one of the periodic boson.
 As discussed
in Ref. (\onlinecite{AO}) [see also Ref. (\onlinecite{FS})], there are
 also eight boundary states which
involve, in addition, the twisted sector. These states are
\begin{equation}
\left|D_o(\phi_0)\pm\right\rangle =2^{-1/2}\left|D(\phi_0)\right\rangle \pm
2^{-1/4}\left|D(\phi_0)_T\right\rangle 
\end{equation}
where $\phi_0$ is at the fixed points of the orbifold, $\phi_0=0,\pi R$,
and similar Neumann like states:
\begin{equation}
\left|N_o(\phi_0)\pm\right\rangle =2^{-1/2}\left|N(\tilde{\phi}_0)\right\rangle \pm
2^{-1/4}\left|N(\tilde{\phi}_0)_T\right\rangle 
\end{equation}
with $\tilde{\phi}_0=0,\pi/2R$.

The partition functions found in Ref.(\onlinecite{AO}) can be written as
\begin{eqnarray}
Z_{D_o(\phi_0)\pm,D_o(\phi_0)\pm}&=&K_I\nonumber\\
Z_{D_o(\phi_0)\pm,D_o(\phi_0)\mp}&=&K_\theta\nonumber\\
Z_{D_o(\phi_0)\epsilon,D_o(\pi R-\phi_0)\epsilon'}&=&{1\over 2}K_{p'}^{(2p')}
\label{diriorb}
\end{eqnarray}
Similarly for the Neumann sector one has
\begin{eqnarray}
Z_{N_o(\tilde{\phi}_0)\pm,N_o(\tilde{\phi}_0)\pm}
&=&K_2^{(2p')}+K_4^{(2p')}
+\ldots+{1\over 2}K_{p'}^{(2p')}+K_I\nonumber\\
Z_{N_o(\tilde{\phi}_0)\pm,N_o(\tilde{\phi}_0)\mp}&=&
K_2^{(2p')}+K_4^{(2p')}
+\ldots+{1\over 2}K_{p'}^{(2p')}+K_\theta\nonumber\\
Z_{N_o(\tilde{\phi}_0)\epsilon,N_o(\pi R-\tilde{\phi}_0)\epsilon'}&=&
K_1^{(2p')}+K_3^{(2p')}+\ldots K_{p'-1}^{(2p')}\label{neumorb}
\end{eqnarray}
Finally, combining Neumann and Dirichlet gives
\begin{equation}
Z_{D_o(\phi_0)\epsilon,N_o(\tilde{\phi}_0)\epsilon'}=K_1^{(8)},\hbox{ resp. }
K_3^{(8)}
\end{equation}
where the choice depends on the signs $\epsilon,\epsilon'$ and the values
of $\phi_0,\tilde{\phi}_0$.

Corresponding boundary states can of course be written, giving rise to  rather
bulky expressions we will avoid here. Rather, we would like to  stress that the
orbifold boundary conditions are all connected by fusion indeed. Starting with
the first partition function in (\ref{diriorb}), we get to  the second one by
fusion with the field $\theta$,   thanks to $\theta\times\theta=I$, so
$D_o(\phi_0)\mp$ follows from $D_o(\phi_0)\pm$ by fusion with $\theta$.
Similarly, starting again with the first partition function in (\ref{diriorb})
and fusing with $\phi_{p'}^{(i)}$ gives the third partition function  using
that $\theta\times \phi_{p'}^{(i)}=\phi_{p'}^{(j)}$, so  $|D_o(\phi_0)\nu_1\rangle $
follows from $|D_o(\pi r-\phi_0)\nu_2\rangle $ by fusion with $\phi_{p'}^{(i)}$.

The Neumann and Dirichlet sectors  are still related through $\sigma^{(i)}$,
or,  also $\tau^{(i)}$, using $\theta\times\sigma^{(i)}=\tau^{(i)}$. Finally,
fusion with $\theta$ and $\phi_{p'}^{(i)}$ gives the remaining partition
functions in the Neumann sector.
As for the continuous orbifold Dirichlet and Neumann boundary conditions, they
are obtained by fusion with the operators $\phi_\lambda$.

One of the lessons illustrated here, is that to obtain all  possible conformal
boundary conditions for the  free boson theory, one needs to consider its $Z_2$
orbifold as well. It is not clear  in general what good ``variant'' of the bulk
theory is needed to explore all possible  conformal boundary conditions for a
given model.
 In the case $c=2$ for example,  we shall see  that fusion with a field of
dimension $1/8$ allows one to go from D to N. This is a natural generalization
of the $c=1$ case,  the $1/8$ field being now a ``double'' twist field for the
bosons $\phi_{1,2}$. We will also see however that fusion with a field of
dimension $2/5$ is also necessary to explore the possible boundary states. The
origin of this field in the two boson language  is still mysterious. Is it some
other sort of twist field?

The discovery of non trivial boundary
conditions should also lead to new bulk theories.
In the $c=1$ case for instance, consideration of the possible
D and N boundary conditions should lead one to the discovery
of the $Z_2$ orbifold,
were this theory not already known. This is also related to the
existence of a $c=1$ theory - the square of the Ising model - which is
not a periodic boson. In the $c=2$ case, the ``embedding''
we will use to describe
partition functions  with boundaries leads similarly to a bulk theory - the
product of Ising, tricritical Ising and Potts -
which does not seem to be any simple sort of two boson
orbifold. Understanding its meaning would be crucial to generalize
our work  to other values of the triangular lattice parameter, $a$.

\section{Quantum Brownian motion and conformal field
theory}
We first discuss the one-dimensional case.
The study of Brownian motion of a classical particle described by the
 simple Langevin
equation
\begeq
M{d^2Q\over dt^2} +\eta {dQ\over dt}+{\partial V\over\partial Q}=\xi(t) \label{langevin}
\endeq
is of course one of the basic topic in non-equilibrium statistical
mechanics. In (\ref{langevin}), $\eta$ is a phenomenological
friction coefficient, $V$ an external potential, $M$ the particle mass,
and $\xi$ is a random force that obeys
\begeqar
\left\langle \xi(t)\right\rangle &=&0\nonumber\\
\left\langle  \xi(t)\xi(t')\right\rangle &=&2\eta T\delta(t-t')
\endeqar
More recently, there has been a great deal of interest in trying to
describe the quantum behavior
of systems for which the classical motion would be described by the
Langevin equation (\ref{langevin}). The most tractable approach
has been to couple the ``particle'' (which might actually represent
something quite different, like the phase in a Josephson junction)
to a bath (an environment) with an infinite number of degrees of freedom,
 which provides both the friction and the fluctuating force.
The simplest example of that approach is the Caldeira Leggett 
model\cite{Caldeira} where
the coordinate $Q$ is coupled linearly to an
infinite set of harmonic oscillators, with the Hamiltonian
\begeq
H={P^2\over 2M}+V(Q)+{1\over 2}\sum_k\left \{ {p_k^2\over m_k}+
m_k\omega_k^2 \left[
q_k-\frac{Q\lambda_k}{m_k \omega_k^2} \right]^2 \right \}.
\label{genedqmhamil0}
\endeq

Here $Q$ and $P$ are the co-ordinate and momentum of the particle;
$V(Q)$ is an arbitrary potential, with $V(Q+a)=V(Q)$.
The exact distribution of coupling constants $\lambda_k$, masses
$m_k$ and
frequencies $\omega_k$ is important {\it for the properties of the particle}
only through the weighted density of states
\begin{equation}
J(\omega)\equiv {\pi\over 2}\sum_k {\lambda_k^2\over
m_k\omega_k} \delta(\omega-\omega_k),~\omega>0.
\end{equation}
In the limit where the number of oscillators is infinite, $J(\omega )$
becomes a continuous function.  In the following we will restrict to the
so called ohmic case $J(\omega)=
\eta\omega$; with this choice, it can be shown
that the Hamiltonian (\ref{genedqmhamil0}) 
when treated classically yields, after
elimination of the bath degrees of freedom,
the Langevin equation (\ref{langevin}).

 By a canonical transformation on the oscillator bath:
 \begeq
 q_k\to {\lambda_k\over m_k}{q_k\over \omega_k^2},\ \
 p_k\to {m_k\over \lambda_k}\omega_k^2p_k,
 \label{cantransf}\endeq
 we may rewrite the Hamiltonian in the form:
\begeq
H={P^2\over 2M}+V(Q)+{1\over 2}\sum_k\left \{{p_k^2\over \tilde m_k}+
\tilde m_k\tilde \omega_k^2 [q_k-Q]^2 \right \},
\label{genedqmhamil} \endeq
where:
\begin{eqnarray}
\tilde m_k&\equiv& \lambda_k^2/m_k\omega_k^4 \nonumber \\
\tilde \omega_k &\equiv & \omega_k \nonumber \\
\tilde \lambda_k &\equiv & \lambda_k^2/m_k\omega_k^2=\tilde m_k
\tilde \omega_k^2 \label{tildem}
\end{eqnarray}
are the masses, frequencies and couplings of the transformed
oscillators.  Henceforth, we drop the tilde notation on
these quantities.
The Hamiltonian  of Eq. (\ref{genedqmhamil}) is  invariant
under the translation:
\begin{equation}
Q\to Q+a,\ \  q_k\to q_k+a.\label{transf}\end{equation}
As the particle moves in the periodic potential it drags a cloud of other
``fictitious particles''
with it which don't feel
the periodic potential $V$ but are bound
to the real particle  by a quadratic interaction.
Due to the translation symmetry of Eq. (\ref{transf}),
 there is a  total momentum, $P_T$, which is conserved
modulo $2\pi /a$:
\begin{equation}
P_T\equiv P+\sum p_k.\end{equation}
Explicitly, a basis of eigenstates of H may be chosen such that
any eigenfunction gets multiplied by the phase $e^{iP_ca}$ under
this translation.  Without loss of generality, the crystal momentum,
$P_c$, may be chosen to lie in the first Brillouin zone, $|P_c|<\pi /a$.

After the canonical transformation of Eq. (\ref{cantransf})
the weighted density of states becomes:
\begin{equation}
J(\omega)\equiv {\pi\over 2}\sum_k \lambda_k \omega_k
 \delta(\omega-\omega_k),~\omega>0.
\label{density}\end{equation}
The properties of the original particle are unaffected by a rescaling
of $\omega_k$ (independently for each value of $k$) provided that
$\lambda_k$ is scaled by the inverse factor such that $\lambda_k\omega_k$
is held fixed.

We note that an alternative way of writing the Hamiltonian is in terms
of the total momentum $P_T$ and transformed oscillator co-ordinates,
\begin{equation}
q_k'\equiv q_k-Q,\end{equation}
which define a transformed canonical set of co-ordinates.  The
Hamiltonian in these co-ordinates becomes:
\begin{equation}
H={(P_T-\sum p_k)^2\over 2M}+V(Q) +{1\over 2}
\sum\left[{p_k^2\over m_k}
+m_k\omega_k^2{q'}_k^2\right].\label{Htransf}
\end{equation}

 A convenient choice for the oscillator
bath is the Fourier modes of a one-dimensional free massless boson
quantum field.  In order to begin
with a discrete set of oscillators, we may define the field theory
with vanishing boundary conditions in a box
of size $2l$, so that:
\begin{equation} \omega_k=k,\ \  k=\pi n/2l, \ \ n=1, 2,
\ldots .\label{ks}\end{equation}
Here we have set an arbitrary ``velocity of light'' to 1.  Strictly
speaking an ultraviolet cut off must be applied, $k<\Lambda$, but
the universal critical behavior will be independent of the cut off.

Note that there is a zero-frequency oscillator  in the limit
$l\to \infty$.
The field couples to the particle at $x=0$ and the boundaries of
the box are at $x=\pm l$.
The oscillator co-ordinates and momenta can be expressed in terms
of the boson creation and annihilation operators as:
\begin{equation} q_k= i\sqrt{k/2}(a_k-a_k^\dagger ),\ \
p_k=(a_k+a_k^\dagger )/\sqrt{2k}. \label{creation}
\end{equation}
We may then define the field and conjugate momentum field on a
fictitious one-dimensional space, $x$, completely unrelated to
the physical space in which the particle moves, by:
\begin{equation} \phi (x)\equiv \sqrt{1\over l}\sum_kp_k\sin k(x+l),
\ \  \Pi (x) \equiv -\sqrt{1\over l}\sum_kq_k\sin k(x+l).\label{phimode}
\end{equation}
 We note that $\phi (x)$ corresponds to a ``non-compact'' boson field, so
no term linear in $x$ is allowed in its mode expansion.

We see that, with this choice of oscillator bath, in order to
obtain ohmic dissipation, we must choose,
$\lambda_k=2\eta /l$. From Eq.~(\ref{tildem}) this implies
that the oscillator
masses are given by:
\begeq m_k=\lambda_k/\omega_k^2=2\eta /l k^2.\endeq
The  fictitious particles become infinitely heavy at zero frequency!

In this
case the Hamiltonian of Eq. (\ref{genedqmhamil}) can be written:
\begeq
H={P^2\over 2M}+ V(Q)
+{1\over 2}\int_{-l}^{l} dx \left[\left( {d\phi \over dx}\right)^2+
(\Pi + \sqrt{2\eta}Q\delta (x))^2\right].
\endeq
It is convenient to separate $\phi (x)$ into its even and odd components
since the odd part decouples from the particle:
\begeq
\phi_{e,o}(x)\equiv [\phi (x)\pm \phi (-x)]/\sqrt{2}.
\endeq
The even field obeys a Neumann b.c., $d\phi_e/dx=0$ at $x=0$.
Thus we may equivalently consider a field defined on the positive axis,
$0\leq x < l$,
with a Neumann b.c. at $x=0$ and a Dirichlet b.c. at $x=l$
 coupled to the particle.
This Hamiltonian is equivalent to Eq. (\ref{genedqmhamil}) with:
\begin{equation} \tilde \omega_k=k, \ \  \tilde \lambda_k=2\eta /l,\ \
 k=\pi (n+1/2)/l, \ \ n=0,1,2,\ldots  \label{evenks}
\end{equation}

  We might expect the translational symmetry
   to lead to the usual band structure in which
the discrete set of eigenstates of given crystal momentum depends
in some smooth way on $P_c$.  However,  when
there is an oscillator, $q_0$, of infinite mass  a dramatic
simplification occurs.  This follows from observing that any eigenfunction
of crystal momentum $P_c$, can be written in the form:
\begin{equation} \Psi = e^{iP_cq_0}\psi ,\end{equation}
where $\psi (Q,q_k)$ is invariant under the translation of Eq. (\ref{transf}).
Since $p_0$ does not appear in $H$, it follows that the periodic function, 
$\psi$, obeys exactly the same Schroedinger equation as the full wave-function,
$\Psi$, independent of $P_c$.  This implies that all eigenstates come in
degenerate sets, where the degenerate members are labelled by $P_c$.
Effectively we can give the infinite mass ``particle'', $q_0$, all the
momentum at no cost in energy.  This
degeneracy is strictly infinite if $Q$ is allowed to range over the entire
real line.  Alternatively, if we restrict $Q$ to an interval of size ${\cal N}a$
then we expect the eigenvalues to form nearly degenerate sets of size
approximately ${\cal N}$, the number of allowed momenta in the Brillouin zone.
 In a localized phase we can identify this
degeneracy with the ${\cal N}$ minima of $V$ at which the particle can
become localized.  However we emphasize that this degeneracy is
an exact property of $H$ (at infinite ${\cal N}$) and does not depend on
which phase the system is in.
We have dwelled on this point here
because we will find that, in the limit where the oscillator frequencies
form a dense set, the groundstate degeneracy can take the more general
form $g{\cal N}$ where $g$ is a universal (generally non-integer) number of $O(1)$
which {\it does} depends on which phase the system is in.  This rather
 unusual behavior of the model is related to its essentially unphysical
nature as a real model of QBM, mentioned in the Introduction.  We elaborate
on this point further in subsection C.

In the field theory representation, the conserved total momentum operator and
transformed oscillator co-ordinates are:
\begeq
P_T=P+\sqrt{\eta}\phi_e (0),\ \  \Pi_e '(x)=\Pi_e (x)+2\sqrt{\eta}Q\delta (x).
\endeq
The transformed Hamiltonian (keeping only the even part) is:
\begeq
H={(P_T-\sqrt{\eta}\phi_e (0))^2\over 2M}+ V(Q)
+{1\over 2}\int_{0}^{l} dx \left[\left( {d\phi_e \over dx}\right)^2+
\Pi_e '^2\right].\label{H2}
\endeq

Much of the literature on QBM proceeds by integrating out the field,
$\phi_e (x)$ to obtain a non-instantaneous effective action for the
particle.  However, we introduce a different approach here in
keeping with previous analyses of quantum impurity problems using
boundary CFT methods.  In our approach we rather absorb the
particle co-ordinate and momentum into a b.c. on the field, in
the effective low energy Hamiltonian.  This has the advantage
that correlation functions of the field are immediately accessible.
For the QBM problem this allows us to study the back-reaction of
the particle on the oscillator bath.
We note that this dichotomy of approaches also exists in
other types of quantum impurity problems.   There are two limits
in which the QBM Hamiltonian reduces to a boundary sine-Gordon
model: the strong and weak corrugation limits.  We consider
each in turn.

\subsection{Strong Corrugation Limit}
 In the ``strong corrugation'' limit
the potential is large compared to the dissipation.  More precisely, we
require the energy bands produced by the potential $V$, ignoring the
dissipation, to have large band gaps compared to the energy scale of
the dissipation, $\eta /M$.  In this limit we expect that band-mixing
effects due to the dissipation will be small and the wave-functions
for the particle-bath system will be linear combinations of wave-functions
in a particular band, $\psi_n(Q;P_c)$.  Here $P_c$ is the crystal
momentum and the wave-functions, $\psi_n$ are periodic:
\begin{equation}
\psi_n(Q+a;P_c)=\psi_n(Q;P_c).
\end{equation}
The solutions of the non-dissipative Schroedinger equation are
$e^{iP_cQ}\psi(Q;P_c)$ and the periodic functions obey the equation:
\begin{equation}
\left\{ {1\over 2M}\left[ -{d^2\over dQ^2}-2iP_c{d\over dQ}+P_c^2\right]
+V(Q)\right\} \psi (Q;P_c)=E\psi (Q;P_c)
\end{equation}
Projecting the Hamiltonian of Eq. (\ref{H2}) into a fixed energy band
reduces it to:
\begin{equation}
H_n\to
\epsilon_n[P_c-\sqrt{\eta}\phi_e (0)]
+\int_{0}^{l} dx {1\over 2}\left[\left( {d\phi_e \over dx}\right)^2+
\Pi_e '^2\right].\label{H_n}
\endeq
Here $\epsilon_n(P_c)$ is the dispersion relation of the $n^{th}$
energy band for the non-dissipative problem.
Note that $\phi_e$ obeys a
N b.c. at $x=0$ and a D b.c. at $x=l$.  The degeneracy discussed
above corresponds to shifting $\phi_e (x)$ by $P_c/\sqrt{\eta}$.
However, this is inconsistent with the vanishing b.c. at $x=l$ so
the degeneracy is lifted by effects of $O(1/l)$ as expected.
 Upon expanding $\epsilon_n$ in
harmonics:
\begin{equation}
\epsilon_n(P_c)=A_n+B_n\cos P_ca + \ldots \label{epn}\end{equation}
and ignoring the (less relevant) higher harmonics,
we   obtain the boundary sine-Gordon model:
\begin{equation}
H=B\cos [aP_c-\sqrt{\eta}a\phi_e(0)]
+\int_{0}^{l} dx {1\over 2}\left[\left( {d\phi_e \over dx}\right)^2+
\Pi_e '^2\right].\label{HSG}
\endeq

This result can be confirmed by considering a (single band)
tight-binding model.\cite{Guinea}  Introducing operators $c_j$ which
annihilate a particle at site $j$, the position operator
is represented as:
\begin{equation} Q=a\sum_jjc^\dagger_jc_j.\end{equation}
The transformation to the total momentum operator then
corresponds to a unitary transformation by the operator:
\begin{equation}
U\equiv e^{-i\sqrt{2\eta}Q\phi (0)}.\end{equation}
Explicitly:
\begin{equation}
U^\dagger [\Pi (x)+\sqrt{2\eta}Q\delta (x)]U=\Pi (x).\end{equation}
The part of the Hamiltonian containing only the particle can
be diagonalized in momentum space.  Defining:
\begin{equation}
c_p\equiv {1\over N}\sum_jc_je^{-ipj},\end{equation}
the particle part of the Hamiltonian is:
\begin{equation}
H_0=\sum_p\epsilon (p)c^\dagger_pc_p.\end{equation}
We find:
\begin{equation}
U^\dagger c_pU = c_{p+\sqrt{2\eta}\phi (0)},\end{equation}
and hence the Hamiltonian is transformed into Eq. (\ref{H_n}),
using $\sqrt{2}\phi (0)=\phi_e(0)$.
In the case of nearest neighbor hopping the dispersion
relation is:
\begeq \epsilon (P)=-2t\cos Pa .\endeq
  The above derivation
generalizes this result slightly to the case of several
well-separated bands with arbitrary dispersion relation.
Of course, for sufficiently weak dissipation and sufficiently
low temperature, we will only be concerned with the lowest band.

It is convenient to define the parameter:
\begeq R\equiv {1\over \sqrt{\eta}a},\label{defR}\endeq
so that the Hamiltonian is invariant under the translation:
\begeq \phi_e\to \phi_e+2\pi R.\endeq
However, as mentioned above, the boson $\phi_e$ is not
compact so we cannot identify the field under this translation.
We return to this point below.

The boundary interaction has scaling dimension
\begeq \Delta = \frac{\eta a^2}{2\pi} =
\frac{1}{2\pi R^2}.\endeq
This can be verified, for example,
by using the N b.c. to replace $\tilde \phi ''(0)$ by
$2\phi_L(0)$ and using the standard result that the scaling
dimension of $\cos \beta \phi_L$ is $\beta^2/8\pi$.
The boundary interaction  is relevant for $\eta a^2<2\pi$.  For $\eta a^2>2\pi$,
we may think of the hopping term as being irrelevant corresponding
to a flow to the localized fixed point.  For $\eta a^2<2\pi$ we
expect a flow to the diffusing fixed point. Note that either increasing
the distance over which the particle must hop or increasing the coupling
to the heat bath promotes localization.

\subsection{Weak Corrugation Limit}
The momentum operator $P_T$ and co-ordinate $Q$, can be absorbed into the
field, $\phi_e$, by transforming to a dual oscillator bath, a transformation
 that
was introduced by Fisher and Zwerger from a different viewpoint.\cite{Fisher}
This is done
by first diagonalizing the oscillator part of the Hamiltonian,
 whose even part is:
\begin{equation}
H_0 \equiv  {\eta [\phi_e (0)]^2\over 2M}
+{1\over 2}\int_{0}^{l} dx \left[\left( {d\phi_e \over dx}\right)^2+
\Pi_e '^2\right].
\endeq
This quadratic Hamiltonian can be diagonalized by solving the classical
Euler-Lagrange equations:
\begeq
\omega^2\phi (x)=\left[ - \frac{d^2}{dx^2}+2 \frac{\eta}{M} \delta (x)
\right] \phi (x).
\label{Sch}\end{equation}
This is just the Schroedinger equation with a $\delta$-function potential.
The odd wave-functions are unaffected by the potential and the even ones
can be expressed in terms of a phase shift, $\delta (k)$:
\begin{equation}
\phi_k (x)\propto \cos [k|x|+\delta (k)],\end{equation}
with frequency $\omega (k)=|k|$.  [Here $k$  runs over the
wave-vectors of the even modes, given in Eq. (\ref{evenks}).]
Substituting into Eq. (\ref{Sch}) we obtain the phase shifts:
\begin{equation}
\cot \delta (k)=- \frac{kM}{\eta} .\label{PS}\endeq
For small $k$, this becomes:
\begeq
\delta (k)\approx \frac{\pi}{2}+\frac{kM}{\eta} + O(k^2).\label{PSapp}\endeq
The allowed wave-vectors are given by:
\begeq
kl+\delta (k)=\pi (n+1/2),\ \  n=1,2,3, \ldots \endeq
For large $l$ ($l \gg M/\eta$) and small $k$ ($k \ll \eta /M$) we obtain:
\begin{equation}
        k\approx \pi n/l, \ \ (n=1,2,3,\ldots )\label{kapprox}
\end{equation}
The mode expansion for $\phi_e(x)$ now takes the form:
\begin{equation}
\phi_e(x)=\sqrt{2\over l}\sum_kp_k\cos [k|x|+\delta (k)],
\label{phiemode}\endeq
and similarly for $\Pi_e(x)$.  [The $p_k$'s are the same operators, up
to a minus sign, that
occur in Eq. (\ref{phimode}) for wave-vectors corresponding to
odd values of $n$ in Eq. (\ref{ks}).]
Noting that, for small $k$,
\begeq \cos (k|x|+\delta )\approx -\sin [k|(x|+M/\eta )],\endeq
we see that the effect of the $\phi_e (0)^2$ term in $H_0$ at low energies,
is to impose, approximately a Dirichlet b.c., $\phi_e =0$ on the even
field.  (More accurately, the low energy field vanishes at $x=-M/\eta$
 but, in the low energy effective theory $\phi (x)$ does not vary
 on such short wavelengths so we ignore this shift.)
It is now convenient to introduce a transformed even field, $\phi '(x)$ obeying
 the D. b.c. at $x=0$:
\begeq
\phi '(x)\equiv -\sqrt{2\over l}\sum_k\sin kx\  p_k.\end{equation}
The even part of $H_0$ may be written in terms of $\phi '$ as simply:
\begin{equation}
H_{0}=\frac{1}{2} \int_0^{l} dx[\left(\frac{d\phi '}{dx}\right)^2+\Pi '^2].
\endeq
We now must consider the term in the Hamiltonian $H$ of Eq. (\ref{H2}) that
couples the field to the particle:
\begeq H_{int}\equiv -(\sqrt{\eta}/M)\phi_e (0)P_T.\endeq
This may be expressed in terms of the $p_k$ operators of Eq. (\ref{phiemode})
using the explicit expression for the phase shift in Eq. (\ref{PS}) as:
\begeq
\phi_e (0) = -M\sqrt{2\over l}\sum_k{kp_k\over \sqrt{\eta^2+(kM)^2}}.
\end{equation}
We see that, at low energies this is given by:
\begeq \phi_e (0) \approx \frac{M}{\eta} \frac{d\phi '}{dx},\endeq
but we do not make this approximation here, proceeding to a more exact
analysis.
The  entire Hamiltonian of Eq. (\ref{H2}) can now be written:
\begeq
H \approx {1\over 2}\sum_k \left[
\left( kp_k+ \sqrt{2 \eta \over l[\eta^2+(kM)^2]}
P_T\right)^2+q_k^2\right]+{P_T^2\over 2\eta l}
+V(Q).\label{He}\endeq
In order to complete the square we have used a crucial identity:
\begin{equation}
{2\eta \over l}\left[{1\over 2\eta^2}+
\sum_{n=1}^\infty {1\over \eta^2+(\pi nM/l)^2}\right]
={1\over M}\coth (l\eta /M)
= {1\over M}+O(e^{-2l\eta /M}) .\label{expsmall}\endeq
and dropped the exponentially small term.  [Note that we are being
a bit cavalier and using the small $k$ result, $k\approx \pi n/l$
for all $n$.  It can be shown that the corrections to this approximation
only affect the exponentially small term in Eq. (\ref{expsmall}).]
It is now convenient to Fourier transform to position space, using:
\begin{equation}
{d\phi '\over dx} = -\sqrt{2\over l}\sum_k kp_k\cos kx .
\end{equation}
To this end it is convenient to introduce the function:
\begin{equation}
f(x)\equiv {2\sqrt{\eta}\over l}\left[{1\over 2\eta}+
\sum_{k} {\cos kx\over \sqrt{\eta^2+(kM)^2}}\right] .
\label{deff}\end{equation}
Here the sum is over $k=\pi n/l$ with $n=1,2,3,\ldots$ $f(x)$ vanishes
exponentially at large $x$.
Up to corrections which are exponentially small in $l\eta /M$, we may
write:
\begin{equation} f(x)={2\sqrt{\eta}\over \pi M}K_0\left(
{\eta x\over M}\right).\end{equation}
Here $K_0$ is a modified Bessel function.
We may thus rewrite $H$ as:
\begin{equation}
H={1\over 2}\int_0^l dx\left\{ \left[ {d\phi '\over dx}-f(x)P_T\right]^2
+\Pi '^2\right\}
+V(Q).\end{equation}
It is now possible to make a further transformation that absorbs $P_T$
into $\phi'$:
\begin{equation}
\phi ''(x) = \phi '(x)+\int_x^ldx'f(x')P_T.\label{phi''}
\end{equation}
 We may now also absorb $Q$ by going over to the dual field.  This
 is defined, as usual by:
 \begin{eqnarray}  {d\phi ''\over dx}&=&\tilde \Pi ''\nonumber \\
{d\tilde \phi ''\over dx}&=&-\Pi ''.\label{tildedef}
\end{eqnarray}
These equations only determine $\tilde \phi ''(x)$ up to a constant term.
This may be fixed by imposing the canonical commutation relations
and Euler-Lagrange equations
on $\tilde \Pi ''(x)$ and $\tilde \phi ''(x)$.  The finite momentum
part of the dual field is:
 \begin{equation}
\tilde \phi '(x)=-\sqrt{2\over l}\sum_k{q_k\cos kx \over k}
\end{equation}
The zero mode term in $\tilde \phi ''(x)$, which we write $\tilde \phi_0''$,
must be chosen to commute with all finite momentum modes of
$\tilde \Pi ''(x)$ but have a non-zero commutator with the zero mode of $\tilde \Pi ''(x)$:
\begin{equation}
\tilde \Pi_0''\equiv
-P_T\int_0^lf(x) \frac{dx}{l}=- \frac{P_T}{l\sqrt{\eta}}.
\label{Pi0}\end{equation}
Using the commutator:
\begin{equation}
[\phi '(x),\tilde \phi '(y)]=i\left[{x\over l}-\theta (x-y)\right],
\end{equation}
which can be checked from the mode expansion of these fields, we find
\begin{equation}
\tilde \phi_0 ''=-\sqrt{\eta}\int_0^ldy f(y)\tilde \phi '(y)
-\sqrt{\eta}Q.\label{phi0}\end{equation}
 Note that $\tilde \phi ''(x)$ obeys N b.c.'s at $x=0$ and $l$.  We
 see that $Q$ has become (part of) the zero-mode of the dual field.  The
 Hamiltonian can be written:
\begin{equation}
H \to {1\over 2}\int_0^ldx \left[ \tilde \Pi ''^2
+\left({d\tilde \phi ''\over dx}\right)^2\right] +V\left[
-\int_0^ldx f(x)\tilde \phi ''(x)
\right].\label{Hfinal}
\end{equation}
 Now consider integrating out the high wave-vector components of
 $\tilde \phi ''(x)$ to obtain a low energy effective Hamiltonian.
 Since $f(x)$ vanishes exponentially on the length scale $M/\eta$,
 it follows that once we have reduced the wave-vector cut off to a
 value much less than $\eta /M$ we may approximate $\tilde \phi ''(x)$
 by its value at $x=0$ inside the integral in the last term
 of Eq. (\ref{Hfinal}),
 leading to the simplified expression:
 \begin{equation}
 H\to {1\over 2}\int_0^ldx \left[ \tilde \Pi ''^2
+\left({d\tilde \phi ''\over dx}\right)^2\right] +V\left[
{-\tilde \phi ''(0)\over \sqrt{\eta}}
\right].\label{Hfinal2}
\end{equation}

The degeneracy mentioned above, corresponds to shifting $\tilde \phi ''(x)$,
or equivalently $Q$,
by a constant $na$ where $a$ is the lattice spacing.
This degeneracy is lifted at finite $l$ as expected.
Now let us  consider a particular choice for $V$:
\begin{equation}
V(Q)=V_0\cos (2\pi Q/a).\label{cos}\end{equation}
Our Hamiltonian then becomes the standard boundary sine-Gordon model:
\begin{equation}
H={1\over 2}\int_0^l dx\left[ \tilde \Pi ''^2
+\left({d\tilde \phi ''\over dx}\right)^2\right] +V_0\cos \left[
2\pi R\tilde \phi ''(0)\right] ,\label{Hfinal3}
\end{equation}
where we have again introduced the parameter $R$ defined in
Eq. (\ref{defR}).
Recall that we obtained the boundary sine-Gordon
model only after integrating out high frequency modes, reducing the cut-off
to $\eta /M$.  This process will, in general, introduce renormalization
of the interactions.  In order to be able to ignore this renormalization,
the original dimensionless coupling constant must be small.  We may estimate
how small it needs to be by considering the renormalization of the boundary
interaction in Eq. (\ref{Hfinal3}).  This boundary interaction has a scaling
dimension of \begeq
\Delta = \frac{2\pi}{\eta a^2}=2\pi R^2 .\endeq
 Thus, upon reducing the cut-off from its
bare value, $\Lambda$ to $\eta /M$, the effective coupling is renormalized to:
\begin{equation}
{V_0\over \Lambda}
\to {V_0\over \Lambda}\left({\Lambda \over \eta /M} \right)^{1-\pi /\eta a^2}.
\end{equation}
Requiring the renormalized dimensionless coupling constant to be small after
reducing the cut off to $\eta /M$ gives the condition:
\begin{equation}
V_0\ll(\eta /M\Lambda)^{1-\pi /\eta a^2}\Lambda .\end{equation}
Thus, this boundary sine-Gordon model is only obtained in the weak corrugation
limit.  The dimensionless parameter, $\sqrt{\eta}a=1/R$ may take any value,
however.

We note that the weak corrugation formulation of the problem
is actually equivalent at $l\to \infty$, under a dual transformation,
to a tight binding model interacting with a heat bath with
a Lorentzian weighted density of states, $J(\omega )$.  This
can be seen by starting from Eq. (\ref{He}) and making the duality
transformation:
\begeq
p_k\rightarrow -q_k/k, \ \  q_k\to kp_k,\ \
P_T\rightarrow -Q, \ \  Q\rightarrow P_T.
\endeq
The dual Hamiltonian is:
\begeq
H \approx {Q^2\over 2\eta l}+V(P_T) +
{1\over 2}\sum_k \left[ \left( q_k+\sqrt{2\eta \over l[\eta^2+(kM)^2]}
Q\right)^2+k^2p_k^2\right].\label{Hdual}\endeq
The last term has the same form as our original  heat bath
term in Eq. (\ref{genedqmhamil0}) with
\begeq \omega_k=k,\ \ 
m_k={1\over k^2},\ \  \lambda_k=-\sqrt{2\eta \over l[\eta^2+(kM)^2]}
\endeq
and hence a weighted density of states:
\begeq J(\omega )={\pi /2}\sum_k{2\eta k\over l[\eta^2+(kM)^2]}
\delta (\omega -k).\endeq
Taking the limit $l\to \infty$ this gives:
\begeq
J(\omega ) = {\eta \omega \over \eta^2+\omega^2M^2}.\endeq
This is ohmic below a cut off scale $\eta /M$:
\begeq
J(\omega) \approx {\omega \over \eta}.\endeq
Note that the dissipation strength parameter, $\eta$ has been
inverted.
  We may also
drop the first term in Eq. (\ref{Hdual}) in the limit $l\to \infty$
so that the particle part of the Hamiltonian is diagonal in
momentum space.  This is then equivalent to a tight binding model
with a dispersion relation:
\begeq \epsilon (P_T)\to V(P_T).\endeq
This gives another way of understanding why a boundary sine-Gordon
model is obtained in both the weak and strong corrugation cases.
We note however, that the $P_T^2/2\eta l$ term, which gave
$\tilde \Pi ''(x)$ a zero mode, plays quite an important role in
the following sub-section.

\subsection{Groundstate Degeneracy and Boson Compactification}
An interesting  quantity in quantum impurity problems is
the zero temperature impurity entropy or its
exponential, the ``groundstate degeneracy'', $g$.  More generally,
 we may define an impurity free energy by
subtracting off the bulk free energy (the term proportional to
$l$), then taking
the length of the bulk system to infinity.  The zero temperature
impurity entropy is defined this way, with the $T\to 0$ limit
taken after the infinite length limit.  Note that if we take $T\to 0$ {\it first},
before taking $l\to \infty$, the degeneracy is necessarily
integer valued since the spectrum of the finite size Hamiltonian
 is discrete.  However, taking the limits in the opposite order,
 we are effectively dealing with a continuous spectrum and it
 is entirely possible to obtain a non-integer (even non-rational)
 ``groundstate degeneracy'' $g$, independent of system size.
$g$ has been argued to be
universal and to always decrease under renormalization group
flow between boundary fixed points.\cite{g-theorem}  Thus it plays
a role in boundary critical phenomena analogous to that of
the conformal anomaly parameter, $c$, in two dimensional bulk
critical phenomena.  A new feature occurs when we consider
$g$ in QBM; it turns out to be proportional to the length
of the interval on which the particle is allowed to move.  We emphasize
that we are taking the length, $l$, of the fictitious line interval used
to define the oscillator bath to $\infty$.  Thus the oscillator
spectrum becomes continuous.  On the other hand, we are considering
a long but finite physical line interval, of length ${\cal N}a$ with
${\cal N} \gg 1$,
on which the particle moves.

To orient ourselves let us first consider the partition function in
 the case $\eta =0$ where the particle is decoupled from the heat bath and the
 periodic potential $V$ is set to zero.
 We only consider the even oscillator modes, with energies
 $\pi (n+1/2)/l$, $n=0,1,2,\ldots$. (Recall that $\phi_e$ obeys a N
 b.c. at $x=0$ and a D b.c. at $x=l$ so that the allowed momenta
 are proportional to the half-integers.)  Thus the oscillator part
 of the partition function is:
 \begin{equation}
 Z_{osc}=\prod_{n=0}^\infty \left[ 1-e^{-\pi (n+1/2)/lT}\right]^{-1}.
 \label{Zos1}\end{equation}
Taking the limit $lT \gg 1$ we obtain:
 \begin{equation}
 Z_{osc}\to {e^{\pi lT/6}\over \sqrt{2}},\label{Zos2}\end{equation}
 corresponding to an oscillator groundstate degeneracy of
 \begin{equation}
 g_{osc}=1/\sqrt{2}.\end{equation}
This must be multiplied by the partition function of the particle.  To make
this well-defined, we place the particle in a box of size ${\cal N}a$,
with periodic
b.c.'s.  In this case the allowed momenta are $P=2\pi m/{\cal N}a$ giving the
full partition function:
\begin{equation}
Z=\sum_{m=-\infty}^\infty \exp \left[-\left({2\pi m\over {\cal N}a}\right)^2
{1\over 2MT}\right]
{e^{\pi lT/6}\over \sqrt{2}}.\end{equation}
If we now take $T\to 0$, holding ${\cal N}$ fixed, the particle has a unique
groundstate with $P=0$ so it makes no contribution to $g$.  We note, however,
that in the  opposite limit, $({\cal N}a)^2MT \gg 1$, the partition function is
proportional to ${\cal N}$ since the particle may be treated classically:
\begin{equation} Z\approx {e^{\pi lT/6}\over \sqrt{2}}a{\cal
N}\int_{-\infty}^\infty {dP\over 2\pi} e^{-P^2/2MT}=
\frac{a{\cal N}}{2}\sqrt{\frac{MT}{\pi}}e^{\pi lT/6}.\end{equation}
Remarkably, this factor of ${\cal N}$ will persist in the partition function as
we take $T\to 0$ at fixed ${\cal N}$, once we couple the particle to the
oscillators.   If we now include the periodic potential $V$, the partition
function behaves the same way at temperatures small compared to the bandwidth.
This follows since we are only concerned with low momentum states in the lowest
energy band, whose dispersion may be approximated by:
\begin{equation}
\epsilon_1(P)\approx \frac{P^2}{2M^*},\end{equation}
for some effective mass, $M^*$.

Let us now couple the particle to the oscillator bath, $\eta >0$, but, for the
moment ignore the potential $V$.  Thus we are considering a freely diffusing
particle.  From the mode expansion of the dual field, $\tilde \phi ''(x)$,
which obeys N b.c.'s at $x=0$, $l$ with the zero mode of Eq. (\ref{Pi0}), or
more directly from Eq. (\ref{He}), we see that the spectrum is given by:
\begin{equation}
E={P_T^2\over 2l\eta}+\sum_{m=1}^\infty {\pi m\over l}n_m.\end{equation}
Here the $n_m$'s are the occupation numbers of the finite momentum oscillator
modes.  In this case the oscillator partition function is given by:
\begin{equation}
Z_{osc}=\prod_{m=1}^\infty [1-e^{-\pi m/lT}]^{-1}\to {e^{\pi lT/6}\over \sqrt{2lT}}.
\end{equation}
Thus the entire partition function, in the limit of large $l$, for the freely
diffusing particle is given by:
\begin{equation} Z_{diff}={e^{\pi lT/6}\over \sqrt{2lT}}\sum_{m=-\infty}^\infty
\exp \left[-\left( {2\pi m\over {\cal N}a}\right)^2{1\over 2l\eta T}\right].
\end{equation}
Now taking the limit $l\to \infty$ {\it first} before taking $T\to 0$,
we may replace the sum over $m$ by a classical integral over $P_T$:
\begin{equation} Z_{diff}= {e^{\pi lT/6}\over \sqrt{2lT}}a{\cal
N}\int_{-\infty}^\infty {dP\over 2\pi}e^{-P^2/2l\eta T}={a{\cal N}\over
2}\sqrt{\eta \over \pi}e^{\pi lT/6}. \label{Zdiff}\end{equation}
We have thus obtained a groundstate degeneracy:
\begin{equation}
g_{diff}={a{\cal N}\over 2}\sqrt{\eta \over \pi}.\label{gdiff}\end{equation}
Of course, the precise value of $g$ depends on precisely how we have
defined the oscillator bath, in particular, the boundary condition at $x=l$
on the original field $\phi (x)$.  However, ratios of $g$ at different
fixed points are expected to be independent of these choices, depending
on the oscillators only through the parameter $\eta$.  This
is related to the fact that the degeneracy may be regarded as a product
of factors for each of the two boundaries of the system.  Comparing
this calculation to the case discussed in the previous paragraph we see
that the degree of freedom associated with $P_T$ has become infinitely
massive in the limit $l\to \infty$ so that the associated partition
function remains in the classical regime all the way down to $T=0$,
yielding the factor of ${\cal N}$.

Now let us consider the effect of the periodic potential, beginning with
the weak corrugation Hamiltonian of Eq. (\ref{Hfinal3}).  This
has a scaling dimension of $2\pi /\eta a^2$.  If the potential
is irrelevant, $\eta a^2<2\pi $, we expect that $g$ remains unchanged.
On the other hand, if it is relevant, $\eta a^2>2\pi $, we expect a
RG flow to a different fixed point.  The nature of this fixed point
may be deduced by assuming that $V_0$ flows to $\infty$, pinning
$\tilde \phi ''(0)$ at its minima.  On the other hand, $\tilde \phi ''$
still obeys N b.c.'s at $x=l$.  Thus its mode expansion takes the form:
\begin{equation}
\tilde \phi ''(x) = \sqrt{\eta}am + \sqrt{2\over l}
\sum_{n=0}^\infty \sin [\pi (n+1/2)x/l]p_k.\end{equation}
The oscillator part of $Z$ is the same as in Eqs. (\ref{Zos1}), (\ref{Zos2})
for the case where the bath is decoupled from the particle,
and the sum over $m$ simply gives a factor of ${\cal N}$,
the number of minima
of the potential where the particle can get localized.  Thus we obtain,
in the localized phase:
\begin{equation}
g_{loc}={{\cal N}\over \sqrt{2}}.\label{gloc}
\end{equation}
Despite all the transformations and the RG flow that went into this result the
answer is intuitively obvious.  The degeneracy is simply ${\cal N}$ times that
of the decoupled bath, reflecting the ${\cal N}$ locations where the particle
can be localized.  Note that for the decoupled bath, the original field $\phi_e$
obeys N b.c. at $x=0$ and D at $x=l$.  On the other hand, at the localized fixed
point the transformed dual field, $\tilde \phi ''$ obeys D b.c. at $x=0$ and N
at $x=l$.

It is a useful check on our results to derive the low energy spectra
 using the strong corrugation form of the Hamiltonian,
Eq. (\ref{H_n}). In this case the simple limit is the one
in which the dispersion relation $\epsilon_n\to 0$.  This
corresponds to the limit of zero hopping between sites when
the particle is localized.  We expect to renormalize to this
fixed point whenever $\epsilon_n$ of Eq. (\ref{epn}) is irrelevant.
This has dimension $\eta a^2/2\pi$ and so is irrelevant
for $\eta a^2>2\pi$.  It is very important that the hopping term
is irrelevant in the strong corrugation formulation whenever the potential
is relevant in the weak corrugation formulation and vice versa.  This
follows since the scaling dimensions are the inverse of each other.
Thus the diffusing fixed point is stable when the localized one
is unstable and vice versa.  This is consistent with our assumption
that the coupling constant flows to $\infty$ when it is relevant.
Ignoring the hopping term, $\epsilon_n$, the partition function
is just that of the decoupled oscillator bath, Eq. (\ref{Zos1}),
 multiplied by the
result of integrating over $P_c$. Note that the crystal momentum,
$P_c$ is restricted to lie in the first Brillouin zone.  The number
of momenta in the first zone when the particle lives on a line
of length ${\cal N}a$ is ${\cal N}$, so we obtain a partition function
of ${\cal N}Z_{osc}$, the same partition function
(and finite size spectrum)
 as we obtained using the weak corrugation formulation.
Now let us consider the case where the hopping term is relevant.
If $\epsilon_n$ renormalizes to $\infty$, $\phi_e(0)$ gets
pinned at one of the minima of $\epsilon_n$ which we can
take to lie at $\phi_e(0)=(2\pi /a)(P_c+m/\sqrt{\eta})$ for
integer $m$.  At $x=l$, $\phi_e$ obeys a simple D b.c.,
$\phi_e(l)=0$.  Thus the mode expansion for $\phi_e(x)$
contains a winding mode in this case:
\begin{equation}
\phi_e(x)={(l-x)\over l}{P_c+2\pi m/a\over \sqrt{\eta}}-\sqrt{2\over l}
\sum_n\sin (\pi nx/l)p_n.\end{equation}
For fixed $P_c$, the partition function is:
\begin{equation}
{e^{\pi lT/6}\over \sqrt{2lT}}\sum_{m=-\infty}^\infty e^{-(P_c+2\pi m/a)^2
/2lT\eta}.\end{equation}
We must further sum $P_c$ over the Brillouin zone.  The result
of the combined  sums is a sum over $P_c=2\pi m/{\cal N}a$ with
$m$ summed over all integers.
Thus we obtain the same partition function as in
Eq. (\ref{Zdiff}), which resulted from the weak corrugation formulation
and hence the same finite size spectrum.

We also note that the same
ratio of $g$ factors could be obtained using a {\it compact} boson
field with D and N b.c.'s. In this case
we would identify:
\begeq \phi_e (x) \leftrightarrow \phi_e (x) +  2\pi R,\endeq
and
\begeq \tilde \phi ''(x) \leftrightarrow \tilde \phi ''(x)+\frac{1}{R}.\endeq
the smallest possible compactification radius consistent with
the BSG interactions.
In fact, although the original field $\phi (x)$ is non-compact, as
we remarked above, the transformed field, $\tilde \phi ''(x)$ can
be thought of as being compact.  This follows from the mode expansion
of Eqs. (\ref{Pi0}) (\ref{phi0}) once we impose periodic boundary
conditions on $Q$.
Since $Q$ is identified with $Q+{\cal N}a$ it follows from
Eq. (\ref{phi0}) that
\begeq  \tilde \phi ''(x) \leftrightarrow
\tilde \phi ''(x)- \frac{{\cal N}}{R}.\endeq
The effective compactification radius for the dual boson is
increased by a factor of ${\cal N}$.
The allowed values of
the zero mode of the conjugate momentum given by Eq. (\ref{Pi0}) become
$\Pi_0'' = -2\pi R/{\cal N}$ the correct values for the
periodic boson radius $R/{\cal N}$.  The N b.c. that we have been discussing
corresponds to an N b.c. on a periodic boson. However the D b.c. is
{\it not} a simple D b.c.  for the periodic boson.  The boundary
sine-Gordon interaction of Eq. (\ref{Hfinal3}) has ${\cal N}$
inequivalent minima.  Thus $\tilde \phi ''(0)$ may take any of
these ${\cal N}$ inequivalent values corresponding to a
``multi-Dirichlet'' b.c.

The corresponding N boundary state is:
\begeq
|N^u(\phi =0) \rangle =g^u_N\sum_m|(0,{\cal N}2\pi m/R) \rangle \rangle
\endeq
with
\begeq g_N^u=\sqrt{{\cal N}\over 2R}\pi^{-1/4}.\endeq
The D boundary state for this compactification radius is:
\begeq |D^u(\tilde \phi = \tilde \phi_0) \rangle =
g_D^u\sum_ne^{i\tilde \phi_0nR/{\cal N}}|(nR/{\cal N},0) \rangle \rangle
\endeq
with
\begeq g_D^u=\pi^{1/4}\sqrt{R\over {\cal N}}.\endeq
The multiple-Dirichlet boundary state, $|MD \rangle $, is
obtained by summing $\tilde \phi$ over the
${\cal N}$ lattice points where the particle can
be localized:
\begeq
        |MD^u \rangle =\sum_{m=1}^{\cal N}|D^u(\phi =mR) \rangle.
\endeq
Using:
\begeq \sum_{m=1}^{\cal N}e^{i2\pi mn/{\cal N}}={\cal N}\delta_{n,{\cal N}p},
\endeq
we thus obtain:
\begeq
|MD^u \rangle =\sqrt{\cal N}|D(\tilde \phi =0) \rangle,
\endeq
proportional to the ordinary D boundary state with radius $R$.  Thus:
\begeq g_{MD}^u=\sqrt{\cal N}g_D=\pi^{1/4}\sqrt{{\cal N}R}.\endeq
Note that both degeneracy factors are simply increased by a
factor of ${\sqrt{\cal N}}$ compared to the ordinary compact D and N
values.  We may use these degeneracy factors for each boundary
to calculated the total degeneracy associated with the partition
functions.  These give us:
\begeqar
g_{diff}=(g_N^u)^2&=&{{\cal N}\over \sqrt{4\pi R}},\\
g_{loc}=g_N^ug_{MD}^u&=&{{\cal N}\over \sqrt{2}}.\endeqar
These are exactly the same values obtained from an explicit
calculation in Eqs. (\ref{gdiff}) and (\ref{gloc}).

We note that this infinite groundstate degeneracy, in the limit
where the particle is diffusing on an infinite line, is related
to the unphysical nature of the Caldeira-Leggett type model for
QBM.  The more physical applications of the model, to the quantum
wire problem, involve compact bosons from the beginning,
and do not have this strange behavior.

In general, the {\it change} in $\ln g$ due to a
change in b.c.'s at one end of the system appears to be insensitive
to the compactness or non-compactness of the boson.
This arises
from the fact that the compactness only appears when one considers
properties depending on both boundaries whereas $g$ is a property
of each boundary separately.
This can be
checked very explicitly in the boundary sine-Gordon case, where changes
in $g$ can be computed perturbatively to a very large order, and
matched against the
thermodynamic Bethe ansatz results \cite{FLS}.
In the perturbation theory, there is no dependence on the compactification
at all orders.

We note that, upon imposing periodicity of the Hamiltonian
under $\phi \to \phi +2\pi R$ or $\tilde \phi \to \tilde \phi
 +1/R$, the correct operator content at the D and N fixed points
is given by the finite size spectrum of the ordinary {\it compact}
theory (radius $R$).
Thus in order to deduce the operator
content (which is consistent with the imposed symmetry)
from the finite size spectrum,
it is convenient to use compact bosons of
radius $R$.
The main effect of the non-compactness
seems to be to increase the groundstate degeneracy in the
partition function by a factor of ${\cal N}$, a phenomena whose
origin can be traced back to the infinite mass oscillator discussed
near the beginning of this section.
In our discussion of the triangular
lattice case we shall use bosons compactified on this lattice.
Most physical quantities, such as the change in $\ln g$,
 are not affected by this
compactification.

Note that in the case when the N (diffusing) b.c. is stable,
$R>1/\sqrt{2\pi}$, $g_N<g_D$ and vice versa.  Thus the ``g-theorem''
is obeyed; the RG flow always serves to decrease $g$.

\section{quantum brownian motion on a triangular lattice}
In this section we first review the phase diagram of the boundary 3-state
Potts model.  We then introduce the triangular lattice QBM problem
and conjecture a phase diagram by analogy with that of the Potts
model.  This is substantiated and studied in more detail in the
following sections.

\subsection{Phase Diagram of the Boundary 3-State Potts Model}
The 3-state Potts model is a natural generalization of the Ising model
to a discrete spin variable that takes on three equivalent values.
(We sometimes refer to these as $A$, $B$ and $C$.)  As such
it is naturally related to QBM on a triangular lattice, as
will become clear in the next subsection.  Here we review the
boundary Potts model phase diagram.
The classical Potts model Hamiltonian contains nearest neighbor
interactions of these spins such that the energy is (-1) when
two neighboring spins are in the same state and zero otherwise.
The critical behavior of the two dimensional Potts model
 is equivalent to that of a one dimensional quantum chain.
In addition to the classical Potts interaction this also
contains a transverse field term which permutes the spins
on each site between the three states with equal amplitudes.
A complete set of conformal invariant b.c.'s for the quantum
Potts chain consists~\cite{AOS,FSPotts} of only four b.c.'s.
One of these is the fixed b.c. in which the quantum spin at
the end of the semi-infinite chain is constrained to take
a fixed value (either $A$, $B$ or $C$).  A second b.c.
is a mixed b.c. where the quantum spin at the end of the
chain is hopping back and forth between two of the states
(eg. $A$ and $B$).  A third b.c. is the free b.c. which
results from simply terminating the bulk Hamiltonian
without otherwise modifying it near the boundary.  Finally
there is a fourth b.c. whose physical interpretation
is not obvious, and which we referred to as the ``new''
b.c.

In  Ref. ({\onlinecite{AOS}) a phase diagram was conjectured
for the boundary Potts model in which both the transverse
and longitudinal fields acting on the spin at the end
of the chain was varied.  As stated above, when the
boundary transverse field is positive (the same sign as in the
bulk at the bulk critical point), the system is at
the free fixed point.  If a boundary longitudinal field is now
turned on which favors the $A$ state, the system
renormalizes to the fixed b.c.  On the other hand, if
this boundary longitudinal field has the opposite sign so
as to equally favor $B$ and $C$ then the system renormalizes
to the mixed b.c. It should be emphasized that the mixed b.c.
corresponds to the system dynamically jumping back and
forth between the $B$ and $C$ states.  In particular, it
is invariant under the $Z_2$ sub-group of $Z_3$ which
interchanges $B$ and $C$.  On the other hand, the broken
symmetry b.c. whose boundary state is sum of $B$ and $C$
fixed boundary states is unstable against an RG flow
to the mixed b.c.
If there is no longitudinal field and
the boundary transverse field is negative, then the
system renormalizes to the ``new'' b.c. The special
case where both types of boundary fields vanish is
a type of degenerate boundary condition for which
the corresponding boundary state is a linear combination
of the three different fixed boundary states.

It turns out that all of these RG flows have direct
analogies in the problem of QBM on a triangular lattice,
which thus provides a particular realization
of the quantum Potts chain.

\subsection{Back to Quantum Brownian Motion}

We consider a natural extension of the 1D QBM model discussed
in Sec. 2 to two dimensions.  The Hamiltonian of Eq. (\ref{genedqmhamil})
is extended by introducing a separate set of oscillators,
$q_k^a$ for each component of the particle co-ordinate vector, $Q^a$,
with an identical set of masses and coupling constants:
\begeq
H={\vec P^2\over 2M}+V(\vec Q) +\sum_k\left[{\vec p_k^2\over 2m_k}
+{m_k \omega_k^2 \over 2}
\left( \vec q_k- \frac{\vec Q\lambda_k}{m_k \omega_k^2}\right)^2 \right].
\end{equation}
In the ohmic case we may again represent the oscillators by
free massless relativistic boson fields, however we now get a separate
boson, $\phi^a(x)$ for each component $Q^a$.  We emphasize that
the bosons still live on a fictitious {\it one dimensional} line.
The derivation of the boundary sine-Gordon model in the strong and
weak corrugation limits carries over directly to the multi-dimensional
case.  In the strong corrugation limit the Hamiltonian is:
\begeq
H_S={1\over 2}\int_0^ldx\left[\vec \Pi^2+\left({d\vec \phi \over dx}\right)^2
\right] +\epsilon [\vec P_c-\sqrt{\eta}\vec \phi (0)].\endeq
We have dropped extraneous $'$ and $e$ notation.
Here $\epsilon (\vec P)$ is the dispersion relation (in the lowest band)
for the dissipationless problem.  $\vec \phi (x)$ obeys an N b.c. at
$x=0$ and a D b.c. at $x=l$.  The c-numbers $\vec P_c$ can take
any values in the first Brillouin zone. 
   Similarly in the
weak corrugation formulation this Hamiltonian is:
\begin{equation}
H_W = {1\over 2}\int_0^l \left[ \tilde{\vec \Pi} ^2
+\left({d\tilde{\vec \phi}\over dx}\right)^2\right] +V\left[
{-\tilde{\vec\phi} (0)\over \sqrt{\eta}}
\right].\label{HDW}
\endeq
(We have again dropped the cumbersome $''$ notation.)
$\tilde{\vec \phi} (x)$ obeys N b.c.'s at $x=0$, $l$.

The observations about boson compactness made in Sec. III carry over
to the two dimensional case.  We henceforth set:
\begeq \eta =1,\endeq
by a rescaling of the lattice spacing.  We label the physical lattice
on which the particle moves $\Gamma^*$.  It is convenient to
impose periodic b.c.'s on the particle co-ordinate, $\vec Q$, so that
we identify:
\begeq \vec Q\leftrightarrow \vec Q+{\cal N}\vec a \label{DefN}\endeq
where $\vec a$ is any vector in $\Gamma^*$ and ${\cal N}$ is a large
integer.    It follows from the derivation in Sec. III of the
boundary sine-Gordon model for QBM that $\tilde{\vec \phi}$ may
be regarded as compactified on this ``coarse lattice'' with
spacing bigger by the factor of ${\cal N}$.  Equivalently, we
may regard $\tilde{\vec \phi}$ as being compactified on the
lattice $\Gamma^*$ with an extra factor of ${\cal N}$ appearing
in the degeneracy, $g$.

We wish to consider a model of QBM in a periodic potential with
hexagonal symmetry.  The tight-binding model that we consider is
a simple generalization of the one introduced in Ref. (\onlinecite{YiKane}).
We consider a triangular lattice generated by the vectors:
\begeq \vec a_1=(a,0),\ \  \vec a_2=(a/2,\sqrt{3}a/2).
\label{a_i}\endeq
This  model exhibits interesting behavior for any value of the
dimensionless parameter $a$ but in this paper we focus primarily
on the value:
\begeq a^2=4\pi /3,\endeq
since, only then, can it be mapped into the Potts model.
The simplest form of the tight-binding Hamiltonian, before adding
dissipation, is:
\begeq H=-t\sum_{\langle i,j\rangle }c^\dagger_ic_j,\endeq
where the sum is over nearest neighbors.

In order to understand the connection with the 3-state Potts model
it is useful to focus on the symmetry transformations that leave
fixed the center of one of the triangles on the lattice.  These
consist of three-fold rotations and reflections, referred to
as 3m in the international crystal nomenclature.  There is actually
a larger point group symmetry, 6mm, when one considers transformations
that hold fixed a lattice point.  However, this is not relevant
for our purposes.  It is convenient to decompose the lattice
into three sublattices, $A$, $B$ and $C$ such that the nearest
neighbours of a point in one sub-lattice are in the other two, as
drawn in Fig. (\ref{fig:lattice}). Note
that a $2\pi /3$ rotation of the lattice maps each point on the
$A$ sublattice into a point on the $B$ sublattice and similarly
$B\to C$ and $C\to A$.  Simlarly a reflection about a line
passing through $A$ points interchanges all $B$ points with $C$
points.  Hence we may think of this point group symmetry as
the permulation group on three objects, $S^3$.  This is
the symmetry of the 3-state Potts model.
We will also
consider an on-site potential which assigns different energies to the $A$,
$B$ and $C$ sub-lattices, $v_A$, $v_B$ and $v_C$, thus breaking
the $S^3$ symmetry.

\begin{figure}
\epsfxsize=10 cm
\centerline{\epsffile{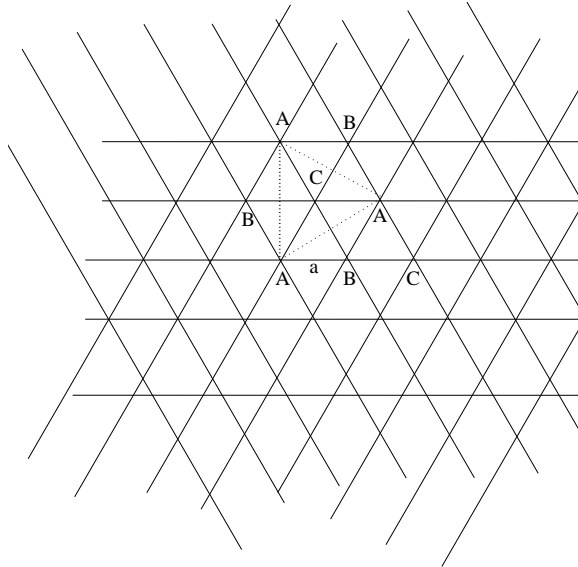}}
\caption{Triangular lattice with A,B and C sublattices marked.  Note that
the A points also form a triangular lattice with spacing $\protect\sqrt{3}a$ and
orientation rotated by $90^0$.  The B and C points form a honeycomb lattice.}
\label{fig:lattice}
\end{figure}

To map out the phase diagram let us begin with the strong corrugation
 formulation in the case $t>0$ and all
$v_i=0$.  $t$ is relevant for this lattice spacing,
(of dimension $x=2/3$) corresponding
to the boundary sine-Gordon Hamiltonian:
 \begin{equation}
 H_S=H_0 + H_{int}
 \end{equation}
 where
 \begin{eqnarray}
 H_0&\equiv& {1\over 2}\int_0^ldx\left[\vec \Pi^2+\left({d\vec \phi
 \over dx}\right)^2
\right]\nonumber \\
H_{int}&\equiv&-t\left[ \cos \sqrt{2\pi}\sqrt{2\over 3}\phi_1+\cos
\sqrt{2\pi}\left(\sqrt{1\over 6}\phi_1+\sqrt{1\over 2}\phi_2\right)+\cos
\sqrt{2\pi}\left(\sqrt{1\over 6}\phi_1-\sqrt{1\over 2}\phi_2\right)\right]
\label{trel}.
\end{eqnarray}
(The fields are all evaluated at $x=0$ but we suppress this argument.  We
have set $\vec P_C$ to 0.)
We expect $t$ to induce an RG flow to
the
perfect mobility phase, corresponding to a D boundary condition,
$\vec \phi (0)=0$.  This fixed point  is stable as we can see by considering
the lowest
dimension potential term with the full symmetry of the lattice, in the
weak corrugation formulation,
which is:
 \begin{equation}
 H_W=H_0+H_{int}
 \end{equation}
 where
 \begin{eqnarray}
H_0&\equiv& {1\over 2}\int_0^ldx\left[\tilde{\vec \Pi}^2+\left(
{d\tilde{\vec \phi}
 \over dx}\right)^2
\right]\nonumber \\
H_{int} &\equiv & -v\left[ \cos \sqrt{4\pi} \tilde \phi_2+\cos
\sqrt{\pi}\left(\tilde \phi_2+\sqrt{3}\tilde \phi_1\right)+\cos
\sqrt{\pi}\left(\tilde \phi_2-\sqrt{3}\tilde \phi_1\right)\right].
\end{eqnarray}
This has dimension 2 and is irrelevant.
Now consider turning on a potential $v_A<0$, which
favors
the $A$ sub-lattice.  We do this in the most symmetric possible way,
preserving a $Z_3$ symmetry of rotation about a lattice point, as
well as a mirror symmetry about a lattice link, the 3m subgroup
of the original 6mm point group.

Choosing the origin to lie on the $A$ sub-lattice,
this
potential is:
 \begin{equation}
H\to H-v_A\left[ \cos \sqrt{2\pi}\sqrt{2\over 3} \tilde \phi_1+\cos
\sqrt{2\pi}\left(\sqrt{1\over 6}\tilde \phi_1+\sqrt{1\over 2}\tilde \phi_2\right)+\cos
\sqrt{2\pi}\left(\sqrt{1\over 6}\tilde \phi_1-\sqrt{1\over 2}\tilde \phi_2\right)\right]
\label{vrel}\end{equation}
This operator has dimension 2/3 at the perfect mobility fixed point,
which corresponds to a D b.c. on $\vec \phi$ or equivalently,
an N b.c. on $\tilde {\vec \phi}$; it is
relevant.
Thus, for $v_A>0$, we expect a flow to a localized fixed point,
corresponding to
a D b.c. on  $\tilde {\vec \phi}$.
The stability of this fixed point can be checked by
observing
that, if $v_A$ flows to infinity, then only hopping between $A$
sub-lattice
points is possible.
 The particle gets localized on one of the $A$ sub-lattice
sites.  This is a stable fixed point since the intra-sublattice
is now $\sqrt{3}a$ and hence the hopping term gives:
 \begin{equation}
H_{int}=-t\left[ \cos \sqrt{4\pi}\phi_2+\cos
\sqrt{\pi}\left(\phi_2+\sqrt{3}\phi_1\right)+\cos
\sqrt{\pi}\left(\phi_2-\sqrt{3}\phi_1 \right)\right].\label{tintra}
\end{equation}
of dimension 2 which is irrelevant.  Note that the dimension of the {\it nearest
neighbor} hopping term is 2/3, smaller by a factor of 3 due to the reduced
distance of the hop and hence relevant.  The effect of the symmetry breaking
term, $v_A$ is to stabilize a localized phase on one of the sublattices.

The analogy with the Potts model is quite transparent.  The
perfect
mobility fixed point corresponds to the free b.c. in the Potts model.  The
potential $v_A$ corresponds to a longitudinal boundary field which
produces a
flow to the fixed ($A$) b.c.  Now consider the case $v_A<0$.  If $v_A\to
-\infty$, the particle is localized on one of the $B$ or $C$ sub-lattice
sites.
Together, these two sub-lattices define a honeycomb lattice. However, for
finite
negative $v_A$, this is not a stable fixed point.  This can be seen by
considering hopping on this honeycomb lattice.  since the nearest neighbor
distance is again $a$, this hopping term has dimension 2/3
and is relevant.  Thus, following the logic of Yi and Kane, there must
be an intermediate mobility fixed point.  In the Potts model we add a
boundary field which favors either the $B$ or $C$ state.  This leads to
a flow to the mixed b.c. of the boundary Potts model.  Thus we see that
the new fixed point found by Yi and Kane corresponds physically to the
mixed boundary fixed point in the Potts model.  In the Potts model we think
of the boundary spin as fluctuating back and forth between $B$ and $C$ states.
In the QBM problem we think of the particle as hopping back and forth between $B$
and $C$ sublattices.  As can be shown explicitly, this state has an intermediate
mobility.  We refer to this at the ``Y'' fixed point, after Yi-Kane.

In fact, we can discover yet another intermediate fixed point in the QBM
problem by pursuing this analogy further.  We now set the symmetry breaking
potentials $v_i$, to $0$ but consider a negative hopping term, $t<0$.  This
would seem to correspond to a negative transverse boundary field in the Potts
model.  This was argued to lead to a flow to another fixed point,
unimaginatively referred to as the ``new'' fixed point in 
Ref. (\onlinecite{AOS})). Thus 
we expect a new fixed point to occur in the QBM problem with $t<0$.  
We refer to this as the ``W'' fixed point.  Such a new fixed point is
possible due to the lack of particle-hole symmetry of this model.  For $t>0$
the lowest energy single-particle state has crystal momentum $\vec p=0$.
However, for $t<0$ the lowest energy single-particle states occur
at two inequivalent crystal momenta: $\vec p=\pm
(2\pi /3a,\pi /\sqrt{3}a)$.  These two lowest energy  states for the
particle are dual to the $B$ and $C$ sub-lattices in the mixed phase.   In
fact, duality in  QBM is easily understood, simply corresponding to:
\begin{equation} \vec \phi \leftrightarrow \tilde {\vec \phi }.\end{equation}
 Comparing
Eqs. (\ref{trel}) and (\ref{vrel}) we see that the flow from localized to
perfect mobility is dual to the flow from perfect mobility to localized on the
$A$ sub-lattice, induced by a positive $t$ or $v_A$ respectively.  Similarly,
the  flow from localized to mixed is dual to the flow from perfect mobility to
new, induced by a negative $t$ or $v_A$ respectively.

If both transverse and
longitudinal boundary fields are set to 0 in the Potts model we get a phase
referred to as $A+B+C$ signifying that the boundary spin remains fixed in any
of its 3 possible states.  The corresponding boundary state is a sum of three
boundary states corresponding to the 3 possible fixed b.c.'s.  Turning on a
positive transverse field produces a flow to the free b.c. whereas a negative
transverse field produces a flow to the new b.c.  The analogue in QBM is to
start with $t=0$, corresponding to the particle being localized at any
lattice site ($A$, $B$ or $C$).  We may, if we wish, include an irrelevant
intra-sublattice hopping, of dimension 2.  Now turning on $t$ induces
a flow to either perfect mobility or new fixed points.  This phase diagram
is summarized schematically in Fig. (\ref{fig:phasediagram}).  Note that,
since only localized behavior is possible on the line $t=0$, two copies
of the $Y$ fixed point must occur, as drawn.  Also note that, although
the flow away from the ``localized on $A$, $B$ or $C$'' fixed point
depends on the sign of $t$, that from the ``localized on $B$ or $C$''
fixed point does not.  This is consistent with the fact that the sign
of the hopping term cannot be changed by a  redefinition of the sign
of the electron operators for a triangular lattice but can be for a
honeycomb lattice.
\begin{figure}
\epsfxsize=10 cm
\centerline{\epsffile{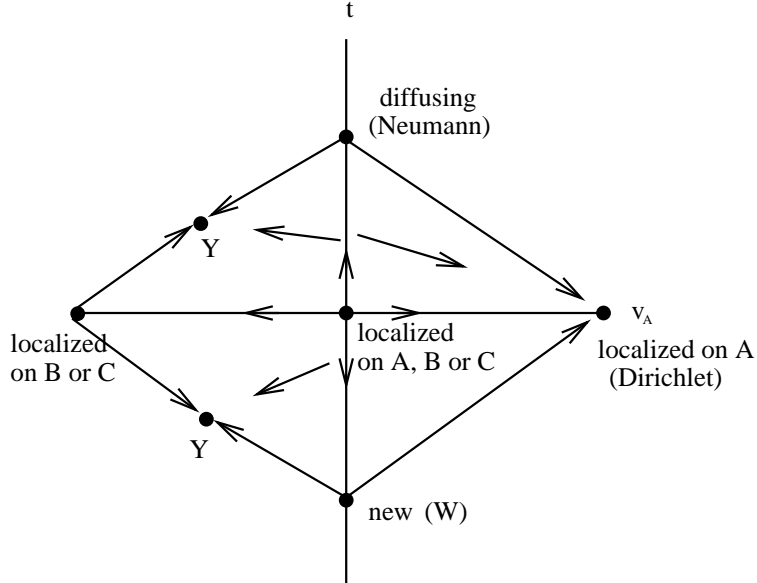}}
\caption{Phase diagram and RG flows of the triangular lattice QBM model
 with hopping strength $t$ and potential $v_A$. Neumann and Dirichlet 
b.c.'s are imposed on the dual fields, $\tilde{\vec \phi}$.
 As pointed out  in Ref. (
\protect\onlinecite{YiKane}) and discussed in Sec. VII,
the ``localized on B or C'' phase corresponds to th 
weak coupling fixed point in the 3-channel Kondo problem.}
\label{fig:phasediagram}
\end{figure}

We also comment briefly on the model with general values of $a$.
The strong corrugation Hamiltonian of Eq. (\ref{trel})
becomes:
\begin{equation}
H_{int}=-t\left[ \cos a\phi_1+
\cos a\left({\phi_1\over 2}+{\sqrt{3}\phi_2\over 2}\right)
+\cos a\left({\phi_1\over 2}-{\sqrt{3}\phi_2\over 2}\right)
\right]
\end{equation}
and the symmetry breaking weak corrugation term in Eq. (\ref{vrel}) becomes:
 \begin{equation}
H\to H-v_A\left[ \cos {4\pi \tilde \phi_1\over 3a}+
\cos \left({2\pi \tilde \phi_1\over 3a}+{2\pi \tilde \phi_2\over \sqrt{3}a}\right)+
\cos \left({2\pi \tilde \phi_1\over 3a}-{2\pi \tilde \phi_2\over \sqrt{3}a}\right)\right]
\label{VAa}\end{equation}
The scaling dimension of the hopping term at the (symmetric) localized
fixed point and of the symmetry breaking potential at the
N fixed point are:
\begeqar \Delta_D&=&a^2/2\pi \\
\Delta_N&=&8\pi /9a^2 .\endeqar
Thus the hopping term is relevant for $a^2<2\pi$ and
$v_A$ is relevant for $a^2>8\pi /9$ indicating that the perfect
diffusion fixed point is unstable in this region.  For $v_A>0$
we again expect a flow to the fixed point where the particle
gets localized on the $A$ sublattice.  The scaling dimension
of the intra-sublattice hopping term is $3a^2/2\pi$ so this
localized phase is stable in this region and we expect
an RG flow from diffusing to ``localized on $A$''.  On the other
hand, if $v_A<0$ we should consider the stability of the
``localized on $B$ or $C$'' phase.  The scaling dimension
of the inter-sublattice hopping term (between the $B$ and $C$
sub-lattices) is $a^2/2\pi$, so neither diffusing nor localized
phases are stable over the range:
\begeq 8\pi /9<a^2<2\pi .\endeq
In this entire range we expect a non-trivial fixed point.  This
can be studied using an $\epsilon$ expansion for
$a^2=8\pi /9+\epsilon$ or $a^2=2\pi -\epsilon$ for
$0<\epsilon \ll 1$.
It can be solved exactly at $a^2=4\pi /3$
as we show in the next section.  In general, however, it
remains an unsolved problem.

It is also instructive to calculate the degeneracy ($g$) factors
at the diffusing ($v_A=0$) and ``localized on $A$'' phases.
We may do this using the results in Sec. IIB for compact bosons,
taking into account the connection of the QBM problem with
compact bosons explained in Sec. III and earlier in this section.
We assume that the space on which the particle moves is periodic
with
\begeq \vec Q\leftrightarrow \vec Q+{\cal N}\vec a_i,\endeq
where the primitive vectors $\vec a_i$ are given by
Eq. (\ref{a_i}).  The unit cell area for the fine lattice with
spacing $a$ is:
\begeq V_0(\Gamma^*)=\sqrt{3}a^2/2.\endeq
(Note that this is {\it twice} the area of a triangle on the
lattice; each triangle contains 1/2 point or three points,
each of which is shared with six triangles.)
Thus the degeneracy factor for the D fixed point where the
particle is localized on any of $A$ $B$ or $C$ sublattices is:
\begeq g_{3D}={{\cal N} \sqrt{2\pi}\over 3^{1/4}a}.\endeq
This is the result for the compactified boson with dual
lattice $\Gamma^*$ from Eq. (\ref{g_D}) multiplied by the
factor of ${\cal N}$.
The degeneracy for the state where the particle is localized
on the $A$ sub-lattice only is:
\begeq g_D=g_{3D}/3,\endeq
and that for the state where the particle is localized on
the $B$ or $C$ sub-lattices is:
\begeq g_{DD}=(2/3)g_{3D}.\endeq
These factors of 1/3 and 2/3 can be understood as simply
corresponding to the number of sites on which the particle
can be localized.  Alternatively, we may think of
the D fixed point as being the same as 3D except that
the lattice constant $a$ is increased by a factor of
$\sqrt{3}$ and the factor ${\cal N}$, defined in
Eq. (\ref{DefN}), is decreased by
a factor of $\sqrt{3}$ in order that the area on which
the particle moves remains fixed.
The degeneracy factor for the N fixed point is:
\begeq g_N={{\cal N}a3^{1/4}\over 2\sqrt{2\pi}}.\endeq
This is obtained from Eq. (\ref{g_N}) for a compact
boson, multiplied by the factor of ${\cal N}$.
Note that the total degeneracy for the phase where the
particle is localized on $A$, $B$ or $C$ sub-lattices is:
\begeq
g_{loc}=g_{3D}g_N={{\cal N}^2\over 2}.\endeq
Following the discussion in Sec. III for the one-dimensional case,
we see that this is simply the number of lattice sites times
the degeneracy for the decoupled oscillator bath.  (The
factor of $1/\sqrt{2}$ that occurs in the one-dimensional case
gets squared in two dimensions.)
We also see that the ratio between fully localized and freely
diffusing fixed points is:
\begeq
{g_{3D}\over g_N} ={4\pi \over \sqrt{3}a^2}.\endeq
Thus the flow from fully localized to freely diffusing is consistent
with the g-theorem only for:
\begeq a^2<{4\pi \over \sqrt{3}}\endeq
On the other hand the ratio of g-factors between the ``localized on $A$''
and freely diffusing fixed points is:
\begeq {g_D\over g_N}={4\pi \over 3\sqrt{3}a^2},\endeq
so the flow between freely diffusing and localized on $A$ is consistent
with the g-theorem for:
\begeq a^2>{4\pi \over 3\sqrt{3}}.\endeq

It is instructive to consider a more general model with even
less symmetry than that of Eq. (\ref{vrel}).  The model of
Eq. (\ref{vrel}) in which we have changed the energy on the $A$
sites, relative to $B$ and $C$ sites still has a 3m point group
symmetry when we consider transformations holding fixed a lattice
point ($A$, $B$ or $C$).  The less symmetric model that we now
consider has no rotational symmetry whatsoever.  This model is
obtained from Eq. (\ref{vrel}) (weak corrugation formulation)
by letting the coefficients of
the 3 terms be different:
 \begin{equation} H_W= H_0-v_1\cos \sqrt{2\pi}\sqrt{2\over 3}
\tilde \phi_1-v_2\cos \sqrt{2\pi}\left(\sqrt{1\over 6}\tilde
\phi_1+\sqrt{1\over 2}\tilde \phi_2\right)-v_3\cos
\sqrt{2\pi}\left(\sqrt{1\over 6}\tilde \phi_1-\sqrt{1\over
2}\tilde \phi_2\right) .\label{nosym}\end{equation}
The behavior of this model becomes obvious in the case
$v_1\neq 0$, $v_2=v_3=0$.  In this case the (relevant) potential
only depends on the co-ordinate $Q_1$.  Thus we expect the
motion of the particle to be localized in the $1$-direction
but freely diffusing in the $2$-direction.  Now there is
only 1 minimum per unit cell for either sign of $v_1$. We may think
of the particle as diffusing freely along vertical lines
on the lattice.  To check the stability of this fixed
point we should consider the  hopping process between vertical lines.  Since
the horizontal distance between points on the $A$ sub-lattice
is  $1/2$ times the lattice
spacing (of the $A$ sub-lattice) we expect this hopping term
to have dimension 1/2 [smaller by 1/4 than the dimension 2 hopping
in Eq. (\ref{tintra}).]  Thus this fixed point is unstable.

\section{Construction of boundary states via a conformal embedding}

\subsection{Conformal embedding}

Let us consider the QBM on the triangular lattice discussed
in Section IV, considering the weak corrugation Hamiltonian
of Eq. (\ref{VAa}).
Because the symmetry among the sublattices A,B and C is
broken by the applied potential,
the generic translation symmetry of the model is
the invariance under translations which do not interchange the sublattices.
We can impose this symmetry on the system, so that we
can use the compactified formulation.
The compactification lattice $\Gamma^*$ is thus a triangular
lattice of nearest neighbor distance $\sqrt{3}a$.
Namely, we introduce the compactification
\begin{equation}
  \tilde{\vec{\phi}} \sim \tilde{\vec{\phi}} +  \Gamma^*,
\end{equation}
where $\Gamma^*$ is a triangular lattice generated by
\begeq (0,\sqrt{3}a), \ \ (3a/2,\sqrt{3}a/2 ).\label{lattice}\endeq
Its dual $\Gamma$ is generated
by $(2/3a,0)$ and $(1/3a, 1/\sqrt{3}a)$.
The allowed vertex operators made from $\vec{\phi}$ are of the form
$\exp{( i \vec{v} \cdot \vec{\phi})}$, where $\vec{v}$ is an element
of the lattice $\Gamma^*$. As mentioned in Sec. IIIC, the operator
content upon compactification corresponds to the possible
boundary perturbations given the imposed symmetry.
Note that the Hamiltonian of Eq. (\ref{VAa}) is the most general one,
up to less relevant operators, respecting the 3m point group symmetry.

For $v_A=0$, the corresponding boundary condition is Neumann
on $\tilde{\vec \phi}$,
while it is Dirichlet on $\tilde{\vec \phi}$ for $v_A \rightarrow \infty$.
However, there are other nontrivial boundary conditions
which are neither Dirichlet or Neumann, as discussed in Sec. IV.
They correspond to nontrivial phases in  QBM.
Our problem, then, is to construct and to classify
the possible boundary states of the $c=2$ CFT.
Possible boundary conditions may be restricted by higher
symmetries, such as current conservation at the boundary.
However, in the case of QBM, apparently the only requirement (at the
RG fixed point) is the conformal invariance on the boundary
and Cardy's consistency conditions.
Even if there are extra symmetries in the bulk, they are not
necessarily respected at the boundary.
This allows some extra boundary conditions which would be
forbidden if the higher symmetry is imposed on the boundary.

Unfortunately, at present we have no systematic understanding
of the boundary states in an irrational CFT,
which has an infinite number of primary fields.
While we do not know the solution to this fundamental problem,
in this paper we analyze some nontrivial boundary conditions
for the special compactification radius $a^2=4\pi /3$.
In this case, the irrational $c=2$ CFT admits a conformal embedding
in terms of rational CFTs.
For such a rational CFT, we can invoke the fusion construction
of  boundary states, to
find non-trivial boundary states for the $c=2$ CFT.

This value of $a^2$ is actually the point where
Yi and Kane~\cite{YiKane} obtained the exact value of
the mobility at a nontrivial
fixed point by a slightly different approach.
Our approach leads to a somewhat more systematic description
of the boundary states, and several additional results
including novel boundary states.

The present $c=2$ boundary CFT admits a conformal embedding
$c= 2 = 1/2 + 7/10 + 4/5 =$ (Ising) + (Tricritical Ising) + (Potts).
Namely, the partition functions on the strip can be expressed 
(in the open string channel), as 
\begin{eqnarray}
 Z_{DD}(q) &=&
(\chi^I_0 \chi^T_0 + \chi^I_{1/2} \chi^T_{3/2})
(\chi^P_0 + \chi^P_3)
\nonumber \\
&&
+ (\chi^I_0 \chi^T_{3/5} + \chi^I_{1/2} \chi^T_{1/10})
(\chi^P_{2/5} + \chi^P_{7/5}),
\label{eq:ZDDdecomp}
\\
  Z_{NN}(q) &=&
(\chi^I_0 \chi^T_0 + \chi^I_{1/2} \chi^T_{3/2})
(\chi^P_0 + \chi^P_3 + 2 \chi^P_{2/3})
\nonumber \\
&&
+ (\chi^I_0 \chi^T_{3/5} + \chi^I_{1/2} \chi^T_{1/10})
(2 \chi^P_{1/15} + \chi^P_{2/5} + \chi^P_{7/5}),
\label{eq:ZNNdecomp}
\end{eqnarray}
where $\chi^{I,T,P}_h$ is the Virasoro character of the weight $h$
for Ising, Tricritical Ising and Potts model, respectively.
(The argument $q$ is omitted.)
The two amplitudes have the same factors for Ising and Tricritical
Ising sector; the only difference is in the Potts sector.  We
emphasize that these are the partition functions for compact bosons
compactified on the lattice, $\Gamma^*$ of Eq. (\ref{lattice}).
They do not quite correspond to the
partition functions occurring in the QBM problem.  Nonetheless,
as explained in Sec IIIC, these are the partition functions
which are related to the boundary operator content.  We remark
that, since the $0$ and $3$ and also $2/5$ and $7/5$ partition
functions only occur added together in Eqs. (\ref{eq:ZDDdecomp})
and (\ref{eq:ZNNdecomp}), $Z_{DD}$ and $Z_{NN}$ can be
expressed in terms of the W-characters of the 3-state Potts
model, invariant under W-symmetry.

In fact, we have no proof of these identities.
However, we have verified, using MATHEMATICA,
that the first several ($40$ to $100$) terms
in the series expansion agree exactly.
In the following, we assume this conformal embedding.
Because of several additional pieces of evidences, we believe
it is indeed correct.
Namely, as we will show later,
we can construct several reasonable boundary states
based on the embedding.

We also note that, the torus partition function of the
$c=2$ theory, even at this particular compactification,
apparently can {\em not} be written in terms of
Ising, tricritical Ising and Potts models.
We are in a somewhat strange situation that
the boundary CFT admits the conformal embedding
while the bulk CFT does not.
Although this is difficult to understand, it does not pose
a real problem in our construction.
The fusion construction is a multiplication by constant
factors of each Ishibashi state appearing in the reference
boundary state, Eq. (\ref{bsfusion}). These constant factors are given by
the matrix elements of the matrix of modular transformations.
Thus, by construction, the obtained boundary states is
a linear combination of the Ishibashi states.
They are well contained in the Hilbert space of the
theory, as long as the original reference state is a
physical boundary state.

In fact, a similar phenomenon already appeared for the Dirichlet and
Neumann boundary conditions in the simpler $c=1$ free boson theory,
as we have discussed in Section II.
There, they are connected by the fusion with a twist operator,
which is contained in the $Z_2$ orbifold theory.
While the free boson theory is not equivalent to the
$Z_2$ orbifold, the latter appears in the discussion of
the boundary conditions in the former.

It is a valuable check to see that the dimensions of boundary
operators at D and N fixed points are reproduced by the conformal
embedding of Eqs. (\ref{eq:ZDDdecomp}) and (\ref{eq:ZNNdecomp}).
We see that at the D fixed point we have only integer dimension
operators.  (The $\Delta =1$ operators presumably correspond to
the exactly marginal operators,
$\partial \tilde{\vec \phi}$, which are not allowed by symmetry.)
On the other hand, at the N fixed points we see that there are
6 relevant operators with $\Delta =2/3$.  These must correspond
to the 3 cosine operators in Eq. (\ref{nosym}) together with
the corresponding sines.  It is interesting to note that 2 of
these operators are purely in the Potts sector while the other
4 are products of  operators in 2 or more sectors.  We expect
that the 2 pure operators are the interaction term in the
Hamiltonian of Eq. (\ref{VAa}) and the same operator obtained
by replacing all cosines by sines, as argued below.

In fact, there is an alternative conformal embedding, also
involving the Potts model, in which the Ising and tri-critical
Ising models are replaced by the $Z_3^{(5)}$ conformal field theory.\cite{FZ2}
(A detailed review of $Z_3^{(p)}$ theory is given in Ref.\onlinecite{FL-W}.)
This may be regarded as a sort of tri-critical 3-state Potts model.
Again there is a W-algebra and only W characters occur in the conformal
embedding.  This alternative conformal embedding is obtained from
Eq. (\ref{eq:ZDDdecomp}) and (\ref{eq:ZNNdecomp}) by replacing the Ising and tricritical
Ising characters by the $Z_3^{(5)}$ characters.
Here we use the (conjectured) identity
\begin{eqnarray}
\chi^I_0 \chi^T_0 + \chi^I_{1/2} \chi^T_{3/2}& = & \chi_0^5+2\chi_2^5 ,\\
\chi^I_0 \chi^T_{3/5} + \chi^I_{1/2} \chi^T_{1/10} &= &
2\chi^5_{3/5}+\chi^5_{8/5}.
\end{eqnarray}
Here $\chi_a^5$ refers to characters in the $Z_3^5$ theory.
We note that, again, this identity is only verified up to finite order with MATHEMATICA.
The equivalence
between the $Z_3^{(5)}$ model and the product of Ising and tri-critical
Ising models is only a partial equivalence.  Certain sums of conformal
characters in the two theories are presumably equal but there is not a complete
equivalence of all characters.

Further insight into the conformal embeddings can be obtained by rewriting
the interaction term in Eq. (\ref{VAa}) in a more symmetric way,
first by reexpressing everything in terms of chiral components,
using $\tilde{\phi}_L=\tilde{\phi}_R$ at the fixed point, and by
introducing the new chiral fields $\Phi_j$:
\begin{eqnarray}
\Phi_1\equiv &{8\pi \over 3a}\tilde \phi_{1L}&\nonumber\\
\Phi_2\equiv&- {4\pi\over 3a}\left(\tilde \phi_{1L}+\sqrt{3}
\tilde \phi_{2L}\right)&\nonumber\\
\Phi_3\equiv&- {4\pi \over 3a}\left(\tilde \phi_{1L}-\sqrt{3}
\tilde \phi_{2L}\right)& \label{change}
\end{eqnarray}
with propagators %
\begin{eqnarray}
\langle \Phi_{i}(z)\Phi_{i}(w) \rangle =&-{16\pi \over 9a^2}\ln(z-w)&\nonumber\\
\langle \Phi_{i}(z)\Phi_{j}(w) \rangle =&\ {8\pi \over 9a^2}\ln(z-w)\ ,&\ \ \ \ \ i\neq j .
\label{propaga}
\end{eqnarray}
The (weak corrugation) interaction term in Eq. (\ref{VAa}) now reads,
recalling $a^2=4\pi/3$,
\begin{equation}
H_{int}=-v_A(\cos\Phi_1+\cos\Phi_2+\cos\Phi_3)=-(v_A/2)
(\Psi_1+\Psi_1^\dagger)\label{hubi}
\end{equation}
where:
\begin{equation}
\Psi_1\equiv \sum_{k=1}^3 e^{i\Phi_k}. \label{intro}
\end{equation}
$\Psi_1$ corresponds to the fundamental $Z_3$ parafermion field of
the Potts model.\cite{EH,Nem}  (This generalizes the representation
 of a Majorana fermion-$Z_2$ parafermion- as $\cos \Phi$.)

 The relation between the $c=2$ free bosonic theory and the Potts model is not
straightforward. It can be formulated in terms of cosets beginning with
$SU(3)_1\times SU(3)_1$.  This conformal field theory arises from
bosonization of critical $SU(3)$ ``spin'' chains (the two factors of
$SU(3)_1$ arising from left and right movers) and the associated boundary
critical phenomena is closely related to that of the triangular
lattice QBM problem.  It will be discussed in a later paper.\cite{SU(3)}  It is
natural to associate the $SU(3)_2$ CFT with the diagonal $SU(3)$ symmetry
of this model.  The remaining coset, $SU(3)_1\times SU(3)_1/SU(3)_2$, with
$c=4/5$, gives the 3-state Potts model.  This is also equivalent
to $SU(2)_3/U(1)$.  Yi and Kane first discussed the non-trivial critical
behavior of the the triangular lattice QBM problem using results
from the 3-channel SU(2) Kondo problem, which is, in turn, related
to the $SU(2)_3$ CFT.  The second factor in the conformal embedding,
$SU(3)_2$ can be further factorized into two $U(1)$ CFT's [the maximal abelian
sub-algebra of $SU(3)$] and a c=6/5
coset, $SU(3)_2/U(1)^2$ which is naturally regarded as the $Z_3^5$ CFT.
Since the $SU(3)_1$ CFT, with c=2, is actually equivalent to two free bosons,
 this conformal embedding essentially expresses 2 free bosons in terms
of the Potts model and the $Z_3^5$ CFT.  One can certainly write the
stress energy tensor for 2 free bosons
as $T=T_1+T_2$  where:
\begin{eqnarray} T_1={1\over 5} \left[ -{1\over 2}\sum_{j=1}^3
\left(\partial\Phi_j\right)^2+ \sum_{j\neq k}
e^{i\left(\Phi_j-\Phi_k\right)}\right]\nonumber\\ T_2={1\over 5}\left[ -{3\over
4}\sum_{j=1}^3 \left(\partial\Phi_j\right)^2 -\sum_{j\neq k}
e^{i\left(\Phi_j-\Phi_k\right)}\right]\label{stressdec} \end{eqnarray}
 such that the short
distance expansion of $T_1$ with $T_2$ is trivial.  Here $\partial$ denotes
$\partial /\partial z$ where $z=\tau +ix$ and $\tau$ is imaginary time.
 $T_1$ is a stress energy
tensor with   central charge $c_1={4\over 5}$ of the coset $SU(3)_1\times
SU(3)_1/SU(3)_2$ or $SU(2)_3/U(1)$, and similarly for $T_2$ with $c_2={6\over 5}$
of the $SU(3)_2/U(1)^{2}$ coset. The parafermion $\Psi_1$ turns out to be
primary  both for $T_1$ and $T_1+T_2$, but it is not the case for most other
operators of interest. Some of the most crucial operators in the Potts model,
like the field  with conformal weight ${2\over 5}$, do not have any known
representation in  the two boson theory, neither as vertex operators nor
generalized twist fields.

 It follows that, strictly speaking, we do not have a conformal embedding. 
 As mentioned before, the situation is reminiscent of what happens at the
 Ising square point
of the $c=1$ theory: while $c=1={1\over 2}+{1\over 2}$, and a decomposition
of the $c=2$ stress energy tensor  analogous to
 (\ref{stressdec}) exists,\cite{Ki} 
the periodic boson
theory cannot be considered as the product of two Ising theories; only the 
$Z_2$ orbifold can. In the absence of an understanding of the orbifold,
one would nevertheless observe that, if the torus partition function of
the periodic boson cannot be expressed in terms of Ising model characters,
boundary partition functions can:
\begin{eqnarray}
Z_{NN}= (\chi^I_0+\chi^I_{1/2})^2\nonumber\\
Z_{DD}={\chi^I_0}^2+{\chi^I_{1/2}}^2\nonumber\\
Z_{ND}=\chi^I_{1/16}(\chi^I_0+\chi^I_{1/2}).
\end{eqnarray}

Again, the existence of a ``boundary embedding'' in our $c=2$ case
suggests that the situation is similar, and that there is some sort
of orbifold version of the bosonic
theory for which there would be a genuine conformal embedding.
Nevertheless, since the
perturbation in (\ref{VAa}) makes sense as a pure Potts operator, it is
natural to expect it to induce a flow purely in the Potts sector of the theory.
Thus we come to the important conclusion that the interaction term in
Eq. (\ref{VAa}) is purely a
Potts operator.  In fact, it is precisely the same boundary operator which
corresponds  to a boundary magnetic field,  $h\delta_{\sigma ,1}$ in the Potts
model where $\sigma =1,2,3$ labels the Potts variable.
The case $v_A<0$ then corresponds to $h<0$, inducing a flow from free to fixed
boundary conditions in the Potts model, while the case $v_A>0$  corresponds to
$h>0$, and should instead induce a flow from free to mixed. This verifies
the handwaving arguments of Sec. IV relating QBM to the Potts model.  Note
however that this connection is only established for the special choice
of lattice spacing $a^2=4\pi /3$.

We also consider the more general QBM model of Eq. (\ref{nosym}) in which
the remaining 3m point group symmetry is broken.  Upon using the conformal
embedding, we now expect that products of Potts with $Z_3^{(5)}$
(or Ising $\times$ tri-critical Ising) operators will appear in the
Hamiltonian.

In the remainder of this section we use the fusion technique to extract
information about the boundary critical points of the QBM model.  By
considering fusion in the Potts sector we can study all the fixed
points that occur in the 3m symmetric model.  We expect that we have
obtained all fixed points with this symmetry yielding the phase diagram
of Fig. (\ref{fig:phasediagram}).
On the other hand, by considering fusion in the
other sectors we obtain a collection of additional fixed points whose
properties we understand in less detail.  In Sec. VI we discuss the
integrability of some of the RG flows between fixed points and thereby
confirm the corresponding ratios of $g$-factors.

\subsection{Fusion in the Potts sector}

The Potts sector in the first term of the amplitudes
$\chi^P_0+ \chi^P_3$ and
$\chi^P_0+ \chi^P_3+2 \chi^P_{2/3}$
are exactly the fixed-fixed and free-free amplitudes
in the Potts model.
It suggests that the Dirichlet and the Neumann
boundary conditions correspond to
the free and fixed boundary conditions
of the Potts model, respectively.

In the (pure) Potts model, we can construct the free boundary state from
the fixed boundary state by the fusion with the weight-$1/8$ primary
operator ${\cal O}_{44}$ (which is absent in the bulk spectrum of the Potts.)
In fact, the fusion with the same operator
in the Potts sector in our conformal embedding
gives the Neumann b.c. from the Dirichlet b.c.
The Potts sector in the first term of~(\ref{eq:ZDDdecomp}) is the
free-free amplitude of the Potts model, and is thus transformed
to the fixed-fixed amplitude by double fusion:
\begin{equation}
\chi^P_0 + \chi^P_3 \rightarrow \chi^P_{1/8} + \chi^P_{13/8}
\rightarrow \chi^P_0 + \chi^P_3 + 2 \chi^P_{2/3},
\end{equation}
where each $\rightarrow$ means the fusion with $1/8$.
On the other hand, the Potts sector of the second term of~(\ref{eq:ZDDdecomp})
is transformed as:
\begin{equation}
\chi^P_{2/5} + \chi^P_{7/5} \rightarrow \chi^P_{1/40} + \chi^P_{21/40}
\rightarrow 2 \chi^P_{1/15} + \chi^P_{2/5} + \chi^P_{7/5}.
\end{equation}
Thus, by double fusion with $1/8$ in the Potts sector,
the Neumann-Neumann amplitude~(\ref{eq:ZNNdecomp}) is obtained
from the Dirichlet-Dirichlet amplitude~(\ref{eq:ZDDdecomp}).

In this fusion construction of the Neumann boundary state,
the ratio of the ground-state degeneracy $g_N/g_D$
is given by the ratio of modular $S$-matrix elements
of the Potts model.
Thus it automatically
agrees with the ratio of Potts degeneracy
$g_{\rm free}/g_{\rm fixed}$.
The correspondence between Dirichlet/Neumann boundary conditions
in the $c=2$ theory and fixed/free boundary conditions
in the Potts model is also consistent with
the integrable field theory approach taken in Section VI.

In a similar way, we can construct another boundary state
for the $c=2$ theory, which corresponds to the mixed boundary
condition of the (pure) Potts model.
In the Potts model, the mixed boundary condition is obtained
from the fixed boundary condition by fusion with the $2/5$-operator
($\epsilon$).
Thus we attempt a fusion with the same operator in the Potts sector
to construct a boundary state from the Dirichlet boundary state.
The Potts sector of the first term in~(\ref{eq:ZDDdecomp})
is transformed as:
\begin{equation}
  \chi^P_0 + \chi^P_3 \rightarrow \chi^P_{2/5} + \chi^P_{7/5}
   \rightarrow \chi^P_0 + \chi^P_3 + \chi^P_{2/5} + \chi^P_{7/5},
\end{equation}
which is the same as the transformation of fixed-fixed amplitude to
mixed-mixed one in the Potts model.
That of the second term is transformed as:
\begin{equation}
  \chi^P_{2/5} + \chi^P_{7/5} \rightarrow
   \chi^P_0 + \chi^P_3 + \chi^P_{2/5} + \chi^P_{7/5}
   \rightarrow
   \chi^P_0 + \chi^P_3 + 2 \chi^P_{2/5} + 2 \chi^P_{7/5}.
\end{equation}
Thus, the amplitude for the nontrivial
boundary state ( $Y$-state) is given by
\begin{eqnarray}
  Z_{YY}(q) &= &
(\chi^I_0 \chi^T_0 + \chi^I_{1/2} \chi^T_{3/2})
(  \chi^P_0 + \chi^P_3 + \chi^P_{2/5} + \chi^P_{7/5}) +
\nonumber \\
&&
(\chi^I_0 \chi^T_{3/5} + \chi^I_{1/2} \chi^T_{1/10})
(   \chi^P_0 + \chi^P_3 + 2 \chi^P_{2/5} + 2 \chi^P_{7/5}),
\end{eqnarray}
where the argument $q$ is omitted.
By the modular transformation to the ``closed string'' channel,
it is expressed as
\begin{eqnarray}
Z_{YY}(\tilde{q}) &=&
  \frac{3 + \sqrt{5}}{4\sqrt{3}}
  [(\chi^I_0 + \chi^I_{1/2})(\chi^T_0+\chi^T_{3/2})+
  2 \chi^I_{1/16} \chi^T_{7/16}]
  (\chi^P_0 + \chi^P_3 + 2 \chi^P_{2/3}) +
\nonumber \\
&&
  \frac{3 - \sqrt{5}}{4 \sqrt{3}}
  [(\chi^I_0 + \chi^I_{1/2})(\chi^T_{3/5}+\chi^T_{1/10})+
  2 \chi^I_{1/16} \chi^T_{3/80}]
  (\chi^P_{2/5} + \chi^P_{7/5} + 2 \chi^P_{1/15}),
\label{eq:ZYYclosed}
\end{eqnarray}
where the omitted argument is now $\tilde{q}$.
The ground-state degeneracy for this state is
$g_Y = \sqrt{(3 + \sqrt{5})/(4 \sqrt{3})}=
2 \cos{(\pi/5)}/\sqrt{2 \sqrt{3}}$.
Again, by construction, the ratio of the degeneracy is equal to
corresponding one in the pure Potts model.
\begin{equation}
  \frac{g_N}{g_Y} = \frac{g_{\rm free}}{g_{\rm mixed}} =
  \frac{\sqrt{3}}{2 \cos{\pi/5}}.
\end{equation}
It turns out that this Y-state is identical to
the nontrivial fixed point found by Yi and Kane~\cite{YiKane}
using a different mapping.
This will be confirmed by the calculation of mobility,
and also by a physical consideration on the relation
to the Potts model.

It has been known that the Potts model admits fixed, mixed
and free boundary conditions as conformally invariant
boundary conditions.
Recently\cite{AOS}, a new boundary condition was found in the Potts model
and related to the mixed one by the duality transformation.
The new boundary state is obtained by fusion with
the operator ${\cal O}_{22}$ of dimension $1/40$
from the fixed boundary state.
Applying the fusion with the same operator to the Dirichlet-Dirichlet
amplitude, we obtain a new boundary state, which we label $W$,
 for the present problem:
\begin{eqnarray}
 Z_{WW}(q) &=&
(\chi^I_0 \chi^T_0 + \chi^I_{1/2} \chi^T_{3/2})
(\chi^P_0 + \chi^P_3 + \chi^P_{2/5} + 2 \chi^P_{2/3}
                + 2 \chi^P_{1/15})
\nonumber \\
&&
+ (\chi^I_0 \chi^T_{3/5} + \chi^I_{1/2} \chi^T_{1/10})
( \chi^P_0 + \chi^P_3 + 2 \chi^P_{2/5} + 2 \chi^P_{2/3} + 4 \chi^P_{1/15}).
\end{eqnarray}
This is another nontrivial boundary state, which corresponds to
the Potts ``new'' boundary state.
The ground-state degeneracy of the $W$-state is given by
$g_W = 2 \cos{(\pi/5)} \sqrt{\sqrt{3}/2}$.

Thus, there is a corresponding boundary state in the present problem
for each boundary state in the Potts model.
The ratios of $g$-factors between the boundary states are identical
to the corresponding ratios in the Potts model, by construction.
Moreover, more boundary states can be constructed in the Potts
model by forming superposition of several boundary states.
Such boundary states contains dimension-zero boundary operator(s)
other than identity. Although they are often unphysical due to
their instability, they can be relevant for some physical
situations with a first-order transition.

In the present problem, the potential minima form a triangular
lattice isomorphic to $\Gamma$ for $v_A>0$.
However, for negative $v_A$, the potential minima form
a hexagonal lattice, which has two points within each unit cell
of $\Gamma$.
In the limit $v_A\rightarrow -\infty$, the corresponding
boundary condition is Dirichlet, but the boundary field is
pinned to one of two inequivalent minima.
This can be represented by ``Double Dirichlet'' boundary state,
which is the superposition of two Dirichlet boundary states
for the two minima.
Considering the correspondence Dirichlet $\leftrightarrow$ fixed
and Neumann $\leftrightarrow$ free, it would be natural if
the double Dirichlet corresponds to the sum of two fixed, say $A$ and $B$
boundary states in the Potts model.
Indeed, the double Dirichlet amplitude can be expressed as follows:
\begin{equation}
 Z_{DD,DD}(q) =
(\chi^I_0 \chi^T_0 + \chi^I_{1/2} \chi^T_{3/2})
(2 \chi^P_0 + 2 \chi^P_3 + 2 \chi^P_{2/3} )
+ (\chi^I_0 \chi^T_{3/5} + \chi^I_{1/2} \chi^T_{1/10})
        (2 \chi^P_{1/15} + 2 \chi^P_{2/5}),
\end{equation}
where the Potts part of the first term is the amplitude
for the superposed state $A+B$.

To summarize, based on the conformal embedding,
we find several boundary states for the $c=2$
problem constructed by fusion in Potts sector
as follows. There is a $c=2$ boundary state
corresponding to each Potts boundary state.

\bigskip

\begin{center}
\begin{tabular}{llc}
$c=2$ model & Potts model & $g$-factor (for $c=2$) \\
\hline \\
Dirichlet & Fixed ($A$) & $1/\sqrt{2 \sqrt{3}}$ \\
Neumann & Free & $\sqrt{\sqrt{3}/2}$ \\
Y(i-Kane) & Mixed ($AB$) & $2 \cos{(\pi/5)} /\sqrt{2 \sqrt{3}}$ \\
Double Dirichlet & $A+B$ & $\sqrt{2/\sqrt{3}}$ \\
W       & New & $2 \cos{(\pi/5)} \sqrt{\sqrt{3}/2}$ \\
\end{tabular}
\end{center}

\subsection{Fusion in Ising/Tricritical Ising sector}

It is also possible to construct other boundary states
by fusion in Ising or tricritical-Ising sectors.

The U-state is obtained from D-state by fusion with
either $1/16$ operator of Ising or $7/16$ operator of Tricritical Ising.
Its diagonal partition function is given by
\begin{equation}
Z_{UU}(q) = (\chi^I_0 + \chi^I_{1/2})
[ (\chi^T_0 + \chi^T_{3/2}) (\chi^P_0 + \chi^P_3)
+ (\chi^T_{1/10} + \chi^T_{3/5}) \chi^P_{2/5}],
\end{equation}
with the degeneracy $g_U = 1/3^{1/4}$.
Similarly, S-state is obtained by fusion with
the same operators from the N-state.
\begin{equation}
Z_{SS}(q) = (\chi^I_0 + \chi^I_{1/2})
[ (\chi^T_0 + \chi^T_{3/2}) (\chi^P_0 + \chi^P_3 + 2 \chi^P_{2/3})
+ (\chi^T_{1/10} + \chi^T_{3/5})
(2 \chi^P_{1/15} + \chi^P_{2/5})],
\end{equation}
with the degeneracy $g_S = 3^{1/4}$.
It turns out that these state represents mixtures of Dirichlet
and Neumann:
imposing Dirichlet in one component and Neumann in the other.

On the other hand, another boundary state, the 
T-state is obtained from D-state by fusion with
$3/80$ operator of Tricritical Ising.
The diagonal amplitude is
\begin{equation}
Z_{TT}(q) = (\chi^I_0 + \chi^I_{1/2})
[ (\chi^T_0 + \chi^T_{1/10} + \chi^T_{3/5} + \chi^T_{3/2}) \chi_I
+ (\chi^T_0 + 2 \chi^T_{1/10} + 2 \chi^T_{3/5} + \chi^T_{3/2}) \chi_{\epsilon}
 ],
\end{equation}
with the degeneracy $g_T = (7 + 3 \sqrt{5}/6)^{1/4}$.
Yet another state, R, is found by fusion with the same operator
from N-state.
\begin{eqnarray}
Z_{RR}(q) &=& (\chi^I_0 + \chi^I_{1/2})
[ (\chi^T_0 + \chi^T_{1/10} + \chi^T_{3/5} + \chi^T_{3/2})
        ( \chi_I + \chi_{\psi} + \chi_{\psi^{\dagger}})\nonumber \\
&& + (\chi^T_0 + 2 \chi^T_{1/10} + 2 \chi^T_{3/5} + \chi^T_{3/2})
(\chi_{\sigma} + \chi_{\sigma^{\dagger}} + \chi_{\epsilon})],
\end{eqnarray}
which has the degeneracy $g_R = [3 (7 + 3 \sqrt{5})/2]^{1/4}$.
The calculation of the mobility, which will be given
later, suggests that
they are mixtures of Y and W, namely
taking the dual of one component
of the bosons in Y or W boundary states.

\subsection{Fusion in the $Z^{(5)}_3$ sector}

We can apply similar lines of arguments as in the previous
section to the $Z^{(5)}_3$ sector
in the conformal embedding of the $c=2$ model.
As we will see, we obtain new boundary states as well as
those found using the Potts sector.
Starting from the Dirichlet boundary state,
we have tried fusion with all primary fields in the $Z^{(5)}_3$ model.
As a result, we found new boundary states,
besides the same ones obtained by Potts fusion.
For simplicity, here we focus on the boundary states with
single dimension-0 boundary operator.

A new boundary state $F$ is obtained by fusion with $Z_3^{(5)}$
operators of dimension $1/9$, $7/9$ or $13/9$
from the $D$ boundary state:
\begin{eqnarray}
Z_{FF} & = &
\chi_I ( \chi^5_0 + 2 \chi^5_2 + 3 \chi^5_{1/2} )
\nonumber \\
& &
+ \chi_{\epsilon} (2 \chi^5_{3/5} + \chi^5_{8/5} + 3 \chi^5_{1/10} ).
\end{eqnarray}
This turns out to represent free diffusion under a
magnetic field,\cite{Callan2}.\
While a detailed discussion of this identification will be
given in a later publication, we will see a signature
of the magnetic field, namely the Hall effect,
in the next subsection.

Another one, $X$ is obtained from the $D$ by fusion with
an operator of dimension $2/45$, $17/45$ or $32/45$.
The amplitude in the open string channel is given by
\begin{eqnarray}
Z_{XX}(q) &=&
        \chi_I (\chi^5_0 + 2 \chi^5_{3/5} + \chi^5_{8/5} + 2 \chi^5_{2}
                + 3 \chi^5_{1/2} + 3 \chi^5_{1/10})
\nonumber \\
&&    + \chi_{\epsilon}( \chi^5_0 + 4 \chi^5_{3/5} + 2 \chi^5_{8/5}
                        + 2 \chi^5_{2} + 3 \chi^5_{1/2} + 6 \chi^5_{1/10})
\end{eqnarray}
By construction, the ground-state degeneracy $g_X$ satisfies
\[
        \frac{g_X}{g_D} = \frac{S^0_c}{S^0_0}
\]
where $S$ is the matrix modular transformations of the $Z_3^{(5)}$ model
and $0$ and $c$ represents the identity operator and the operator
used in the fusion.
$S^0_0 = \sqrt{(5-\sqrt{5})/10}/6$ and
$S^0_c = \sqrt{(5 + \sqrt{5})/10}/3$ gives
$g_X = 4 \cos{(\pi/5)} g_D = 2 g_Y$.
Fusion with an operator of dimension $3/5$ or $8/5$ gives
the same $Y$-state previously obtained by fusion in Potts sector.

On the other hand, starting from the Neumann boundary state,
we obtain two additional boundary states $V$ and $Z$ by fusion in
the $Z_3^{(5)}$ sector.
No other states are found by fusion in $Z_3^{(5)}$ on other
known states.
To summarize, there are four new boundary states $F$,$X$, $V$
and $Z$ obtained by fusion in $Z_3^{(5)}$ sector.
As we will see later, these four states exhibits finite
Hall mobility, and are presumably related to
RG fixed points of QBM under a magnetic field.

\subsection{Mobility and nature of the boundary states}

Now let us calculate the mobility in the QBM problem.
It helps us to
identify the physical natures of the constructed boundary states.
The mobility is defined as the linear response coefficient
of the velocity of the particle to the external force.
The external $\vec F$ acting on the particle is
represented by
the additional term to the Hamiltonian, $-\vec F \cdot \vec Q$.
The linear response of the velocity of the
particle to this force is given by the Kubo formula
\begin{equation}
\langle \frac{dQ^a}{dt} \rangle =
        - F^b \int dt \langle  \frac{dQ^a(0)}{dt}  Q^b(t) \rangle .
\end{equation}
Thus the static mobility tensor can be expressed\cite{YiKane} as
\begin{equation}
\mu_{ab} = \lim_{\omega \to 0}
\frac{\pi}{\omega}
\langle \frac{dQ^a}{dt}(\omega) \frac{dQ^b}{dt}(-\omega) \rangle .
\end{equation}

Using the weak corrugation formulation we can identify
$\vec Q$ with $\vec{\tilde{\phi}} (0)$,
and the velocity operator $d \vec{Q}/dt$ with
 $d \vec{\tilde{\phi}} (0) / dt$, which corresponds to
the current operators.
Thus the mobility is deduced from the two-point
boundary correlation function of the current
$j_a = J_a + \bar{J}_a$ ($a=x,y$), where
$J_a$ and $\bar{J}_a$ are $a$-component of
the holomorphic and antiholomorphic currents.
The mobility corresponds to the conductance in the
problem of a tunneling between quantum wire\cite{Kane-Fisher,WongAffleck},
which can be also mapped to essentially the same
boundary CFT problem.
Note that we now have a two-component
boson, and thus the current has two components.

The time dependence of the correlation function is basically
insensitive to the boundary, although some coefficients do depend
on the boundary state.
As in Ref. (\onlinecite{WongAffleck}), the $J_a J_b$ and
$\bar{J}_a  \bar{J}_b$ correlations are insensitive
to the boundary conditions,
and can be fixed by an appropriate normalization
of the current operator.
On the other hand, while the $J_a \cdot \bar{J}_b$ correlation obeys
always the same power law, the amplitude depends on the boundary
condition.
The amplitude $A_{ab}$ is determined by
the boundary state~\cite{CardyLewellen} as
\begin{equation}
  A_{ab} =  \frac{\langle 0 | J_{a,1} \bar{J}_{b,1} | B \rangle}{
             \langle 0 | B \rangle},
\label{eq:Afrombs}
\end{equation}
where $|0 \rangle$ is the ground state, $|B \rangle$ is the boundary
state under consideration, and $J_{a,1}$ ($\bar{J}_{b,1}$)
is the level-1 annihilation part of the holomorphic (antiholomorphic)
current.
By a suitable normalization of the current operator, the mobility
$\mu_{ab}$
can be written as
\begin{equation}
  \mu _{ab}= \frac{\delta_{ab} + A_{ab}}{2} .
\label{eq:mufromA}
\end{equation}

Since the N boundary condition
corresponds to the delocalized phase, we normalize the
mobility in the N phase to be unity,
fixing $A^N_{ab} = \delta_{ab}$ for the Neumann boundary condition.
On the other hand, the D boundary condition corresponds to
the localized phase and the mobility should vanish, thus
$A^D_{ab} = - \delta_{ab}$.
In our conformal embedding approach,
the two level-$1$ states are identified as
\begin{eqnarray}
J_{x,-1}  | 0 \rangle =
| 0 \rangle_I |3/5 \rangle_T |2/5 \rangle_P, \\
J_{y, -1} | 0 \rangle =
| 1/2 \rangle_I |1/10 \rangle_T | 2/5 \rangle_P,
\label{eq:currentV}
\end{eqnarray}
since they are the only two
combination of Virasoro primaries with weight $1$.

The coefficients $A_{ab}$ of these states in the boundary state
are changed by fusion.
The changes are given by the modular $S$-matrix elements.
Thus we can calculate the coefficients and consequently
the mobility of the new boundary state constructed by fusion.
Since the both states contains the same Potts primary state
$|2/5 \rangle_P$,
fusion in the Potts sector gives $A_{ab} = A \delta_{ab}$.
Thus the mobility is also diagonal and isotropic:
$\mu_{ab} = \mu \delta_{ab}$.

The $N$-state is obtained by fusion with the operator of
dimension $1/8$ in the Potts sector from
the $D$-state, and the amplitudes is given by
\begin{equation}
        \frac{A^N}{A^D} =
        \frac{S^{2/5}_{1/8}}{S^{2/5}_0} \frac{S^0_0}{S^0_{1/8}}
        = -1,
\end{equation}
where $S$ is the matrix of modular transformations for the Potts model.
This is indeed consistent with the
physical consideration $\mu^D = 0$ and $\mu^N =1$.

Similarly, the Y-state is obtained by fusion with $2/5$-operator
in the  Potts sector and thus
the amplitude for the Y-state $A_Y$ is given by
\begin{equation}
        \frac{A^Y}{A^D} =
        \frac{S^{2/5}_{2/5}}{S^{2/5}_0} \frac{S^0_0}{S^0_{2/5}}
        = - \frac{3 - \sqrt{5}}{2}.
\end{equation}
The mobility $\mu$ is then
\begin{equation}
\mu^Y = (5 - \sqrt{5})/4 = 2 \sin^2{\pi/5}.
\end{equation}
This agrees with the result obtained for the
nontrivial fixed point by Yi and Kane,\cite{YiKane}
implying that our $Y$-state is identical to their fixed point
obtained by a different mapping.

The W-state is obtained by fusion with $1/40$-operator in the
Potts sector:
\begin{equation}
        \frac{A^W}{A^D} =
        \frac{S^{2/5}_{1/40}}{S^{2/5}_0} \frac{S^0_0}{S^0_{1/40}}
        = + \frac{3 - \sqrt{5}}{2}.
\end{equation}
Thus $\mu^W = (\sqrt{5} - 1)/4$.
This is equal to $1 - \mu^Y$; this is related to
the duality between the $Y$-state and $W$-state.

Next let us consider the fusion in the Ising/tricritical Ising sector.
Since the two components of the currents are related to
the Ising/tricritical Ising sector in an asymmetric manner,
in general the amplitude $A_{ab}$ is asymmetric
($A_{xx} \neq A_{yy}$) although it is diagonal ($A_{xy} =0$).
For example, the U-state is obtained from D-state
by fusion with the $1/16$ primary in the Ising sector.
Thus, the coefficients are given by
\begin{eqnarray}
A_{xx}^U  &=& \frac{S^0_{1/16}}{S^0_0} \frac{S^0_0}{S^0_{1/16}} A_{xx}^D
        = -1,
\\
A_{yy}^U  &=& \frac{S^{1/2}_{1/16}}{S^{1/2}_0}
\frac{S^0_0}{S^0_{1/16}} A_{yy}^D
        = 1,
\end{eqnarray}
where $S$ represents the matrix of modular transformations 
of the Ising model.
As a result, the mobility in the U-state is anisotropic:
$\mu^U_{xx} = 0, \mu^U_{yy}=1$ ($\mu^U_{xy}=0$).
This represents perfect mobility in $y$-direction and
complete localization in the orthogonal $x$-direction.
This would naturally correspond to a ``mixed'' b.c.
of D and N. Indeed, this is the case as we demonstrate
in the following.

A similar calculation gives the mobility for the S-state
as $\mu^S_{xx} = 1, \mu^S_{yy}=0$ ($\mu^S_{xy}=0$).
One might think that U and S are equivalent upon a space rotation,
because their mobilities are the same if we exchange $x$ and $y$
directions.
However, they are not equivalent because the underlying
lattice $\Gamma$ is not invariant under  $\pi/2$ rotations.
The inequivalence can be also seen in the following
construction of the mixed D/N boundary state.

As discussed in Section IIIB, the mixed D/N state can
be constructed from~(\ref{eq:bcinv-r}) with an orthogonal matrix $R$.
Let us choose
\begin{equation}
        R = \left( \begin{array}{cc}
                        -1 & 0 \\
                        0 & 1
                \end{array} \right) .
\label{eq:RforU}
\end{equation}
This would correspond to D (localized) in $x$-direction and
N (free diffusion) in $y$, as in the U-state.
Then the allowed zero modes are given by
the integer-coefficient linear combinations of
\begin{eqnarray}
\left[ \vec{u}_1 , \vec{v}_1 \right] &=&
        \left[ (\frac{1}{\sqrt{3 \pi}} ,0), (0,0) \right] , \\
\left[ \vec{u}_2 , \vec{v}_2 \right] &=&
        \left[ (0,0), (0,2 \sqrt{\pi} ) \right].
\end{eqnarray}
Thus, the new lattice $\tilde{\Gamma}$ introduced in
Section~IIIB is a rectangular lattice with the unit cell
of the size $\frac{1}{\sqrt{3 \pi}} \times \frac{1}{\sqrt{\pi}}$.
The diagonal amplitude for this state is given by
\begin{equation}
Z(q) = \left(\frac{1}{\eta(q)}\right)^2 \sum_{\vec{v} \in \tilde{\Gamma}^*}
                 q^{\vec{v}^2/(2 \pi)}.
\end{equation}
This actually agrees with the diagonal amplitude $Z_{UU}(q)$,
implying that the U-state is identical to the constructed mixed D/N state.
Of course, the $g$-factor also agrees:
$g = \sqrt{\pi V_0(\tilde{\Gamma})} = 3^{-1/4} = g_U$.

On the other hand, the S-state would correspond to
\begin{equation}
        R = \left( \begin{array}{cc}
                        1 & 0 \\
                        0 & -1
                \end{array} \right) .
\end{equation}
In this case, the bases for the allowed zero-modes is given by
\begin{eqnarray}
\left[ \vec{u}_1 , \vec{v}_1 \right] &=&
        \left[ (0, \frac{1}{\sqrt{\pi}}), (0,0) \right] , \\
\left[ \vec{u}_2 , \vec{v}_2 \right] &=&
        \left[ (0,0), (2 \sqrt{3 \pi},0) \right].
\end{eqnarray}
This is inequivalent to the previous case~(\ref{eq:RforU}),
because the lattice $\Gamma$ is not invariant under a rotation
by $\pi/2$.
Now the new lattice $\tilde{\Gamma}$ is again a rectangular
lattice, but with a different size $1/\sqrt{\pi} \times \sqrt{3/\pi}$.
This gives $g = \sqrt{\pi V_0(\tilde{\Gamma})} = 3^{1/4} = g_S$.
Moreover, its diagonal partition function agrees with $Z_{SS}$,
implying that the S-state is in fact the mixed D/N state.

We can construct other mixed D/N states, by considering
other (infinitely many) possible rotation matrices $R$.
The fusion construction based on the present conformal
embedding does not give those states.
It does not mean that these states do not exist.
It rather means that, only a part of the possible b.c.'s
can be accessible by the conformal embedding which
reduces the number of conformal towers effectively.
In general, we do not know how far we can reach by
a particular conformal embedding. From  physical considerations, however,
we think that the fixed points appearing in the QBM model
introduced in Section~IV have been exhausted.

The T-state is obtained from the D-state by fusion
with $3/80$ primary in the tricritical Ising sector.
The coefficients are given by
\begin{eqnarray}
A_{xx}^T  &=& \frac{S^{3/5}_{3/80}}{S^{3/5}_0}
        \frac{S^0_0}{S^0_{3/80}} A_{xx}^D
        = \frac{3 - \sqrt{5}}{2},
\\
A_{yy}^T  &=& \frac{S^{1/10}_{3/80}}{S^{1/10}_0}
        \frac{S^0_0}{S^0_{3/80}} A_{yy}^D
        = \frac{-3+\sqrt{5}}{2} .
\end{eqnarray}
Thus the mobility is given by
$\mu^T_{xx} = \mu^Y$ and $\mu^T_{yy} = \mu^W = 1 - \mu^Y$.
The T-state seems to be a mixture of Y and W boundary states.

By a similar calculation, we find the mobility in the
R-state as $\mu^R_{xx} = \mu^W$, $\mu^R_{yy} = \mu^Y$.
Namely, the mobility in the R-state is equivalent to the
T-state after $\pi/2$ rotation in the $xy$-plane.
They are again not equivalent, presumably reflecting the
fact that the lattice $\Gamma$ is not invariant.
T- and R- states are apparently given by ``taking dual''
of Y or W for only one component of the boson.
However, we do not know how they can be realized
in text of QBM.
Possibly they correspond to some anisotropic QBM.
Below we summarize the properties of all the boundary states
obtained by fusion in Potts, Ising or tricritical Ising sectors.

\bigskip

\begin{center}
\begin{tabular}{lccc}
Boundary state &  $g$-factor  & $\mu_{xx}$ & $\mu_{yy}$ \\
\hline \\
D & $1/\sqrt{2 \sqrt{3}}$ & 0 & 0\\
N & $\sqrt{\sqrt{3}/2}$ & 1 & 1 \\
Y & $2 \cos{(\pi/5)} /\sqrt{2 \sqrt{3}}$
        & $(5 - \sqrt{5})/4$ & $ (5 - \sqrt{5})/4$ \\
W & $2 \cos{(\pi/5)} \sqrt{\sqrt{3}/2}$
        & $(-1 + \sqrt{5})/4$ & $(-1 + \sqrt{5})/4$ \\
R & $ [3 (7 + 3 \sqrt{5})/2]^{1/4}$
        & $(-1 + \sqrt{5})/4$ & $(5 - \sqrt{5})/4$ \\
S & $ 3^{1/4}$ & $1$ & $0$ \\
T & $ (7 + 3 \sqrt{5}/6)^{1/4}$
        & $ (5 - \sqrt{5})/4$ & $(-1 + \sqrt{5})/4$ \\
U & $ 1/3^{1/4}$ & $0$ & $1$ \\
\end{tabular}
\end{center}

\bigskip

In the embedding using $Z_3^{(5)}$, the level-1 ``current'' states
corresponds to the product state
$|\frac{3}{5} \rangle_5 | \frac{2}{5} \rangle_P$,
where $|\frac{3}{5} \rangle_5$ is the primary state
of the $Z_3^{(5)}$ theory with weight $3/5$.
In fact, there are two such primary states in the $Z_3^{(5)}$ theory,
which are complex conjugates.
Let us denote one of them as $| \frac{3}{5}^* \rangle$.
In this approach, the natural identification would be
\begin{eqnarray}
\frac{1}{2} ( J_{x,-1} + i J_{y,-1} ) | 0 \rangle &=&
        |\frac{3}{5} \rangle_5 | \frac{2}{5} \rangle_P, \\
\frac{1}{2} ( J_{x,-1} - i J_{y,-1} ) | 0 \rangle &=&
        |\frac{3}{5}^* \rangle_5 | \frac{2}{5} \rangle_P .
\end{eqnarray}
Assuming this correspondence,
for a boundary state constructed from the D-state
by fusion with the $Z_3^{(5)}$ operator $\phi$,
we find
\begin{equation}
A_{ab} = - \delta_{ab} {\rm Re} A - \epsilon_{ab} {\rm Im} A,
\end{equation}
where $\epsilon_{ab}$ is the antisymmetric tensor
$\epsilon_{xy} = - \epsilon_{yx} = 1, \epsilon_{xx}=\epsilon_{yy} = 0$.
Here $A$ is defined as
\begin{equation}
A = \frac{S^{3/5}_{\phi} S^0_0}{S^{3/5}_0 S^0_{\phi}},
\end{equation}
where $S^i_j$ is the matrix of modular transformations of the
$Z_3^{(5)}$ theory.
Thus, when the matrix of modular transformations acquires an imaginary
part, there is a non-vanishing off-diagonal mobility.
This corresponds to the Hall effect.
Physically, such an effect is expected in  QBM under
a magnetic field.

The $Y$-state is obtained by fusion with the $8/5$-operator in the $Z_3^{(5)}$
sector. Thus
\begin{equation}
A_Y =
\frac{S^{3/5}_{8/5}}{S^{3/5}_0} \frac{S^0_0}{S^0_{8/5}}
= - \frac{3 - \sqrt{5}}{2}.
\end{equation}
This gives the same result as we have obtained
by fusion in the Potts sector.
The off-diagonal mobility vanishes in this case.

On the other hand,a similar calculation on the F-state reads
\begin{equation}
A_F =
\frac{S^{3/5}_{1/9}}{S^{3/5}_0}  \frac{S^0_0}{S^0_{1/9}}
= \frac{-1 - \sqrt{3} i}{2} .
\end{equation}
This complex amplitude gives a non-vanishing off-diagonal mobility:
\begin{equation}
\mu_{xx} = \mu_{yy} = \frac{3}{4} ,
\hspace*{2cm}
\mu_{xy} = - \mu_{yx} = \frac{\sqrt{3}}{4}.
\end{equation}
By a fusion with the conjugate operators, we can
also obtain the boundary state with opposite sign
of the off-diagonal mobility.

The other three states $X$,$V$ and $Z$ also exhibit the
non-vanishing off-diagonal mobility.
Thus they should also correspond to a critical behavior
under a magnetic field.
The mobility in these states is summarized as follows.

\bigskip

\begin{center}
\begin{tabular}{lccc}
Boundary state &  $g$-factor  & $\mu_{xx}$ & $\mu_{xy}$ \\
\hline \\
F & $\sqrt{\frac{2}{\sqrt{3}}}$ & $\frac{3}{4}$ &  $\pm \frac{\sqrt{3}}{4}$ \\
X & $[ \frac{14}{3} + 2 \sqrt{5} ]^{1/4}$ &
$\frac{1+\sqrt{5}}{8}$ & $\pm \frac{3 \sqrt{3} - \sqrt{15}}{8}$ \\
V & $[ 6 (7 + 3 \sqrt{5} ) ]^{1/4}$ & $\frac{7-\sqrt{5}}{8} $ &
        $\pm \frac{ 3 \sqrt{3} - \sqrt{15}}{8}$ \\
Z & $ \sqrt{2 \sqrt{3}}$ & $\frac{1}{4}$ & $\pm \frac{\sqrt{3}}{4}$
\end{tabular}
\end{center}

\subsection{Structure of the non-trivial boundary state}

We have obtained the explicit expression of the amplitude
for the boundary state $Y$.
It gives some insight
into the structure of the nontrivial boundary state.

In general, a conformally invariant
boundary state should be a linear combination of
Ishibashi states. Each Ishibashi state is constructed from a
spinless primary field that appears in the bulk theory.
Thus, the closed string channel amplitude~(\ref{eq:ZYYclosed})
must be consistent with the bulk spectrum.

In $c=1$ free boson theory with a generic compactification radius,
each bosonic Fock space build on a vacuum corresponds to an irreducible
representation of the Virasoro algebra (except for zero winding-number
sector.) However, in $c=2$, a bosonic Fock space is always reducible
and contains an infinite number of irreducible representations
 of the Virasoro algebra.
Thus, there are infinitely more Ishibashi states other than those
built on a bosonic vacuum.
Nevertheless, the $c=2$ Dirichlet and Neumann boundary states have
simple expressions in terms of bosons, and their (self-)amplitudes
are written as $\sum_{h} \tilde{q}^h / \eta(\tilde{q})^2$.
Roughly speaking, the factor $1/\eta(\tilde{q})^2$ means the boundary
state is made of whole bosonic Fock space.

However, the non-trivial boundary state is not made of such ``bosonic''
boundary states. This can be seen as follows.
Expressing the diagonal amplitude~(\ref{eq:ZYYclosed}) as a sum of
the character $\tilde{q}^h/ \eta(\tilde{q})^2$ of the Heisenberg algebra,
\begin{eqnarray}
Z_{YY}(\tilde{q}) &=&
\left( \frac{g_Y}{\eta(\tilde{q})} \right)^2    
\left[ 1
+ (21-9 \sqrt{5}) \tilde{q}^{1/6}
+ \frac{9(3-\sqrt{5})}{2} \tilde{q}^{1/2}
+ (16 - 6 \sqrt{5}) \tilde{q}^{2/3} \right. \nonumber \\
&&
\left.
+ (5 - 3 \sqrt{5}) \tilde{q}
+ 9 (3 - \sqrt{5}) \tilde{q}^{7/6}
+ 3 (7-3\sqrt{5}) \tilde{q}^{3/2}
+ (5 - 3 \sqrt{5}) \tilde{q}^{5/3} + \ldots
\right].
\end{eqnarray}
This involves negative coefficients at least for $\tilde{q}$
and $\tilde{q}^{5/3}$.
While any given partition function can be written as an infinite sum of the
Heisenberg character, the coefficients reveal the nature of the boundary state.
If the boundary state is a linear combination of the bosonic boundary
states, all the coefficients must be positive.

Thus,
this is a highly nontrivial boundary state which cannot
be constructed by a generalization such as eq.~(\ref{eq:bcinv-r})
or as in Ref. (\onlinecite{Callan2}) of the Dirichlet or Neumann.
To our knowledge, this is the first proof that such a non-bosonic
boundary state does exist in a free boson field theory.
Similar non-triviality can be shown for W,R,T,X and V states.

On the other hand, since the boundary state is a linear combination
of Ishibashi states, the diagonal amplitude must be a linear combination
of $c=2$ Virasoro characters with positive coefficients.
In fact, the amplitude~(\ref{eq:ZYYclosed}) is expressed as
\begin{equation}
  4 \sqrt{3} Z_{YY}(\tilde{q}) =
   (3 + \sqrt{5})\chi^2_0 + 6 (3 - \sqrt{5})\chi^2_{1/6}
 + 18 \chi^2_{1/2} + (18-2\sqrt{5}) \chi^2_{2/3} + \ldots ,
\end{equation}
with positive coefficients, where
the $c=2$ Virasoro characters for weight $h$ are given by
$\chi^2_h = q^{h-1/12}/\prod_{n=1}^{\infty}(1-q^n)$ for $h>0$ and
$\chi^2_0 = (q^{-1/12} - q^{11/12})/\prod_{n=1}^{\infty}(1-q^n)$ for $h=0$.
Moreover, all the primary weights correspond to
those of spinless fields in the bulk spectrum.
These are guaranteed by the fusion construction.

Thus this nontrivial boundary state is certainly a consistent
boundary state for the
$c=2$ free boson theory, although it appears to be very
complicated in terms of the bosons.
A possible direction to extend the construction of the
nontrivial state for different lattices $\Gamma^*$
is to write down the boundary state in the bosonic language and
then guess a generalization of the boundary state for other lattices.
However, so far we have been unable to develop this approach.

\section{Integrable Field Theory Analysis}
\subsection{Integrable flows away from the
perfect mobility (Neumann) fixed point}
Here we consider the Hamiltonian of Eq. (\ref{nosym}) for small
$v_1$, $v_2$, $v_3$.  This model  turns out to be integrable
for general values of $v_1$, $v_2$, $v_3$ as discussed in detail in
Ref. (\onlinecite{SS}).
This integrability is, in a distant
sense, related to the existence of the conformal embedding;
more precisely,  it
is the  underlying parafermionic algebra satisfied by the
vertex operators  of
(\ref{VAa}) at  that particular radius that ensures the existence of non
local conserved currents.\cite{SS}

For general values of $v_1$, $v_2$, $v_3$ the scattering theory describing
(\ref{nosym}) is simply made up of three left moving and three right moving
massless particles with  mass parameters $m_1,m_2,m_3$, energy and momenta 
being parametrized by rapidities $e=\pm p=m_j e^\theta$.  Scattering between
left and right movers is trivial since the theory is scale invariant in the
bulk, while scattering among left or among right movers is described by  pure
CDD factors:  $S_{k,k\pm1}=i \tanh\left({\theta\over 2}-{i\pi\over 4}\right)$,
$S_{1,3}=1$. Only the  first particle  then scatters non trivially on the
impurity, with a boundary R matrix again given by a CDD factor, $R=i
\tanh\left({\theta-\theta_B\over 2}-{i\pi\over 4}\right)$. Here, $\theta_B$
parametrizes the impurity energy scale $T_B\propto e^{\theta_B}\propto v_i^3$.
The exact dependence of $T_B$, as well as the dependence of the  mass
parameters upon the $v_i$'s are complicated and will not be needed in the
following, except for special cases.

The main result of the integrable approach is that the impurity free energy
can be computed via the thermodynamic Bethe ansatz (TBA).\cite{SS} If we
parametrize  the filling fractions  of the particles  by pseudo energies
$\epsilon_j$ as $\hbox{f}_j={1\over 1+e^{\epsilon_j/T}}$, the latter are
solutions of a set of integral equations
\begin{equation}
\epsilon_j=m_je^\theta-T\sum_k N_{jk}
\int {d\theta'\over 2\pi}{1\over \cosh(\theta-\theta')}
\ln\left(1+e^{-\epsilon_k(\theta')/T}\right),\label{tba}
\end{equation}
Here, $j,k$ take values $1,2,3$ and can be represented as the nodes of an $A_3$
Dynkin diagram with incidence matrix $ N_{jk} $ (crossed nodes indicate the
existence of a source term in the TBA equations)
\bigskip
\noindent
\vskip.2cm\hskip-3cm\centerline{\hbox{
\rlap{\raise28pt\hbox{\hskip.38cm
$\bigotimes$------$\bigotimes$------$\bigotimes$}}
\rlap{\raise39pt\hbox{$\hskip .3cm m_1\hskip.5cm m_2\hskip.5cm
m_3$}}}
}
\smallskip

\noindent  The impurity  free energy reads then
\begin{equation}
F_{imp}=-T\int{d\theta\over 2\pi}{1\over \cosh(\theta -\theta_B)}
\ln\left(1+e^{-\epsilon_1(\theta)/T}\right).\label{tbaf}
\end{equation}

Formula (\ref{tbaf}) gives us quick access to  the change of boundary entropies
between the UV fixed point (Neumann boundary conditions, $T/T_B=\infty$) and
the possible infrared (IR) fixed points ($T/T_B=0$) our model can flow to. Rather than
entropies, we will use the ground state degeneracy, $g=e^{S/T}$.  The change of
g-factor is  then neatly expressed as
\begin{equation} {g_{UV}\over
g_{IR}}=\left({1+e^{-\epsilon_1(-\infty)/T}\over
1+e^{-\epsilon_1(\infty)/T}}\right)^{1/2}.\label{entrch} \end{equation}  When
$\theta\to -\infty$, the source terms just drop  from the equations, the
$\epsilon$'s go to constants  that obey the system, setting
$x_j=e^{-\epsilon_j(-\infty)/T}$,
$$ x_j=\prod_k \left(1+x_k\right)^{N_{jk}/2}.$$
This is solved right away by $x_1=x_3=2, x_2=3$.

Setting similarly $y_j=e^{-\epsilon_j(\infty)/T}$, the system obeyed
by the $y_j$'s depends on the source terms: different possibilities
can arise. If $m_1\neq 0$, the source
term for $\epsilon_1$ diverges, and therefore $y_1=0$. If $m_1=0$ and
$m_2\neq 0$, in the IR the system is simply $\epsilon_1=0$ so $y_1=1$.
Finally, if $m_1=0$ and $m_2=0$, the IR system is
$$
y_1=y_2=(1+y_1)^{1/2}
$$
with solution $y_1={1+\sqrt{5}\over 2}$. We thus obtain the possible ratios
of degeneracies
\begin{eqnarray}
{g_{UV}\over g_{IR}}=&\sqrt{3}&,\ \ \ \ \ m_1\neq 0\nonumber\\
{g_{UV}\over g_{IR}}=&\sqrt{{3\over 2}}&,\ \ \ \ \ m_1=0,m_2\neq 0\nonumber\\
{g_{UV}\over g_{IR}}=&{\sqrt{3}\over 2\cos{\pi\over 5}}&,\ \ \ \ \
m_1=m_2=0 \label{possibrati}
\end{eqnarray}
with $2\cos{\pi\over 5}=\sqrt{3+\sqrt{5}\over 2}$.

We obtained the first ratio in Secs. III and IV when discussing   the flow
from  N boundary conditions to  D boundary conditions  (i.e. freely diffusing
to localized) in the case $v_1=v_2=v_3>0$.   For general $v_i$, the potential
of Eq. (\ref{nosym}) has a unique minimum (within each unit cell) and the TBA
equations in the first case describe the flow from N to D in the general case
where the three-fold symmetry is broken. This is confirmed by the analysis of
the central charge associated with (\ref{tba}) which gives generically $c=2$,
hence indicating a flow within the whole two boson theory. (i.e. not purely
within the Potts sector.)  The third ratio was also obtained previously  when
we discussed the flow from N to Y in the case $v_1=v_2=v_3<0$. Since this ratio
is obtained only when $m_1=m_2=0$, it follows that the flow from N to Y occurs
only in the three-fold symmetric case. Meanwhile, the analysis of the central
charge associated  with (\ref{tba}) gives only $c={4\over 5}$, indicating that
the flow takes place within the Potts sector of the theory only.

Finally the second ratio is somewhat trivial: it corresponds to a flow from
Neumann to Dirichlet boundary conditions for the field $\tilde \phi_1$ while
the field $\tilde \phi_2$ remains at the Neumann fixed point.  This corresponds
to the $U$ boundary state discussed in Sec. IV.  This is obtained by setting
$v_2=v_3=0$, while $v_1$ can take either sign.  The analysis of the central
charge associated with (\ref{tba})  gives $c=1$, confirming that the flow takes
place in the $c=1$ sector only.

\subsection{$W_3$ integrable flows}

Integrable flows can be used to explore more thoroughly the phase space
and the possible fixed points in our problem.

The integrability of  the flow from free to fixed or mixed boundary conditions
in the Potts model, and therefore, from Neumann  to Dirichlet or the Y fixed
point in the $c=2$ theory, is ultimately related with  the fact that  the $Z_3$
parafermionic theory (the three state Potts model) can be represented as the
coset  ${SU(2)_3\over U(1)}$. The field $\Psi_1$ is then the ``adjoint''
operator in this construction.\footnote{In a general  coset construction, one
usually calls adjoint \cite{genref} the operator obtained by taking the
identity representation for the algebras in the numerator, and the adjoint for
the algebra in the denominator.} We observe that  this theory  can also be
represented as the coset ${SU(3)_1\times SU(3)_1\over SU(3)_2}$. The  adjoint
operator for this coset now has dimension $\Delta={2\over 5}$, and physically
corresponds  to the energy operator of the Potts model. It can be shown to
define an integrable perturbation  that preserves the $W_3$ symmetry: more
precisely, there is a natural deformation  of the current $W_3$ that is
conserved, at least perturbatively, away from the UV fixed point, if this fixed
point itself preserves $W_3$ symmetry. It is natural  to assume that this
extends all the way to the IR fixed point, which therefore  should have $W_3$
symmetry too.

In our problem, the operator with $\Delta={2\over 5}$ is present in the
spectrum for mixed boundary conditions in the Potts model, or, equivalently, at
the Y fixed point in the $c=2$ theory.  In the Potts model it corresponds to
applying a magnetic field at the mixed ($AB$) fixed point which breaks the
remaining $Z_2$ symmetry and favors the $A$ state.  This corresponds to a flow
from mixed to fixed.  In the QBM problem  we expect this operator to produce an
RG flow from Y to localized fixed points; if Y corresponds to the particle
being on $A$ and $B$ sub-lattice then the flow is to a state where the particle
is localized on an $A$ site.  This RG flow can also be studied using
integrability.

For the more general coset ${SU(3)_1\times SU(3)_k\over SU(3)_{k+1}}$,
perturbed by the operator of weight $\Delta=1-{3\over k+4}$, which preserves
$W_3$ symmetry, $S$ matrices and $R$ matrices are known \cite{Ahn},  and  the
impurity free energy can be computed once again using the thermodynamic Bethe
ansatz. It reads
\begin{equation}
F_{imp}=-T\int {d\theta\over 2\pi}
\left[G_1(\theta-\theta_B)\ln\left(1+e^{-\epsilon_{11}/T}\right)
+G_2(\theta-\theta_B)\ln\left(1+e^{-\epsilon_{21}/T}\right)\right].
\label{freeenergy}
\end{equation}
Here, as before,  $T_B$ is an impurity energy scale, $T_B\propto e^{\theta_B}$,
 $G_1$ and $G_2$ are known kernels,
and the $\epsilon$ are pseudo energies, solutions of the
equations:
\begin{equation}
\epsilon_{ij}=\delta_{j1} me^{\theta}-T \sum_k
N_{ij,ik}G_1*\ln\left(1+e^{-\epsilon_{ik}/T}\right)-T\sum_l
N_{ij,lj}G_2*\ln
\left(1+e^{\epsilon_{lj}/T}\right).\label{tbaeqs}
\end{equation}
Here, $ij$ are line and column labels on the following diagram (the cross on
the $(1,1)$ node stands for the mass term in (\ref{tbaeqs}))
\bigskip
\noindent
\hskip-10cm\vskip.12cm\centerline{\hbox{
\rlap{\raise28pt\hbox{\hskip.38cm
$\bigcirc$------$\bigcirc$----------$\ldots$----$\bigcirc$--------$\bigcirc$}}
\rlap{\raise39pt\hbox{$\hskip .3cm 1\hskip4.6cm
k$}}
\rlap{\raise14pt\hbox{\hskip.24cm\Big|$\hskip.95cm\Big|\hskip1.3cm\hskip1.cm
\Big|\hskip1.1cm\Big|$}}
$\bigotimes$------$\bigcirc$----------$\ldots$----$\bigcirc$--------$\bigcirc$}}
\bigskip
\bigskip
\noindent
$i=1,2$, $j=1,\ldots, k$, and $N_{ij,kl}$ its incidence matrix.

We will not need the exact form of the kernels in what follows, and shall
concentrate on  the ground state degeneracies in the UV and in the IR. From
the previous formulas, we find the simple result (the $y_{j1}$ all
vanish here)
\begin{equation}
{g_{UV}\over g_{IR}}={1\over
2}\left[\ln(1+x_{11})+\ln(1+x_{21})\right],\label{ratioofg}
\end{equation}
where $x_{ij}=e^{-\epsilon_{ij}/T}$ is solution of the
$\theta\to-\infty$ limit
of the TBA system:
\begin{equation}
x_{ij}=\prod_k \left(1+x_{ik}\right)^{N_{ij,ik}/2}\prod_l
\left(1+{1\over x_{lj}}\right)^{-N_{ij,lj}/2}.\label{xeqs}
\end{equation}

For the case  $k=1$ of interest here, the system reduces to
$x=x_{11}=x_{21}$,
$x={\sqrt{x}\over \sqrt{1+x}}$ or $x=2\cos{\pi\over 5}$. Hence
\begin{equation}
{g_{UV}\over g_{IR}}=2\cos{\pi\over 5}.\label{pottsratio}
\end{equation}
This is the ratio   $g_{\hbox{mixed}}/g_{\hbox{fixed}}$ in the Potts model, or
the ratio  $g_Y/g_D$ for the two boson system. We thus expect that perturbation
by the operator $\Delta={2\over 5}$  does not lead to  a new fixed point, but
induces a flow from Y to D in the two boson system. Let us stress that this is
a rather  abstract statement: although we know  the operator with
$\Delta={2\over 5}$ is present in the spectrum  (because of the partition
functions analysis), we are not able to represent it in terms of bosonic
operators, since we do not have an entirely clear picture of  what the Y
boundary conditions mean for the bosons.

\section{Generalization to higher dimensions and relations with the Kondo model}
\subsection{Generalization to higher dimensions}

Many of our results can be generalized to the case of $n-1$ bosons,
corresponding to QBM on an $(n-1)$-dimensional lattice.  Yi and Kane found
a generalization of the $Y$ fixed point for all $n>3$ using a mapping
onto the $n$-channel $SU(2)$, overscreened Kondo model.  They also
observed
that the 3-dimensional  ($n=4$) case corresponds to QBM on a diamond lattice.
The conformal embedding is now naturally understood in terms of an $SU(n)$
``spin'' chain.  We may decompose the product of left and right moving
$SU(n)_1$ factors into $SU(n)_2$ $\times$ a $Z_n$ parafermion.  We may then
divide out the maximal abelian sub-algebra,  $U(1)^{n-1}$ from $SU(n)_2$.
Taking into account that $SU(n)_1$ is equivalent to $n-1$ free bosons, we see
that we have constructed a conformal embedding:
\begin{equation}
n-1={2(n-1)\over n+2}+{n(n-1)\over n+2}.\end{equation}
The first component of the embedding is the theory of $Z_n$ parafermions, that
can also be considered as the coset ${SU(2)_n\over U(1)}$, with $c={2(n-1)\over
n+2}$. The fundamental parafermion in this theory has weight ${n-1\over n}$,
and  can be bosonized using a system of $n$ bosons satisfying (for the chiral
components)
 \begin{eqnarray} \langle \Phi_{i}(z)\Phi_{i}(w)\rangle =&-2{n-1\over
n}\ln(z-w)&\nonumber\\
\langle \Phi_{i}(z)\Phi_{j}(w)\rangle =&{2\over n}\ln(z-w),\ &\ \ \ i\neq
j,\label{propagai} \end{eqnarray}
as
\begin{equation} \Psi_1=\sum_{k=1}^n
e^{i\Phi_k}.\label{moreboso} \end{equation}
The embedding then follows from the
decomposition $T=T_1+T_2$, where
\begin{eqnarray} T_1={1\over n+2} \left[ -{1\over
2}\sum_{j=1}^n \left(\partial\Phi_j\right)^2+ \sum_{j\neq k}
e^{i\left(\Phi_j-\Phi_k\right)}\right]\nonumber\\ T_2={1\over n+2}\left[
-{n\over
4}\sum_{j=1}^n \left(\partial\Phi_j\right)^2 -\sum_{j\neq k}
e^{i\left(\Phi_j-\Phi_k\right)}\right] \end{eqnarray}
such that the short distance expansion of $T_1$ with $T_2$ is trivial, $T_1$ is
a stress energy tensor with   central charge $c_1=2{n-1\over n+2}$ of the coset
$SU(n)_1\times SU(n)_1/SU(n)_2$ or $SU(2)_n/U(1)$, and similarly for $T_2$ with
$c_2={n(n-1)\over n+2}$ of the $SU(n)_2/U(1)^{n-1}$ coset. $T=T_1+T_2$ of
course is the  stress energy tensor of the initial $(n-1)$ boson theory.

It turns out that the $n-1$ boson theory with
boundary perturbation
\begin{equation}
-\sum_{j=1}^n v_j\cos\Phi_j,\label{newbdr}
\end{equation}
is integrable \cite{SS}, the particular case where all the
$v_j$'s are equal corresponding again to a perturbation of
the form $\Psi_1+\Psi_1^\dagger$. The impurity free energy reads,
generalizing (\ref{tbaf})
\begin{equation}
F_{imp}=-T\int {d\theta\over 2\pi} {1\over\cosh(\theta-\theta_B)}
\ln\left(1+e^{-\epsilon_n(\theta)/T}\right),\label{morefbdr}
\end{equation}
where the $\epsilon$ satisfy the same TBA equations as (\ref{tba}), but the
incidence  matrix is the one of the following diagram
\bigskip
\noindent
\hskip-5cm\vskip.4cm\centerline{\hbox{\rlap{\raise28pt\hbox{$\hskip5.6cm
\bigotimes\hskip.25cm n-1$}}
\rlap{\lower27pt\hbox{$\hskip5.5cm\bigotimes\hskip.3cm n$}}
\rlap{\raise15pt\hbox{$\hskip5.3cm\Big/$}}
\rlap{\lower14pt\hbox{$\hskip5.2cm\Big\backslash$}}
\rlap{\raise15pt\hbox{$1\hskip1cm 2\hskip1.5cm \hskip.8cm n-3$}}
$\bigotimes$------$\bigotimes$-- -- --
--$\bigotimes$-- -- --$\bigotimes$------$\bigotimes$\hskip.5cm $n-2$
}}

\bigskip
\noindent
Here again, $T_B\propto v_j^n$,  and the $m_j$ are dimensionless
numbers functions of the couplings $v_j$.

In the UV, the system of equations for the $x$'s is solved easily with
$x_j=(j+1)^2-1$, $j=1,\ldots,n-2$; $x_{n-1}=x_n=n-1$. If  $m_n$
does not vanish, one has $y_n=0$, and then we obtain the ratio
\begin{equation}
{g_{UV}\over g_{IR}}=\sqrt{n}.\label{moreratio}
\end{equation}
This generalizes the first ratio in (\ref{possibrati}), and corresponds to a
flow  from Neumann  to Dirichlet boundary conditions in the $n-1$ bosons
problem, where at the D fixed point, the field lies on a $n-1$ dimensional
lattice generalizing $\Gamma^*$.

When $m_n$ vanishes, the ratio of degeneracies can take various values
depending on which of the other $m_j$'s vanish. A particularly interesting case
is when  all $m_j$ but  $m_{n-1}$ vanish. This corresponds to a perturbation
with all $a_j=0$, $j\neq 1$, and $a_1>0$. In that case, one gets a  ratio 
\begin{equation} {g_{UV}\over
g_{IR}}={\sqrt{n}\over 2\cos{\pi\over n+2}},\label{evenmoreratio} \end{equation}
generalizing the case $n=3$; we denote the corresponding  fixed point by $Y_n$.
These fixed points were discovered by Yi and Kane.\cite{YiKane}

Let us discuss  the other cases.  If $m_{n-1}$ still does not vanish, and  in
addition some of the other  $m_j$'s don't vanish ($j\leq n-2$), the ratio is of
the form  ${g_{UV}\over g_{IR}}={\sqrt{n}\over 2\cos{\pi\over n+2-k}}$,
indicating a flow to a fixed point  encountered previously for a lower value of
$n$, $Y_{n-k}$. When  $m_{n-1}$  does vanish, while some of the other $m_j$'s
don't ($j\leq n-2$), one gets  a ratio of the form ${g_{UV}\over
g_{IR}}=\sqrt{n\over  n-k}$.  This corresponds to  an IR fixed point where $k$
components of the field have D boundary  conditions, the other ones having N.
We see therefore that, as in the $n=3$ case, no other fixed point is reached,
besides the various $D,N$ combinations, and the $Y_k$ fixed points.

As before, the perturbation of N boundary conditions by the adjoint operator
in  the coset ${SU(2)_n\over U(1)}$ does not preserve the $W_n$ symmetry. A
perturbation  preserving this symmetry is the one with the adjoint in the coset
picture ${SU(n)_1\times SU(n)_1\over  SU(n)_2}$. The perturbing field now has
dimension $\Delta=1-{n\over n+2}$, and the perturbation is integrable.  The
boundary free energy reads as in (\ref{freeenergy}), with now a sum over $n-1$
pseudo energies. The TBA diagram is as in the $SU(3)$ case, but the ``base''
is the Dynkin diagram of $SU(n)$, ie $A_{n-1}$ instead of $A_2$. For more
general cosets ${SU(n)_1\times SU(n)_k\over SU(n)_{k+1}}$,  the diagram looks
as follows
\vskip1cm
\noindent
\centerline{\hbox{
\rlap{\raise60pt\hbox{\hskip.38cm
$\bigcirc$------$\bigcirc$----------$\ldots$----$\bigcirc$--------
$\bigcirc$\hskip.6cm $n-1$}}\rlap{\raise28pt\hbox{\hskip.38cm
$\bigcirc$------$\bigcirc$----------$\ldots$----$\bigcirc$--------$\bigcirc$}}
\rlap{\raise75pt\hbox{$\hskip .3cm 1\hskip4.6cm
k$}}
\rlap{\raise14pt\hbox{\hskip.24cm\Big|$\hskip.95cm\Big|\hskip1.28cm\hskip1.cm
\Big|\hskip1.1cm\Big|$}}
\rlap{\raise50pt\hbox{\hskip.16cm.$\hskip.95cm.\hskip1.3cm\hskip1.cm.
\hskip1.1cm.$}}
\rlap{\raise40pt\hbox{\hskip.04cm.$\hskip.95cm.\hskip1.3cm\hskip1.cm.
\hskip1.1cm.$}}
\hskip-.24cm$\bigotimes$------$\bigcirc$----------$\ldots$----
$\bigcirc$--------$\bigcirc$\hskip.6cm $1$}
}

\bigskip
\noindent
with the obvious generalization for the equations (\ref{tbaeqs}) and
(\ref{freeenergy})
that the row labels now run from 1 to $n-1$. In particular the ratio of
boundary entropies in the UV and IR now reads
\begin{equation}
{g_{UV}\over g_{IR}}={1\over 2}\sum_{i=1}^{n-1} \ln\left(1+x_{i1}\right).
\label{genei}
\end{equation}
For $k=1$ we have
\begin{equation}
x_{i1}=\prod_j \left(1+{1\over x_{j1}}\right)^{-N_{i1,j1}/2}.\label{geneii}
\end{equation}
The solution of this system is ${1\over x_j}=\left({\sin j\pi/n+2\over
\sin\pi/n+2}\right)^2-1$. By some simple manipulations one finds then
\begin{equation}
{g_{UV}\over g_{IR}}={g_Y\over g_D}=2\cos {\pi\over n+2}.\label{geneiii}
\end{equation}
This corresponds to a flow from $Y_n$  to D in the $n-1$ bosons language,
generalizing (\ref{pottsratio}).

As in the $n=3$ case, the key ingredient to understand the phase diagram
is the set of boundary conditions and flows in the $Z_n$ parafermions theory
$SU(2)_n/U(1)$. These theories have known lattice model or quantum spin chains
realizations \cite{FZ}, and it is possible to show, using results in
Ref. (\onlinecite{misc}), that the N and D fixed points correspond
to free and fixed boundary
conditions, while the operator $\Psi_1+\Psi_1^\dagger$ is the most relevant
order parameter, and corresponds again to a longitudinal  boundary  field. The
generalization of the mixed boundary conditions for the ($Z_3$) Potts model to
the $Z_n$ case has not been investigated, to our knowledge.

\subsection{Relation to the Kondo problem}
The emergence of parafermions is also closely related with the n-channel Kondo
problem. Recall\cite{Kondo-review} that this problem involves originally $2n$
Dirac fermions,  which can be bosonized in terms of the current algebras
$SU(2)_n\times SU(n)_2\times U(1)$. Only the spin currents interact with the
boundary spin, leading to a flow entirely  within the $SU(2)_n$ sector, from
the weak coupling fixed point in the UV to the  non Fermi liquid Kondo fixed
point in the IR. The ratio of degeneracy factors is  $g_{UV}/g_{IR}=2/(2\cos
{\pi\over n+2})$.

As is well known, the strong coupling Kondo fixed point does not depend on the
anisotropy, and can be reached as well starting from an anisotropic Kondo
interaction. There is a particularly simple choice of this anisotropy which
corresponds to  the Toulouse limit in the $n=1$ case, or the Emery Kivelson
limit in the  $n=2$ case, where an additional $U(1)$ decouples, and the flow
takes place entirely  within the $SU(2)_n/U(1)$ sector, ie the parafermionic
theory. The perturbation Hamiltonian then reads  \begin{equation}
H_{int}=v\left(\Psi_1^\dagger \sigma^-+\Psi_1\sigma^+\right) \end{equation}   One can
then argue from the integrable analysis that  the IR fixed point - the strong
coupling Kondo fixed point - coincides with the $Y_n$ fixed point identified
previously, the generalization of  the ``mixed'' boundary conditions of the
Potts model to the $Z_n$ case. The UV fixed point - the weakly coupled Kondo
fixed point - does not have a simple interpretation in terms  of the $Z_n$
variables. For the case $n=3$, it coincides with the ``localized on B or C''
fixed point discussed in section IV.  This correspondance was used extensively
by Yi and Kane.\cite{YiKane}

\subsection{Another relation to the Kondo problem: 
a remark on the marginal case}

Returning to the two boson model ($n=3$), we note that 
the case $a^2={8\pi\over 9}$ can also be studied by integrability
techniques, using 
an unexpected mapping onto the four-channel Kondo model. To do this,
recall 
 that it is possible to represent the currents of the $SU(2)_4$ algebra
(that has $c=2$) with two bosons as follows:\footnote{H.S. thanks P. Fendley and N. Warner 
for discussions on this point.}
\begin{eqnarray}
{\cal
J}_x=&\sigma_x\cos(\sqrt{2\pi}\tilde{\phi}_{1L}+\sqrt{6\pi}\tilde{\phi}_{2L})\nonumber\\
{\cal J}_y=&\sigma_y\cos(\sqrt{2\pi}\tilde{\phi}_{1L}-\sqrt{6\pi}\tilde{\phi}_{2L})\nonumber\\
{\cal J}_z=&\sigma_z\cos(\sqrt{8\pi}\tilde{\phi}_{1L})\label{boson}.
\end{eqnarray}
A similar representation was used by Fabrizio and Gogolin \cite{FG} recently;
these authors however did not explicitly consider 
the $\sigma$ operators in their
representation (cocycles), which will play a crucial role in what
follows, though it
doesn't affect their results.
The four channel anisotropic Kondo problem has a boundary interaction
\begin{equation}
H_{int}=J_\bot\left[\tau_x {\cal J}_x(0)+\tau_y{\cal
J}_y(0)\right]+J_{//}
\tau_z {\cal J}_z(0)\label{bdract},
\end{equation}
and is  integrable  for any value of the anisotropy.

Now let us use our bosonization. The boundary action reads then, using
the boundary conditions $\tilde{\phi}_{iL}=\tilde{\phi}_{iR}$ in the UV to introduce
the non chiral fields,
\begin{equation}
J_\bot\left[\tau_x
\sigma_x\cos(\sqrt{\pi\over 2}\tilde{\phi}_1+\sqrt{3\pi\over
2}\tilde{\phi}_2)(0)+\tau_y\sigma_y\cos(\sqrt{\pi\over 2}\tilde{\phi}_1-\sqrt{3\pi\over 2}\tilde{\phi}_2)(0)\right]+J_{//}
\tau_z \sigma_z\cos(\sqrt{2\pi}\tilde{\phi}_1(0)).\label{bdracti}
\end{equation}
Consider now the expansion of the boundary free energy in powers of
$J_\bot,J_{//}$. for every insertion of $\cos\sqrt{2\pi}\tilde{\phi}_1$, we need
one insertion of $\cos(\sqrt{\pi\over 2}\tilde{\phi}_1+\sqrt{3\pi\over 2}\tilde{\phi}_2)$ and
one of $\cos(\sqrt{\pi\over 2}\tilde{\phi}_1-\sqrt{3\pi\over 2}\tilde{\phi}_2)$, contributing a term
$\tau_x\sigma_x\tau_y\sigma_y\tau_z\sigma_z=-1$ to the spin part.
It follows that 
all the spin terms actually disappear, and that the boundary free energy is the
same as the one of the model (\ref{bdracti}) with no spin variables:
%
\begin{equation}
H_{int}= -2J_\bot\cos\sqrt{\pi\over 2}\tilde{\phi}_1\cos\sqrt{3\pi\over
2}\tilde{\phi}_2-J_{//}
\cos\sqrt{2\pi}\tilde{\phi}_1.
\end{equation}
This is nothing but the weak corrugation Hamiltonian of 
section 4 (eq. (4.17)), this time for $a^2={8\pi\over 9}$, where the perturbation is 
marginal.

Using the TBA for the anisotropic Kondo model, the
impurity free energy is simply expressed as
\begin{equation}
F_{imp}=-T\int{d\theta\over2\pi}
{1\over
\cosh(\theta-\theta_B)}\ln\left(1+e^{-\epsilon_1(\theta)/T}\right),
\label{mmoreimp}
\end{equation}
where $T_B\propto e^{\theta_B}$ is the Kondo temperature, and $\epsilon_1$
is a
pseudo energy, solution of the TBA system
\begin{equation}
\epsilon_j=m\ \delta_{4,j}e^\theta- T\sum_k N_{jk}\int {d\theta'\over
2\pi}
{1\over\cosh(\theta-\theta')}\ln\left(1+e^{-\epsilon_k(\theta')/T}\right).
\end{equation}
Here, the labels $k$ run over the nodes of the diagram
\vskip.5cm
\noindent
\centerline{\hbox{\rlap{\raise28pt\hbox{$\hskip-2cm\hskip5.5cm\bigcirc\hskip.25cm t-1$}}
\rlap{\lower27pt\hbox{$\hskip-2cm\hskip5.4cm\bigcirc\hskip.3cm t$}}
\rlap{\raise15pt\hbox{$\hskip-2cm\hskip5.1cm\Big/$}}
\rlap{\lower14pt\hbox{$\hskip-2cm\hskip5.0cm\Big\backslash$}}
\rlap{\raise15pt\hbox{$\hskip-2cm1\hskip1cm 2\hskip1.3cm 4\hskip.8cm t-3$}}
\hskip-2cm$\bigcirc$------$\bigcirc$-- -- --
--$\bigotimes$-- -- --$\bigcirc$------$\bigcirc$\hskip.5cm $t-2$ }}

\bigskip
\noindent
and we have restricted to the simplest values of the anisotropy such that
${t-1\over t}=1-J_{//}$.

Provided $J_\parallel\geq 0$, the perturbation generates 
a flow (the situation here is quite different from the 
one boson theory, where the boundary cosine interaction is 
truly marginal). Independently of the value of $J_\parallel$, one finds
$x_1=e^{-\epsilon_1(-\infty)/T}=3$, while
$y_1=e^{-\epsilon(\infty)/T}=2$.  The ratio ${g_{UV}\over
g_{IR}}={2\over \sqrt{3}}$, the well
known value for the 4 channel Kondo model. This coincides with
the ratio ${g_N\over g_D}$  (4.30) for our  values of the coupling constant.For $J_\parallel\leq 0$  (the equivalent of 
$v_A<0$ in section 4),
meanwhile, the Kondo  perturbation is irrelevant, and no flow is generated 
Hence, we conclude that for $\Delta_N=1$ (4.19), there is no new
fixed point, 
and presumably the  Y
fixed point
becomes identical with the N one in the limit $a^2={8\pi\over 9}$. 
An $\epsilon$-expansion around this marginal point was developed by 
Kane and Fisher.\cite{Kane-Fisher}

\section{Conclusion}

We have established  the general connection between 
a simple model for QBM and boundary
CFT with considerable attention paid to the issue of boson compactness.  
For a simple Hamiltonian we have conjectured a complete phase diagram, 
which is analogous to that of the boundary Potts model. These results
were obtained by using a sort of ``boundary embedding'' involving
the Potts model and the fusion approach to generating boundary states.  
We have also mentioned other fixed points which occur with more
general Hamiltonians and which arise from fusion in the other
sectors of the conformal embedding.  We have discussed
the integrability of  some of the RG flows.

We note
that the $Y$ b.c. is more dynamical or ``quantum'' than D or N.  
Within the weak corrugation formulation, while the potential term
$v_A$ of Eq. (\ref{VAa}) is relevant, it does not simply 
``pin'' the boson fields at one of its minima; rather they fluctuate
between the two inequivalent minima.  This is only possible
for a range of lattice spacing, $8\pi /9<a^2<2\pi$ where a semi-classical
analysis fails.  

We have given an explicit demonstration that other, highly non-trivial
b.c.'s are possible for 2
free periodic bosons besides D and N and their variants.  However, since this
demonstration rests on the conformal embedding and fusion,  and since we don't
have a conformal embedding for general values of the radius parameter, $a$, our
construction only works at one special value of $a$.  A more straightforward
understanding of these b.c.'s, directly in terms of the bosons eludes us.
Thus this general problem, known since 1992\cite{Furusaki,Kane-Fisher}
remains open.  

\acknowledgements
We would like to thank M.P.A. Fisher, A.W.W. Ludwig, C. Schweigert,
A. Zawadowski and J. B. Zuber
for helpful discussions.  This research was begun while all three authors
were visitors at the Institute for Theoretical Physics,
Santa Barbara, in July, 1997.
H.S. also acknowledges the hospitality of
the SPhT Saclay.
This research is supported in part by the NSF under grant No. PHY-94-07194,
NSERC of Canada  (I.A.), Grant-in-Aid
from Ministry of Education, Science, Sports and Culture
of Japan (M.O.), and by the DOE and
the NSF under the NYI program (H.S.).

\end{document}